\documentclass[a4paper,11pt]{article}

\usepackage{jheppub,amsmath,amssymb,amsfonts,amsxtra, mathrsfs, makeidx,graphics,graphicx,amsthm,epsfig, youngtab,bm,longtable,float,tikz}
\usepackage[margin=1cm]{caption}
\usetikzlibrary{positioning}
\newcommand{\ba}{\begin{array}}
\newcommand{\ea}{\end{array}}
\newcommand{\bi}{\begin{itemize}}
\newcommand{\ei}{\end{itemize}}
\def\vec#1{\bm{#1}}
\def\bea#1\eea{\allowdisplaybreaks \begin{align}#1\end{align}}
 \newcommand{\ben}{\begin{enumerate}}
\newcommand{\een}{\end{enumerate}}
\newcommand{\bean}{\begin{eqnarray*}}
\newcommand{\eean}{\end{eqnarray*}}
\newcommand{\eref}[1]{(\ref{#1})}

\newcommand{\nn}{\nonumber}

\newcommand{\PE}{\mathop{\rm PE}}

\newcommand{\BC}{\mathbb{C}}

\newcommand{\BZ}{\mathbb{Z}}

\newcommand{\comment}[1]{}

\newcommand{\CL}{{\cal L}}

\newcommand{\CM}{{\cal M}}

\newcommand{\CW}{{\cal W}}
\newcommand{\CN}{{\cal N}}

\newcommand{\CH}{{\cal H}}
\newcommand{\CZ}{{\cal Z}}

\newcommand{\ie}{{\it i.e.}}
\newcommand{\eg}{{\it e.g.}}

\newcommand{\ud}{\mathrm{d}}

\newcommand{\cN}{\mathcal{N}}
\newcommand{\ff}{\mathcal{F}^\flat}

\newcommand{\Secref}[1]{Section~\ref{#1}}
\newcommand{\secref}[1]{Sec.~\ref{#1}}

\newcommand{\tabref}[1]{Tab.~\ref{#1}}

\newcommand{\figref}[1]{Fig.~\ref{#1}}
\renewcommand{\eqref}[1]{(\ref{#1})}

\title{Mirror Symmetry in Three Dimensions via Gauged Linear Quivers}

\author[a]{Anindya Dey,}
\author[b]{Amihay Hanany,}
\author[c]{Peter Koroteev,}
\author[d]{and Noppadol Mekareeya}
\affiliation[a]{Theory Group, Department of Physics and Texas Cosmology Center\\ The University of Texas at Austin, Austin, TX}
\affiliation[b]{Theoretical Physics Group, The Blackett Laboratory, Imperial College London\\ London, United Kingdom}
\affiliation[c]{Perimeter Institute for Theoretical Physics \\Waterloo, ON}
\affiliation[d]{Theory Group, Physics Department, CERN\\ Geneva, Switzerland}
\emailAdd{anindya@physics.utexas.edu}
\emailAdd{a.hanany@imperial.ac.uk}
\emailAdd{pkoroteev@perimeterinstitute.ca}
\emailAdd{noppadol.mekareeya@cern.ch}

\abstract{Starting from mirror pairs consisting only of linear (framed A-type) quivers, we demonstrate that a wide class of three-dimensional quiver gauge theories with $\CN=4$ supersymmetry and their mirror duals can be obtained by suitably gauging flavor symmetries. Infinite families of mirror pairs including various quivers of $D$ and $E$-type and their affine extensions, star-shaped quivers, and quivers with symplectic gauge groups may be generated in this fashion. We present two different computational strategies to perform the aforementioned gauging procedure -- one of them involves $\mathcal{N}=2^{*}$ classical parameter space description, while the other one uses partition functions of the $\mathcal{N}=4$ theories on $S^3$. The partition function, in particular, turns out to be an extremely efficient tool for implementing this gauging procedure as it readily generalizes to arbitrary size of the quiver and arbitrary rank of the gauge group at each node. For most examples of mirror pairs obtained via this procedure, we perform additional checks of mirror symmetry using the Hilbert series.}

\preprint{UTTG-36-13, TCC-031-13, CERN-PH-TH/2013-279}

\begin{document}
\maketitle

\section{Introduction and Main Results}\label{Sec:Intro}
In this work we discuss mirror symmetry \cite{Intriligator:1996ex, deBoer:1996mp, Hanany:1996ie} for $\mathcal{N}=4$ three-dimensional quiver gauge theories. These theories have been extensively studied in the literature for various types of quivers \cite{Hanany:1996ie,Hanany:1997gh, Hanany:1999sj, Gaiotto:2008ak, Dey:2013nf,Gaiotto:2013bwa}. Among other interesting things, such theories provide a rich laboratory for studying dualities in supersymmetric QFTs. For three dimensional $\mathcal{N}=4$ theories mirror symmetry is a particularly important duality, which involves two or more theories with completely different UV description flowing to the same superconformal point in the IR. Our aim in this paper is to demonstrate that  mirror symmetry for a wide class of $\mathcal{N}=4$ quiver gauge theories is connected in a very interesting fashion to mirror symmetry in linear quivers. 

  Three-dimensional mirror symmetry interchanges Coulomb and Higgs branches of the theory. Clearly this is a very nontrivial mapping. The Higgs branch, where the gauge group is generically broken completely, is a hyper-K\"{a}hler quotient given by the zero locus of the triplet of  $\mathcal{N}=4$ D-terms divided by the gauge group. The metric on the Higgs branch is protected against quantum corrections. On the Coulomb branch, where the gauge group is broken to its maximal torus, a generic classical point is characterized by the scalar vevs of the triplet of scalars in a $\mathcal{N}=4$ vector multiplet and the dual scalar. It is a hyper-K\"{a}hler manifold whose metric receives large quantum corrections. The equivalence of the Higgs branch of theory A with the Coulomb branch of theory B under mirror symmetry immediately implies that the FI parameters of theory A must be linearly related to the $\mathcal{N}=4$ mass parameters of theory B \cite{deBoer:1996mp}. This linear relation between the two sets of quantities is known as the ``mirror map'' and constitutes one of the fundamental pieces of information associated with a given mirror pair.

It was pointed out fairly early \cite{Hanany:1996ie} that mirror symmetry is a direct consequence of S-duality. Therefore reading off the data of the dual of a theory which admits a Hanany-Witten description (branes plus perturbative objects like orbifolds, orientifolds etc.) is, in principle, a solved problem. However, even in this category of examples, the answer may not be very satisfactory -- the S-dual configuration may give rise to a so-called ``bad'' or ``ugly'' quiver which, if treated naively, does not flow to a unitary theory in the infrared. For ``bad'' theories there is however a resolution: the RG flow organizes itself in such a way that a proper number of matter fields acquire minimal R-charges and therefore become effectively free. The theory with those matter multiplets removed is no longer ``bad''. Note, however, that a ``good'' dual of a ``bad'' theory (3d version of the Seiberg duality \cite{Aharony:1997gp,Giveon:2008zn,Yaakov:2013fza}) may also be problematic to identify. For the large class of quiver gauge theories, which do not admit any brane description \cite{Gaiotto:2008ak}, the identification of the mirror dual becomes much more intricate.

The main players in our story are parameter spaces of their supersymmetric vacua $\CL$ \cite{Nekrasov:2009uh,Dimofte:2011ju,Dimofte:2011py} and their ``quantizations'' -- three-sphere partition function $\CZ_{S^3}$ \cite{Kapustin:2009kz, Kapustin:2010xq, Hosomichi:2010vh,Hama:2011ea,Hama:2010av} and Hilbert series $H(t)$ \cite{Gadde:2011uv,Hanany:2010qu,Benvenuti:2010pq,Cremonesi:2013lqa} on the Coulomb branch and the Higgs branch of a given theory. 

In \cite{Gaiotto:2013bwa}, mirror symmetry in $\mathcal{N}=4$ quiver gauge theories of the linear $A_L$ type was analyzed after mass-deforming the original theories to $\mathcal{N}=2^*$ (by turning on mass deformations conjugate to the diagonal U(1) subgroup of the $SU(2)_L \times SU(2)_R$ R-symmetry) and compactifying on a circle. The parameter spaces of the supersymmetric vacua $\CL$, which can be thought of as symplectic Lagrangian submanifolds inside the complex vector space of all canonical mass parameters, were identified for all linear quiver theories and their mirror duals.

The parameter space $\CL$ of massive vacua is one of the basic protected quantities of a theory. There are certainly more sophisticated gadgets which are extensively used in the literature, namely partition functions on various 3-manifolds \cite{Kapustin:2009kz, Hama:2011ea} and 
superconformal indices of different kinds \cite{Kim:2009wb, Imamura:2011su,Krattenthaler:2011da, Kapustin:2011jm}. In particular, partition function on a round sphere turn out to be an extremely effective tool for studying dualities in three dimensions. For example, mirror symmetry in a large class of affine D-type quiver gauge theories was analyzed in \cite{Dey:2013nf, Dey:2011pt} using partition functions of such theories on round sphere.

Another important object that can be used to check three dimensional dualities like mirror symmetry is the Hilbert series -- a generating function which counts chiral operators on the moduli spaces of gauge theories with respect to some specific $U(1)$ charge. Explicit formulae for Hilbert Series on the Higgs branch have been known for quite sometime \cite{Hanany:2010qu,Benvenuti:2010pq}. Recently, analogous formulae for the Coulomb branch of $\mathcal{N}=4$ theories were found \cite{Cremonesi:2013lqa}. Comparison of the Higgs branch Hilbert series of a given theory and the Coulomb branch Hilbert series of the mirror  gives yet another way to check the mirror symmetry.

The theme of this paper, however, is slightly different from the body of work \cite{Kapustin:2010xq, Dey:2013nf, Dey:2011pt, Cremonesi:2013lqa} where much emphasis was placed on checking mirror symmetry for various families of quiver gauge theories. In this work we demonstrate that a large class of quiver gauge theories and their mirror duals, including various avatars of $D$ and $E$ type quivers and their affine extensions, star-shaped quivers and quivers with $Sp(N)$ gauge groups, may be constructed by starting from a mirror pair of linear quivers and gauging appropriate global symmetries on one side of the duality. The operation of gauging flavor symmetries in a linear quiver to obtain a more complicated quiver is relatively straightforward. However, one needs to understand the resultant ``ungauging" on the other side of the duality to derive the correct mirror using this procedure. We present two concrete computational strategies for implementing this gauging/ungauging procedure - one of them uses the $\mathcal{N}=2^{*}$ classical moduli space description while the other uses partition functions of the $\mathcal{N}=4$ theories on $S^3$. The method which uses the $S^3$ partition function is particularly convenient since it generalizes easily to arbitrary size of the quiver and arbitrary rank of the gauge group. In addition, the partition function method gives a straightforward recipe to derive the mirror map for a given pair of mirror duals obtained via this gauging procedure. For most examples of mirror pairs constructed in the fashion described above, we perform additional checks of mirror symmetry using Hilbert series.

The paper is organized as follows. In \Secref{Sec:N4setup} we shall review how several families of three-dimensional linear quiver gauge theories with $\CN=4$ supersymmetry arise from brane constructions and how the mirror symmetry acts on them via the S-duality. We shall also review the parameter space of massive vacua for $A_L$ quivers with canonical mass deformations including the $\CN=4$ supersymmetry breaking mass parameter. Finally, we shall introduce the basics of $S^3$ partition function and Hilbert series that will be needed in the rest of the paper. Some key illustrative examples of the gauging method will be presented in \Secref{Sec:GaugingBasic}. The reader who is familiar with the basics of mirror symmetry in three dimensions may start reading the paper directly from \secref{Sec:GaugingBasic}. The rest of the manuscript from \Secref{Sec:Dquivers} through \Secref{Sec:Spmirrors} consists of detailed derivations of the corresponding mirror pairs using the gauging procedure. 

\subsection{Open Questions}
Some aspects of the 3d mirror symmetry were left beyond the scope of the present paper. We would like to name a few of them below. We hope to address some of these problems in the near future.

One important class of theories missing from our analysis are quiver gauge theories which follow from brane constructions involving $O3$ planes. The present paper only deals with $O5$ mirrors. Including $O3$ planes will allow us to study quivers with orthogonal/symplectic gauge groups in addition to the examples we have covered here. Embeddings of $SO$ groups inside unitary groups should be realized on the level of the parameter space of supersymmetric vacua and the partition function, very much along the lines of \secref{Sec:Spmirrors}, where the analogous embedding for symplectic groups was discussed.

Our computations of Coulomb branch Hilbert series in this paper are performed along the lines of \cite{Cremonesi:2013lqa}. There is, however another form of the Coulomb branch series, namely the one involving Hall-Littlewood polynomials. These two methods together provide an efficient way to compute the Coulomb branch Hilbert series for a large class of theories including those with non-Lagrangian mirrors. These computations will be addressed elsewhere.

We also leave the discussion of implementations of gauging/ungauging to the dual integrable models for future work. Recall that each 3d quiver with $\CN=2^*$ supersymmetry corresponds to a XXZ spin chain of certain length with certain number of Bethe roots at each level of nesting \cite{Nekrasov:2009uh}. In \secref{Sec:Spmirrors} we show that upon a non-Abelian gauging the Coulomb branch of the mirror theory changes dramatically, in particular a quiver `tail' shrinks down to a single node. It would be nice to interpret this phenomenon using the spin chain language, i.e. what happens with the higher level excitations and with the spin chain S-matrix. 

In this work we only regard quiver theories with $\CN=4$ supersymmetry, which is softly broken to $\CN=2^*$. It would be interesting to consider more generic $\CN=2$ quiver theories. Hopefully, some of the results can be easily obtained from our construction by taking certain degenerate limits such that some matter fields will get decoupled. Another modification of our scenario may be carried out by introducing (untwisted) superpotential couplings in the UV Lagrangian of the quiver theory. We do believe that for  specific superpotential deformations our results can be applied almost directly without significant changes.

\subsection{Summary Tables}
Here we present a summary of some of the important quivers we discuss in this paper together with their mirror duals. We refer the reader to the main text for the details on notations and conventions. 

There are several tables below:  \tabref{Tab:StarD} and \tabref{Tab:StarD1} list star-shaped quivers and D-type quivers,  \tabref{tab:QuiversSummary2} lists E-type and uneven star-shaped quivers\footnote{One of the quivers in the second table does not have any global symmetry; we thereby assume that its Coulomb branch is defined as a $U(1)$ quotient of the products of all its gauge groups.}, and  \tabref{tab:QuiversSummary3} shows mirrors for $Sp(N_c)$ theories. Some notations: numbers inside circle nodes denote ranks of unitary gauge groups, numbers inside box nodes denote ranks of global symmetry groups, `A' in rows four and five designate matter transforming in antisymmetric power of the fundamental representation of the group it is charged under.

We refer to mirror duals in these table as `A-model' and `B-model' which should be simply understood as a way of labeling the dual theories. We emphasize that this terminology is in no way connected to the 2d (homological) mirror symmetry.

Note that most of the mirror duals from the table below are already known.\footnote{The newly discovered ``good'' mirrors for double framed $\hat{D}$ quivers are displayed in Tab.~\ref{Tab:StarD1}.} In this work we focus more on viewing the physics of these quivers through the prism of linear quivers and their mirrors rather than establishing new mirror pairs. As we show later in the text that for each quiver from the table there is a direct connection between its BPS protected quantities (parameter space of SUSY vacua and $S^3$ partition function) and similar BPS objects for some linear quivers. We however admit that we do not possess an exhaustive classification of all quivers of this type (which can be obtained by gauging global symmetries of some linear quiver). It is a challenging task to provide such classification. 

\begin{center}
\begin{table}[h]
\begin{tabular}{|c|c|c|}
\hline
 A-model & B-model  & Location in the text \\
\hline \hline 
\includegraphics[scale=0.3]{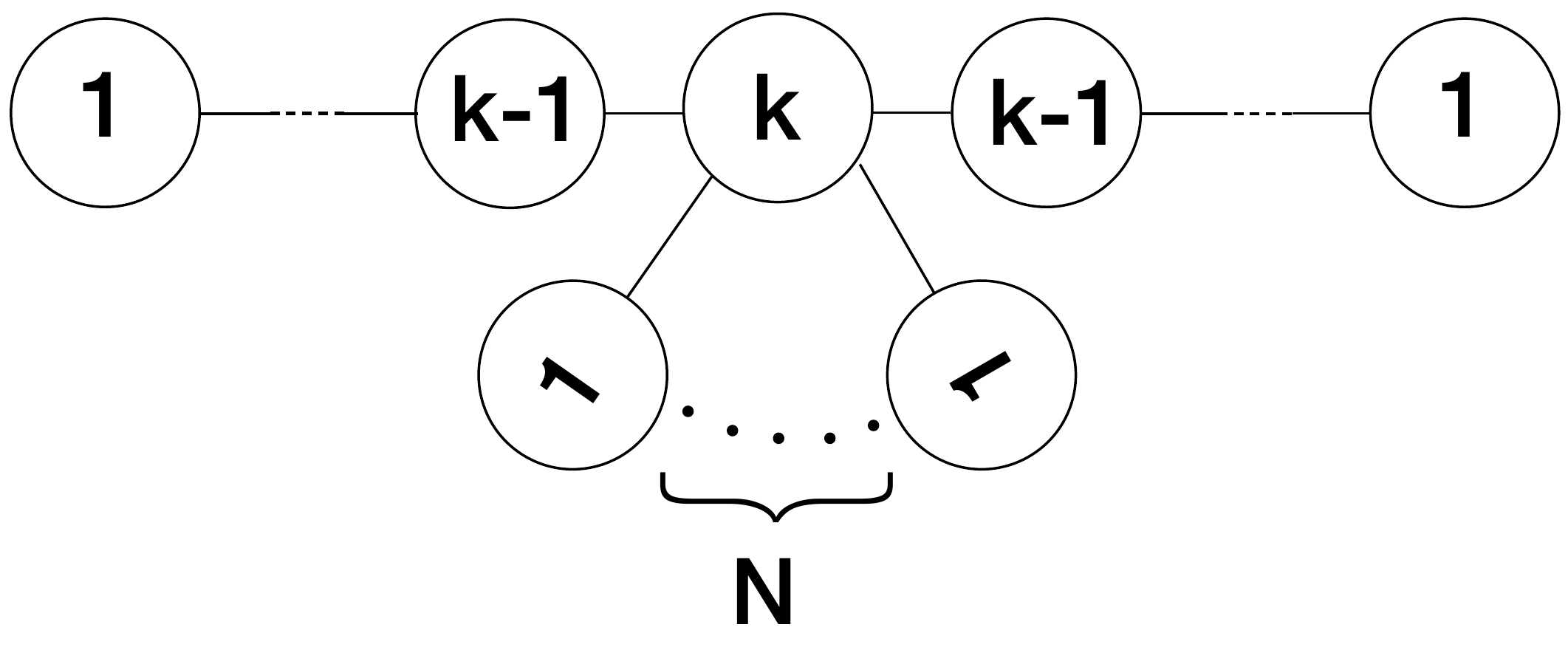} & \includegraphics[scale=0.25]{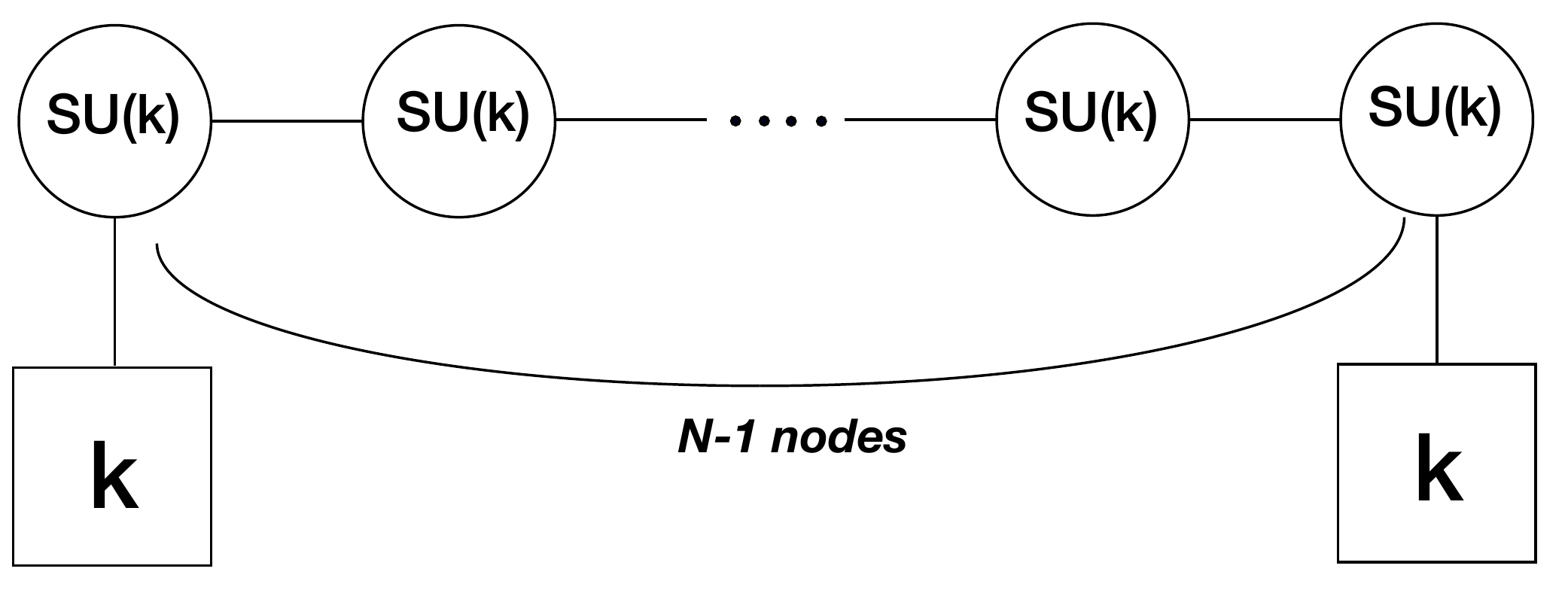} & \secref{star-quiver}, \figref{fig:quiver121new} \\
\hline
\includegraphics[scale=0.3]{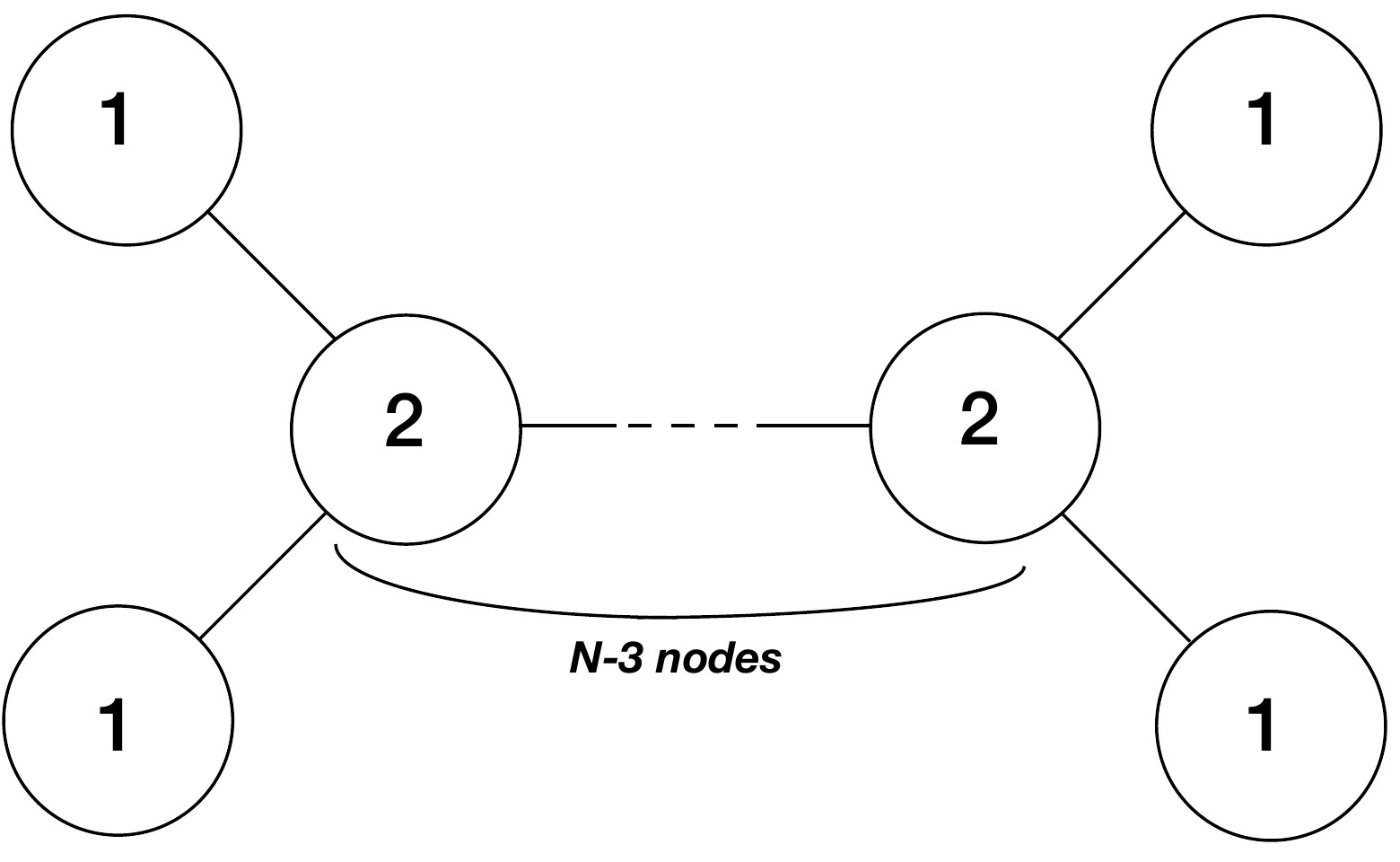} & \includegraphics[scale=0.5]{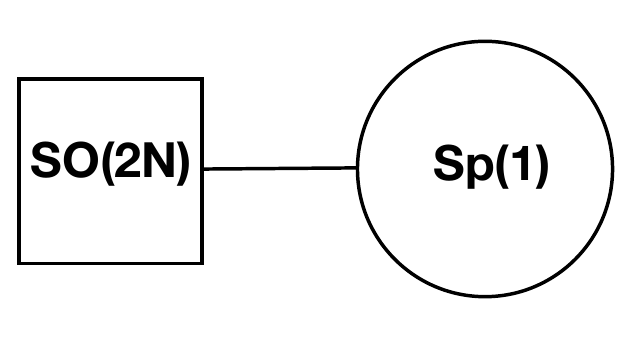} & \secref{sec:DhatSing} \\
\hline
\includegraphics[scale=0.3]{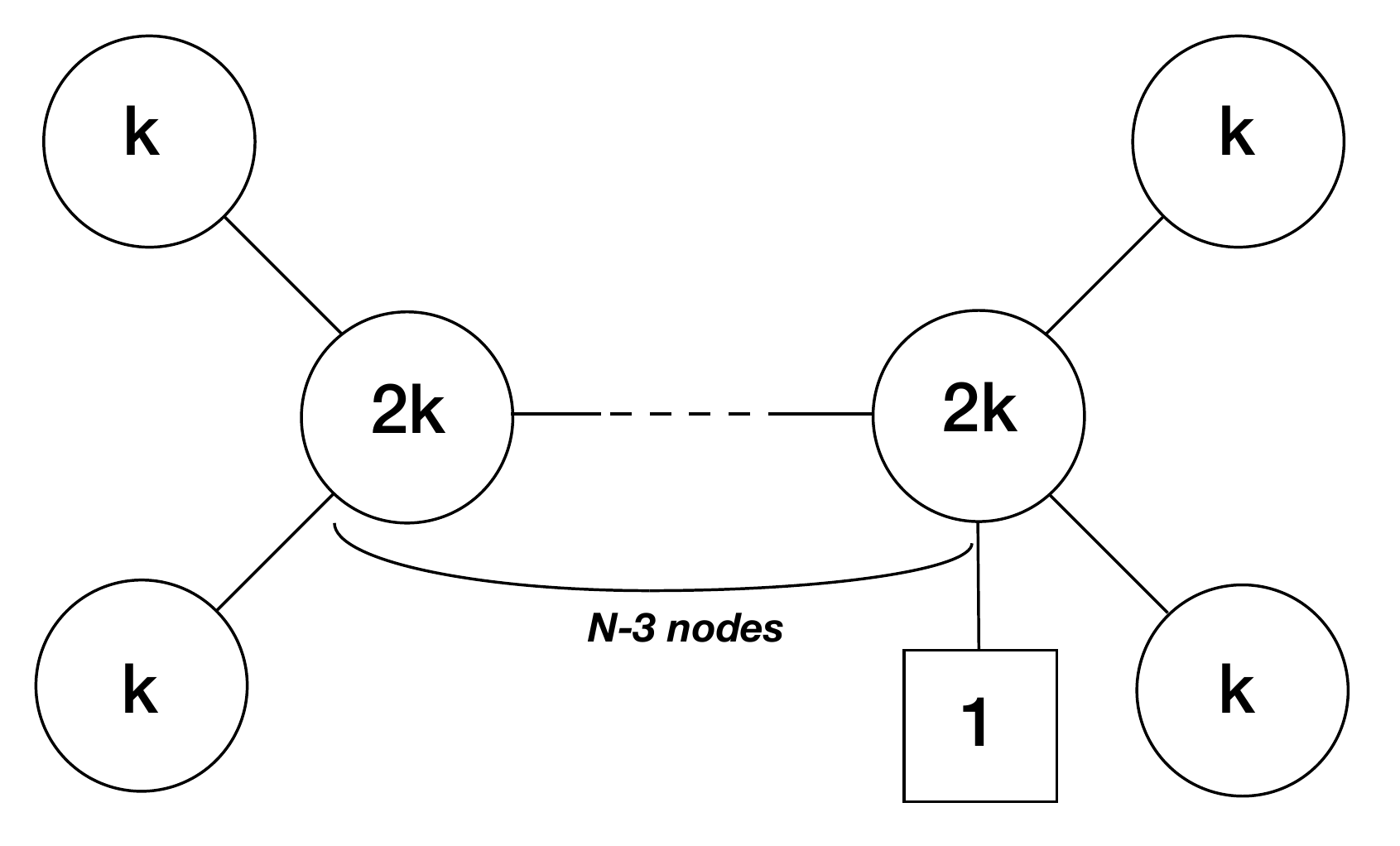} & \includegraphics[scale=0.5]{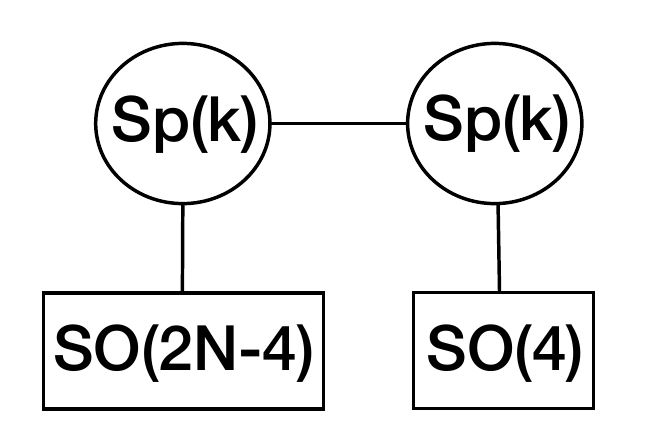} & \secref{sec:DhatSing}, \figref{fig:QuiverflavoredD41} \\
\hline
\includegraphics[scale=0.3]{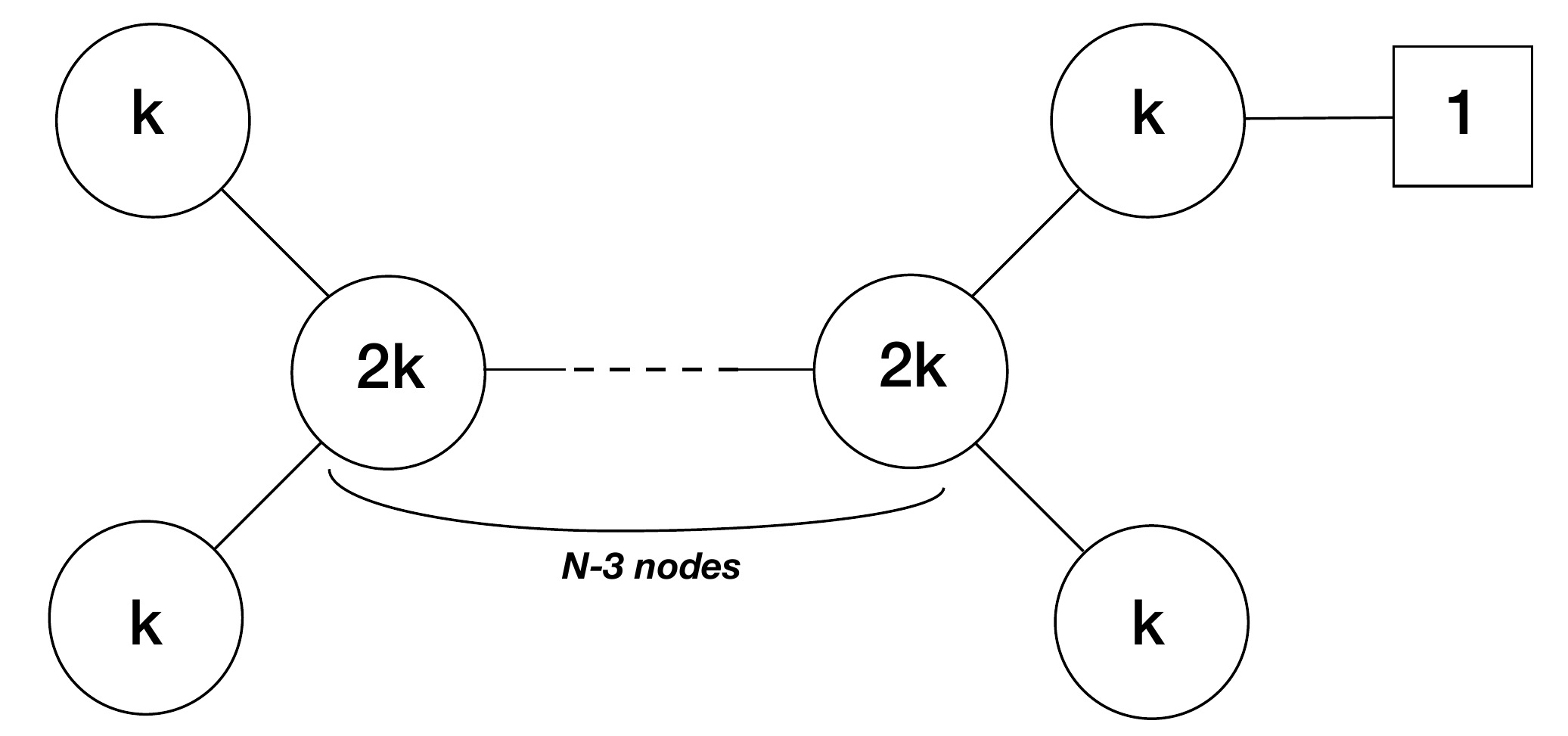}    & \includegraphics[scale=0.5]{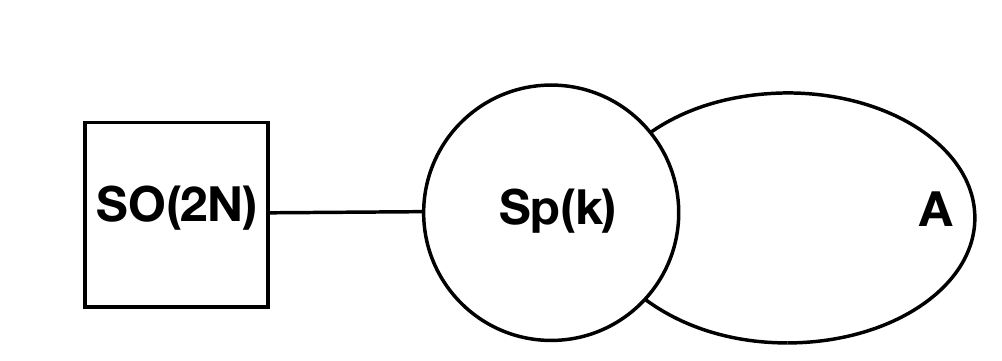} & \secref{sec:DhatSing}, \figref{fig:QuiverflavoredD42}\\
\hline
\includegraphics[scale=0.3]{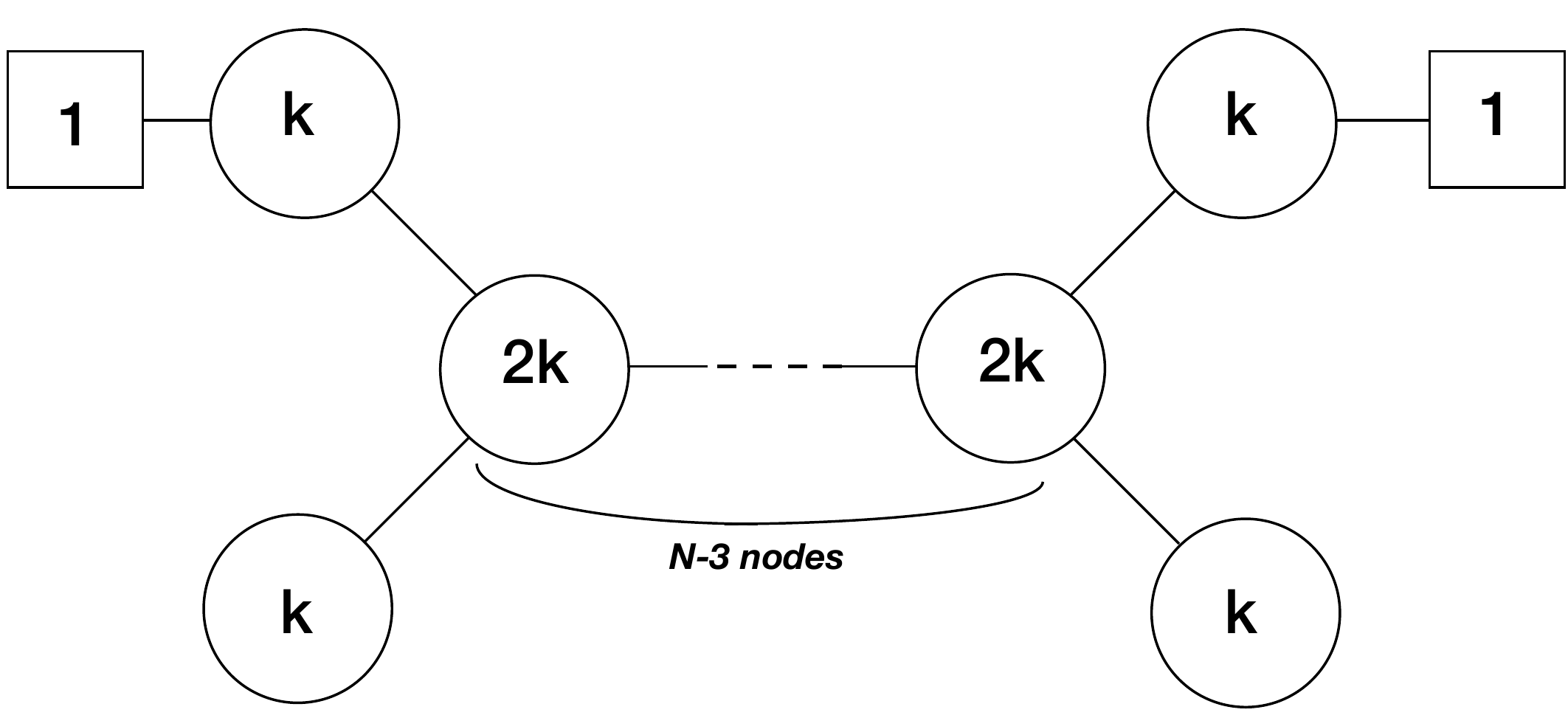}    & \includegraphics[scale=0.5]{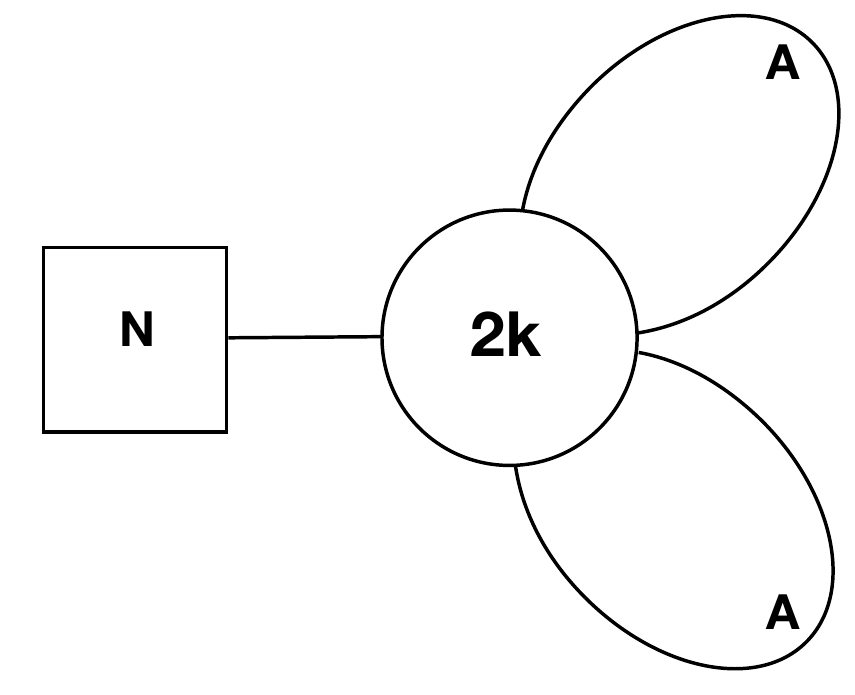} & \secref{star-quiver}, \figref{fig:QuiverflavoredD43}, (c)\\
\hline 
\includegraphics[scale=0.3]{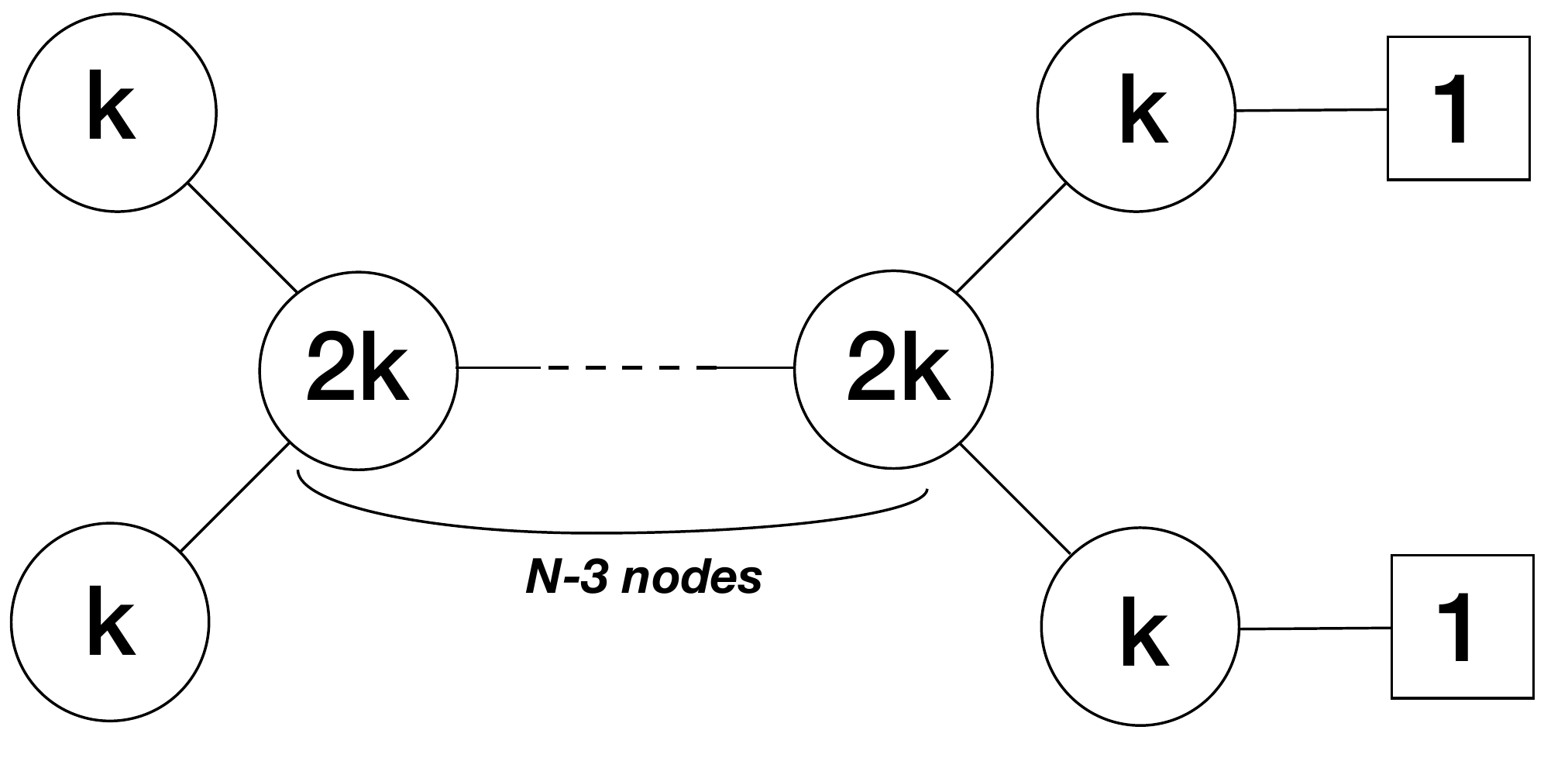}    & \includegraphics[scale=0.5]{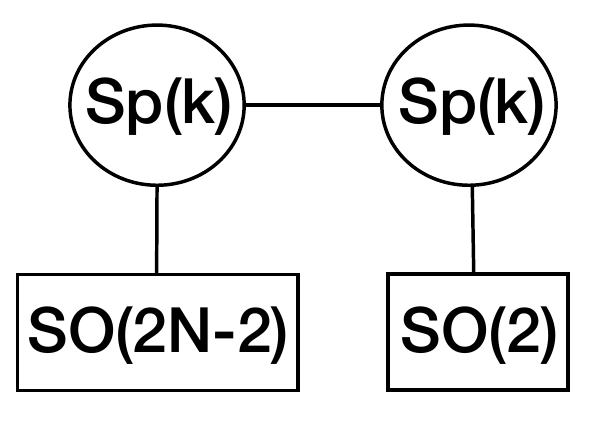} & \secref{star-quiver}, \figref{fig:QuiverflavoredD43}, (b) \\
\hline
\end{tabular}
\caption{Summary table of star and D-shaped quivers and their mirrors.}
\label{Tab:StarD}
\end{table}
\end{center}

\begin{center}
\begin{table}
\begin{tabular}{|c|c|c|}
\hline
 A-model & B-model  & Location in the text \\
\hline \hline 
\includegraphics[scale=0.3]{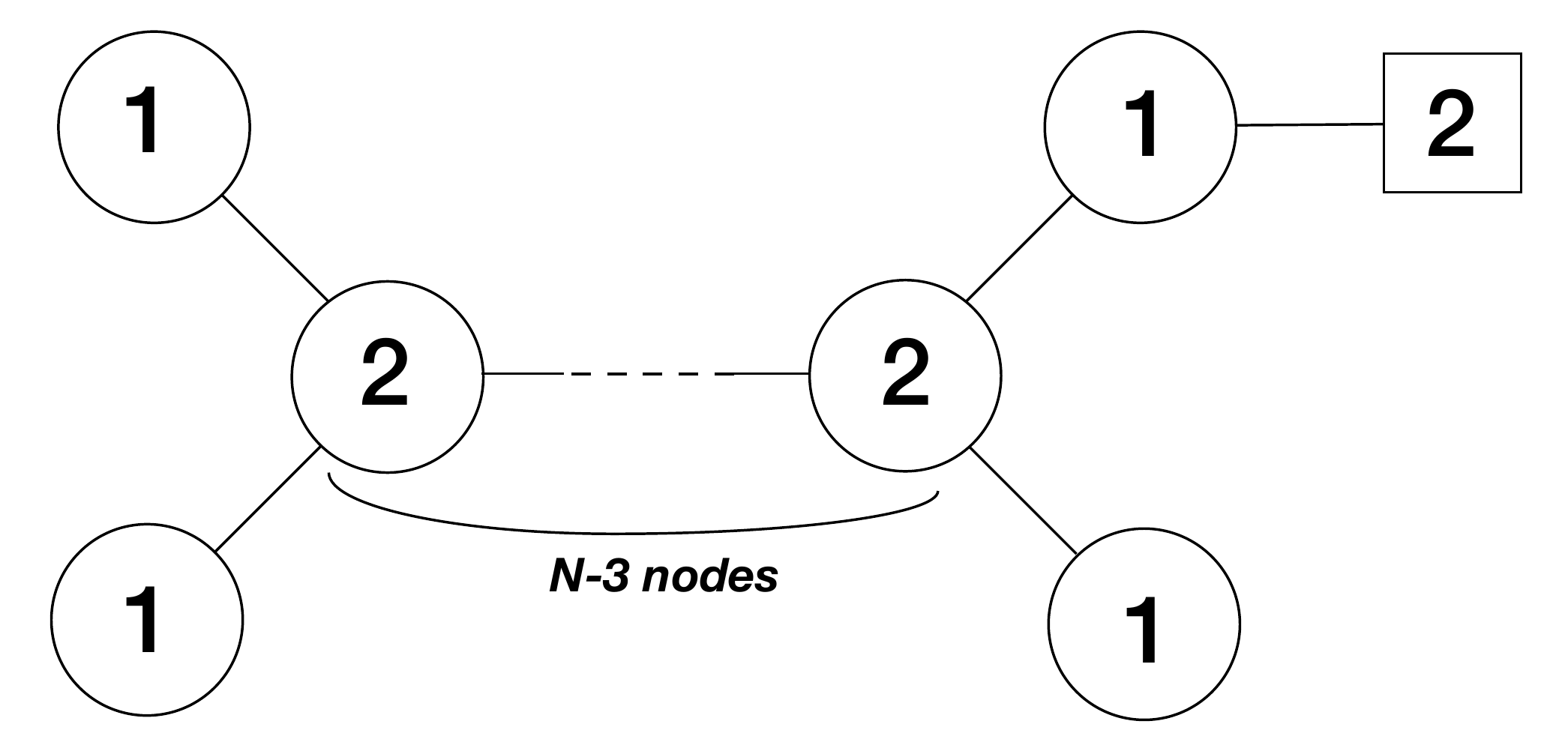}    &\includegraphics[scale=0.35]{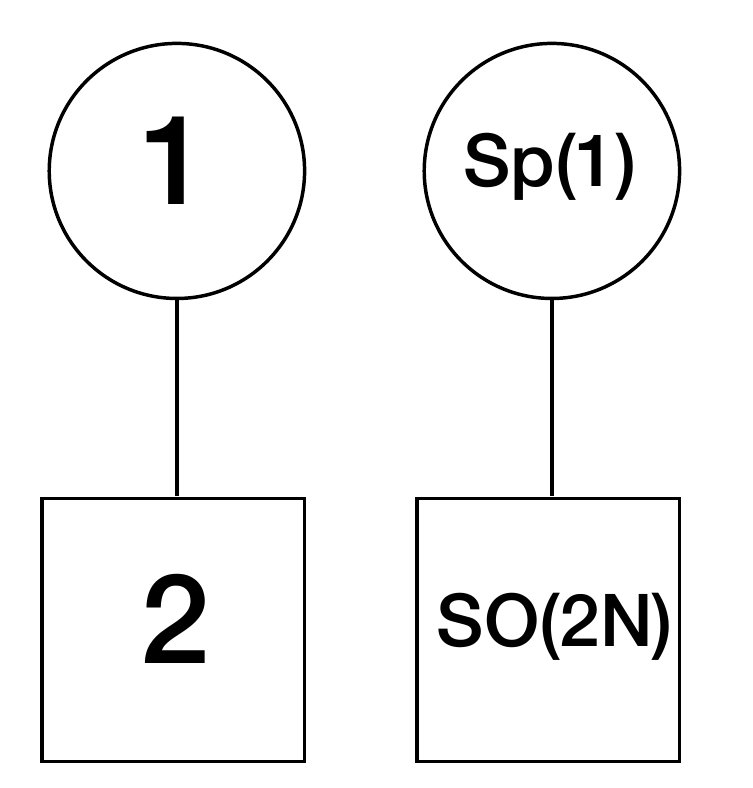} & \secref{sec:DhatDouble}, \figref{fig:QuiverflavoredD44}\\
\hline
\includegraphics[scale=0.3]{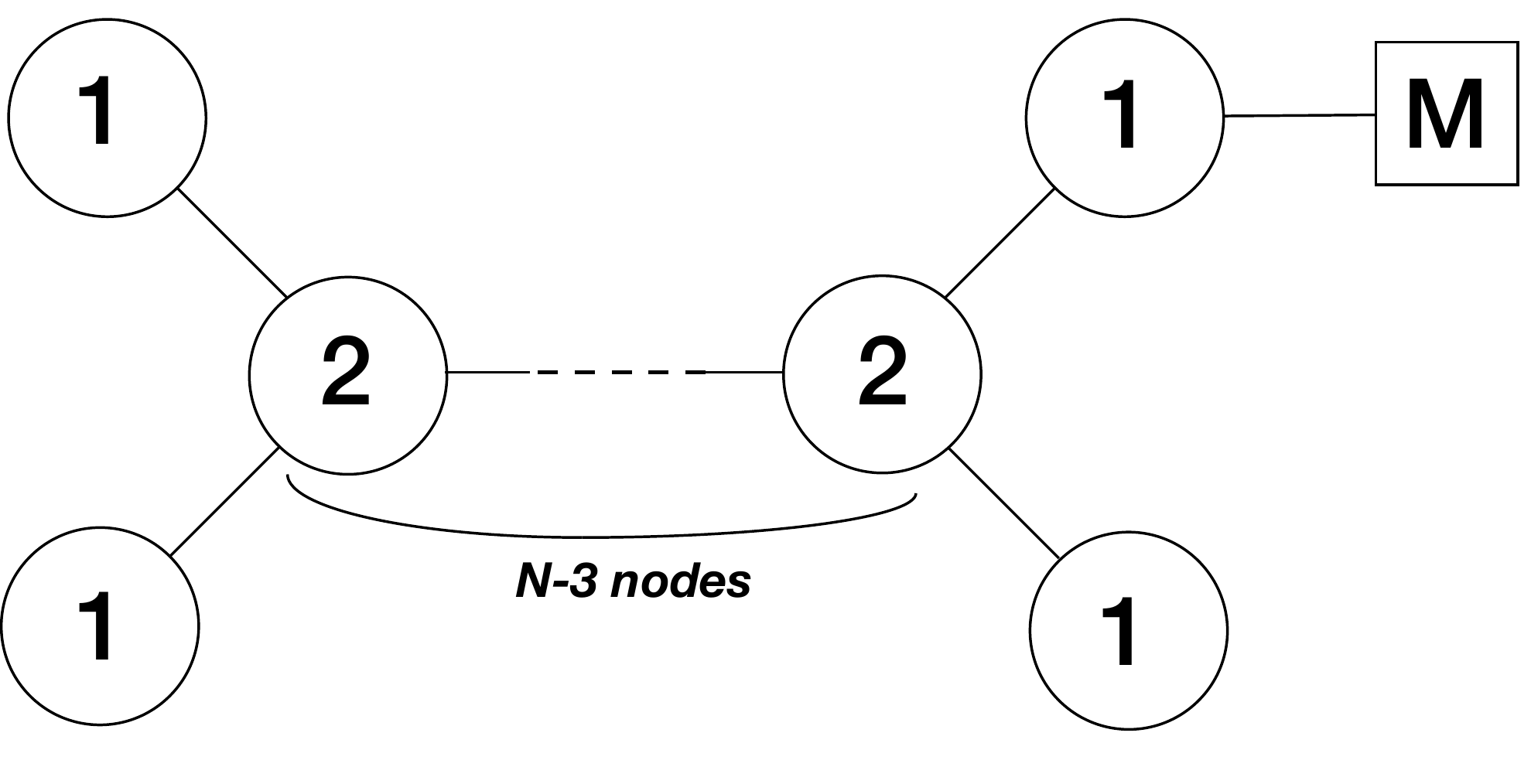}    & \includegraphics[scale=0.3]{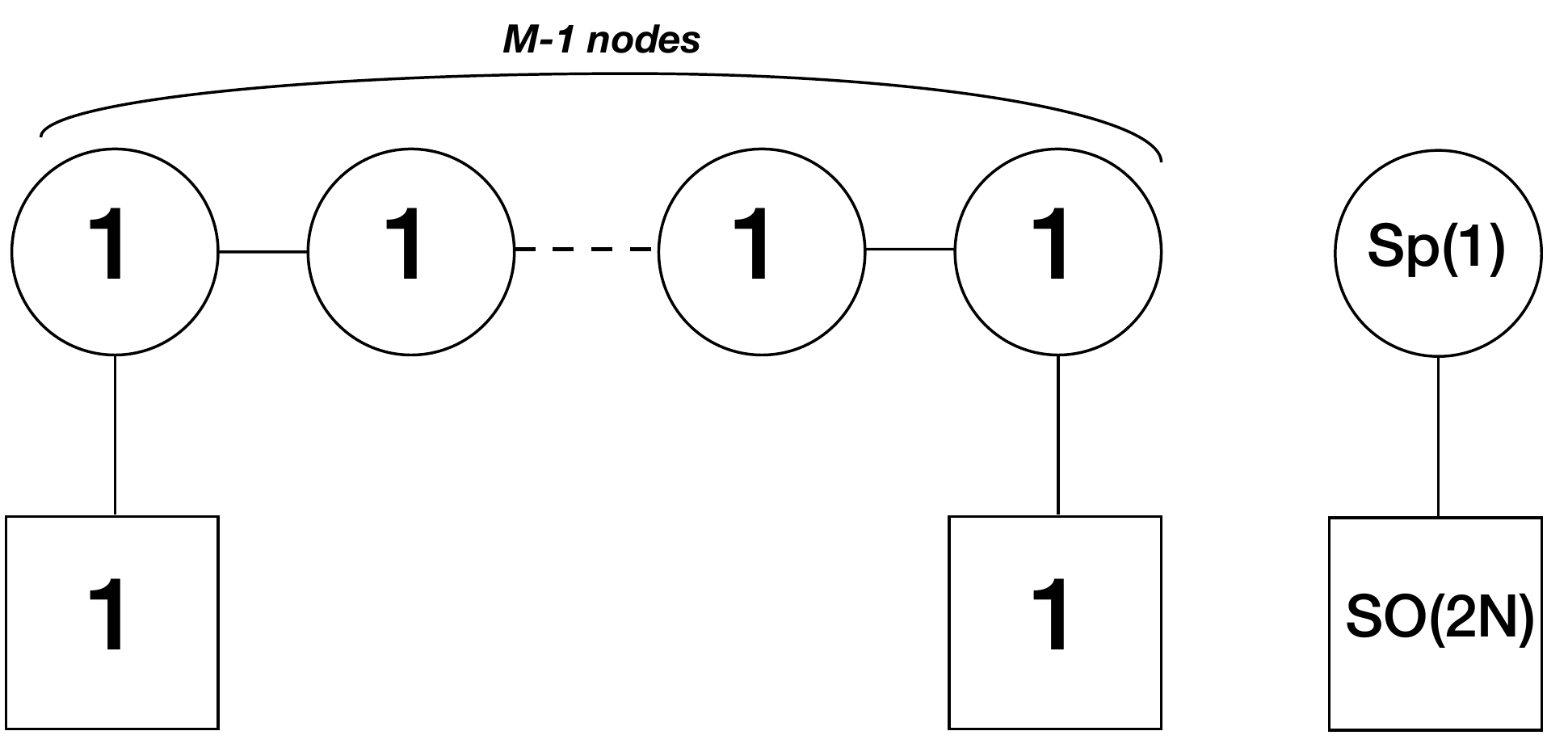} & \secref{sec:DhatDouble}, \figref{fig:Dk1NfM}\\
\hline
\end{tabular}
\caption{Summary table of D-shaped quivers and their mirrors (continued).}
\label{Tab:StarD1}
\end{table}
\end{center}

\begin{center}
\begin{table}
\begin{tabular}{|c|c|c|}
\hline
 A-model & B-model  & Location in the text \\
\hline \hline
\includegraphics[scale=0.8]{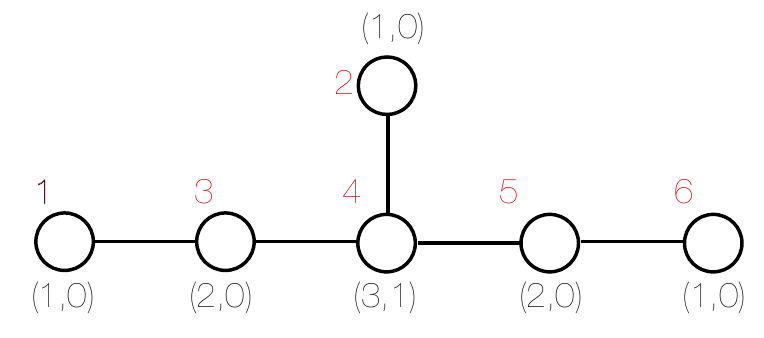} & \includegraphics[scale=0.3]{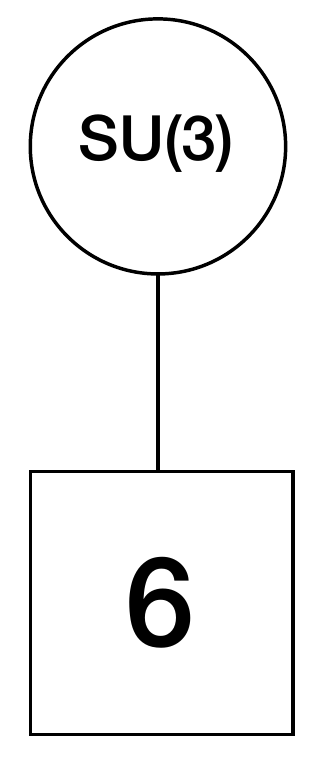} & \secref{Sec:Emirrors}, \figref{fig:E6} \\
\hline
\includegraphics[scale=0.7]{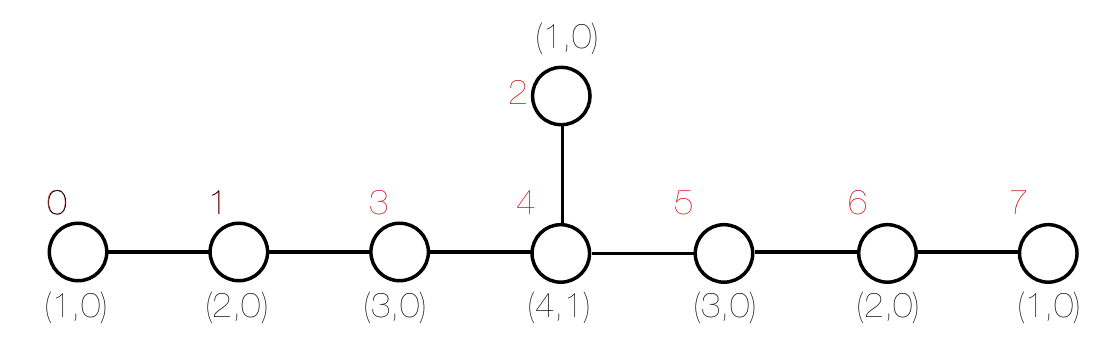} & \includegraphics[scale=0.3]{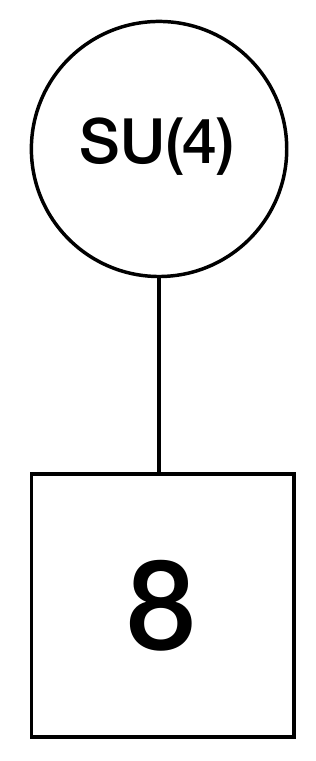} & \secref{Sec:Emirrors}, \figref{fig:E7Aff} \\
\hline
\includegraphics[scale=0.6]{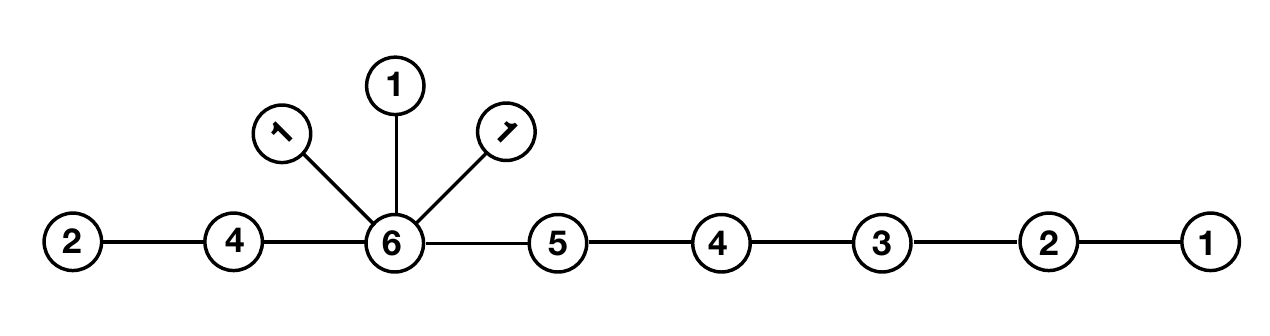} & \includegraphics[scale=0.3]{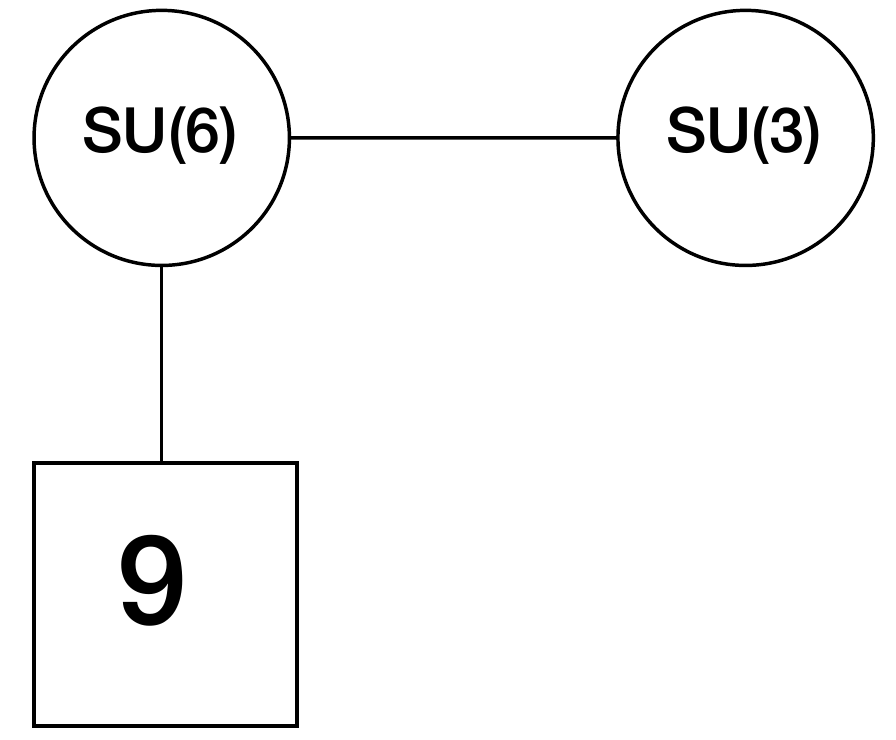} & \secref{Sec:Emirrors}, \figref{fig:E8Aff3} \\
\hline
\end{tabular}
\caption{Summary table of star and E-shaped quivers and their mirrors.}
\label{tab:QuiversSummary2}
\end{table}
\end{center}

\begin{center}
\begin{table}
\begin{tabular}{|c|c|c|}
\hline
 A-model & B-model  & Location in the text \\
\hline \hline
\includegraphics[scale=0.4]{QuiverMirrorO5SpNcNfflv} & \includegraphics[scale=0.4]{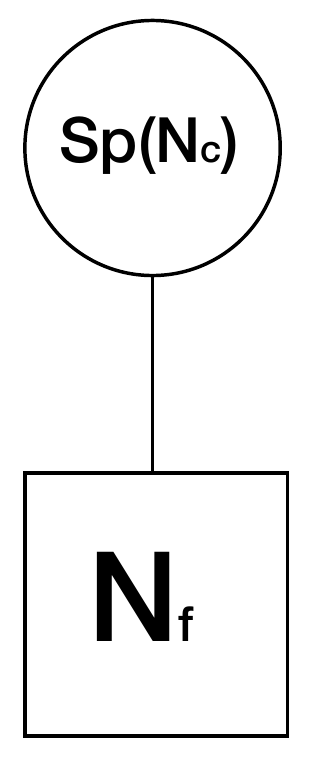} & \secref{Sec:Spmirrors}, \figref{fig:Quiver42etc} \\
\hline
\end{tabular}
\caption{$Sp(N_c)$ gauge theory on the right with its mirror dual quiver.}
\label{tab:QuiversSummary3}
\end{table}
\end{center}

\section{$\CN=4$ Quivers, Mass Deformations and Mirror Symmetry}\label{Sec:N4setup}
Our goal is to understand infrared physics of $\CN=4$ and $\CN=2^\ast$ three dimensional quiver theories which are formulated for quivers of every allowed shape. Recall that in three dimensions there is more freedom than, say, in four dimensions, where, in the subclass of balanced quivers, only (extended) $ADE$-shaped quivers are allowed. Such quivers describe asymptotically conformal theories in the IR; integrating out matter multiplets one can easily obtain asymptotically free theories. However, in three dimensions the corresponding inequality for the linking numbers has the opposite sign. For example, 3d SQCD with gauge group $N_c$ and $N_f$ fundamental hypermultiples  has to obey $N_f\geq 2N_c$ (so-called ``good'' quiver) in order to prevent the runaway of the vacua. Actually, theories with $N_c\leq N_f\leq 2N_c$ are also admissible, but their infrared physics is the same as the theory with $N_f-N_c$ colors and $N_f$ flavors. In what follows, unless otherwise specified we will assume that the stronger constraint $N_f\geq 2N_c$ is satisfied.

\begin{figure}[H]
\begin{center}
\includegraphics[scale=0.5]{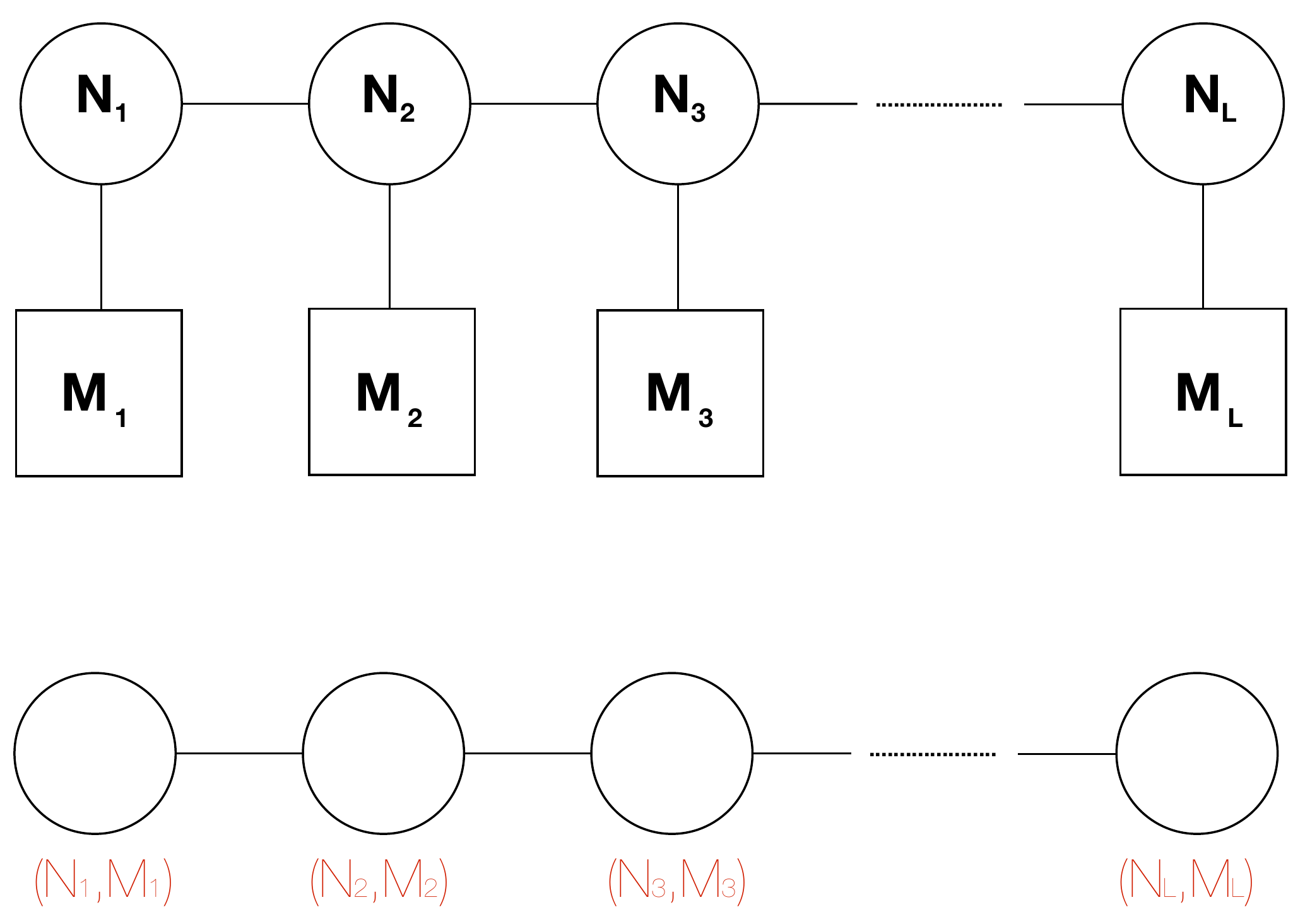}
\caption{Two different notations of linear quiver $A_L$ with labels $(N_1,M_1)_1(N_2,M_2)_2\dots (N_L,M_L)_L$. $N_i$ and $M_i$ are color and flavor labels of the $i$-th node respectively.}
\label{fig:alquivdif}
\end{center}
\end{figure}

We start our analysis with linear $A_L$ quiver theories (see \cite{Gaiotto:2008ak,Gaiotto:2013bwa} for details, here we provide only a minimal review) and then we shall develop an approach to study quivers of other shapes.
The ``goodness'' condition for linear $A_L$ quiver \figref{fig:alquivdif} with color labels $N_i$ and flavor labels (framings) $M_i$ reads
\begin{equation}
\Delta_i := N_{i+1}+N_{i-1}+M_i-2N_i \geq 0\,.
\label{eq:DeltaLin}
\end{equation}
Several notations for quiver varieties are currently used in the literature. In \figref{fig:alquivdif} we list two of them which will be used in our paper interchangeably. These are so-called quanternionic representations of quivers. Each link corresponds to a hypermultiplet in (bi)fundamental representation of the gauge groups it connects. Complex quiver representations reflect each chiral multiplet separately.
A Nakajima quiver variety \cite{zbMATH00727829} is defined as a cotangent bundle to the space of the above quanternionic quiver representation followed by a hyper-K\"{a}hler quotient with respect to the gauge group action. In physics language this construction describes Higgs branches of quiver theories. In the above example of the $U(N_c)$ SQCD with $N_f$ fundamental hypermultiplets the quiver variety (Higgs branch) is isomorphic to the contingent bundle of the complex Grassmannian 
\begin{equation}
\text{Higgs} = T^{\ast}(U(N_f)//U(N_c)) = T^{\ast}G_{N_f,N_c}\,.
\end{equation}
Its quanternionic dimension is $N_c(N_f-N_c)$. 

In the infrared, moduli space of the theory has a Higgs branch and a Coulomb branch. On the Coulomb branch, one has an Abelian theory whose gauge group is the maximal torus of original gauge group of the quiver $U(1)^{\sum N_i}$ while on the Higgs branch the gauge group is generically broken completely. The description of its moduli space depends on the amount of supersymmetry the theory possesses. In an $\CN=4$ theory both the Higgs and the Coulomb branch are singular varieties and the type of singularity can be understood from the quiver itself. For example,  it is well-known that  the Higgs branch for affine ADE quivers has the corresponding ADE singularity. Since mirror symmetry exchanges Coulomb and Higgs branches, the Coulomb branch of the mirror dual of such quivers will also have the corresponding singularity.

 For instance, consider a $U(1)$ gauge theory with $M$ electrons which is mirror dual to a  $A_M$ quiver. The Higgs branch of the latter is the Abelian orbifold $\mathbb{C}^2/\mathbb{Z}_M$ and therefore from mirror symmetry one expects the Coulomb branch of the former to be $\mathbb{C}^2/\mathbb{Z}_M$. Using Hilbert Series analysis, one can readily check that this is indeed the case \cite{Cremonesi:2013lqa}. The analysis is certainly more involved for non-Abelian gauge groups, but there is a canonical way to derive the representation content of the chiral ring on the Coulomb branch\cite{Cremonesi:2013lqa}. As we have just mentioned, Coulomb branches of such theories are singular and for Hilbert Series computation one does not need to consider the resolutions of these singularities. For $S^3$ partition function computations, on the other hand, we require that these singularities are at least partially resolved. Therefore for each framing and for each node of the quiver one has to turn on  resolution parameters (real masses and Fayet-Iliopoulos parameters respectively)  compatible with $\CN=4$ supersymmetry. On top of that, in order to make the parameter space of vacua of the theory (this is another variety we still need to define) nonsingular, we shall introduce another mass which will break the supersymmetry from $\CN=4$ to $\CN=2^*$ \cite{Gaiotto:2013bwa}.

\subsection{Brane Construction and Mirror Symmetry}
\begin{figure}[H]
\begin{center}
\includegraphics[scale=0.5]{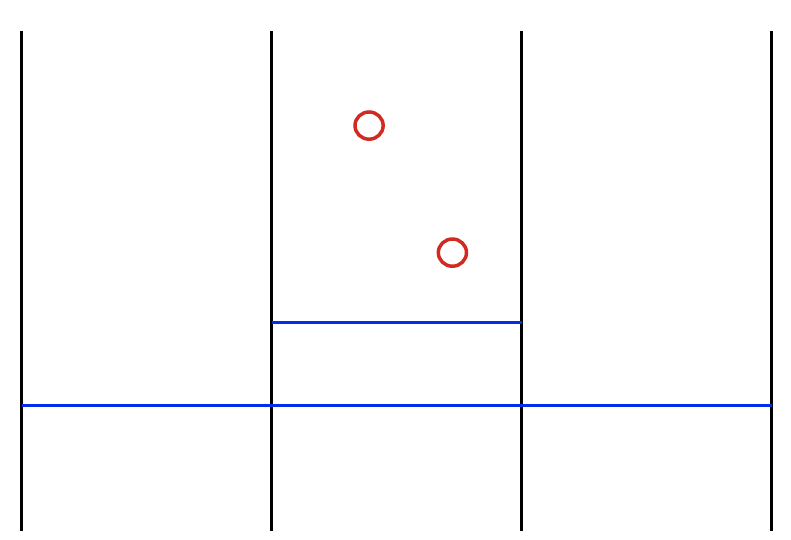}
\caption{Brane construction for $(1,0)(2,2)(1,0)$ quiver. In this figure and later on, red ovals denote D5 branes, horizontal blue lines show D3 branes and vertical black lines designate NS5 branes. In a configuration where none of D3 branes end on D5 branes, the number of D3 branes contained in a given NS5 chamber gives the rank of the corresponding gauge group.}
\label{fig:quiver121f2HW}
\end{center}
\end{figure}

Linear quiver theories can be conveniently formulated using brane constructions of Hanany-Witten type \cite{Hanany:1996ie}. Hanany-Witten type brane setups have been extensively used in string theory and there are many detailed reviews in the literature; here we merely provide a prompt summary. The setup involves D3, NS5 and D5 branes which coincide in the worldvolume directions of the three-dimensional theory and are oriented  in the complementary seven directions of Type IIB string theory such that the system preserves eight real supercharges- see the table below.
\begin{center}
\begin{tabular}{|c|c|c|c|c|c|c|c|c|c|c|}
\hline
         & 0 & 1 &  2 & 3 & 4 & 5 & 6 & 7 & 8 & 9 \\
\hline \hline
NS5 & x &  x &  x &   &   &  &    &  x     &  x & x  \\
\hline
D5    & x &  x & x  &   & x  & x  & x &    &     &    \\
\hline
D3    & x &  x & x & x &    &     &    &   &      &    \\
\hline
\end{tabular}\label{tab:HWbranes}
\end{center}
For example for quiver with labels $(1,0)(2,2)(1,0)$, which we will be using in the next section (see \figref{fig:quiver121new}), we can draw brane diagram shown in \figref{fig:quiver121f2HW}.
Let us now look at the field theory content in more details. Scalar fields parametrizing Higgs and Coulomb branches of the theory form a pair of $SU(2)$ triplets. $\CN=4$ SCFTs admit canonical mass deformations for flavor symmetries of Higgs and Coulomb branches. Therefore there are two types of mass deformations, also $SU(2)$ triplets -- real masses $m^A_i$ on the Higgs branch of the theory and Fayet-Iliopoulos (FI) parameters $t^Z_a$ on the Coulomb branch of the theory, here $A,Z=1,2,3$. The $SU(2)$ symmetry is in fact geometrical, indeed the two symmetry algebras $\mathfrak{su}(2)\sim\mathfrak{so}(3)$ are represented via rotations of $456$ directions for the Higgs branch R-symmetry and $789$ directions for the Coulomb branch R-symmetry. For each $i$ the values of $m_i^A,\, A =4,5,6$ and $t_a^Z,\, Z=7,8,9$ give the coordinates of $i$th D5 brane inside $\mathbb{R}^3_{456}$ and $i$th NS5 brane inside $\mathbb{R}^3_{789}$. Because of the translational symmetry of $\mathbb{R}^3$ all coordinates should be counted modulo the overall shift. In fact,  only the differences $t^Z_a-t^Z_{a-1}$ have  actual physical meaning as FI parameters for the corresponding gauge groups in the linear quiver. Sometimes it is convenient to impose a center of mass constraint on them as well as on the masses for each gauge node of the quiver but we shall refrain from doing so in this paper. It will turn out, somewhat surprisingly, that keeping all the mass deformations unconstrained has some advantage when one works with S-duality and mirror symmetry.

The mirror symmetry in three dimensions can be easily understood via S-duality of the above brane construction. Under S-duality NS5 branes turn into D5 branes and vice versa, D3 branes remain self dual. To read off the dual gauge theory from the S-dual brane system, one needs to move the D5 and NS5 branes appropriately with possible creation/annihilation of D3 branes required to keep the {\it linking numbers} of the individual 5-branes invariant \cite{Hanany:1996ie,Gaiotto:2008ak}. Because Dirichlet and Neveu-Schwarz branes are interchanged Higgs and Coulomb branches are to be swapped together with the $SU(2)$ R-symmetries. In the following sections of the paper we will be using various examples of mirror dual quiver theories, but for now let us consider the mirror for the theory depicted in \figref{fig:quiver121f2HW}. It is an $A_1$ quiver with labels $(2,4)$ or 3d $U(2)$ SQCD with four fundamental hypermultiplets. Indeed, if we switch the NS5 and D5 branes in \figref{fig:quiver121f2HW} and move  NS5 branes to the boundaries of the picture, invariance of {\it linking numbers} for various 5-branes will dictate that the four D5 branes lie inside the NS5 chamber and two D3 branes end on these NS5 branes. This is clearly the Type IIB description for the $A_1$ quiver $(2,4)$. 

Note that we can easily generalize the prescription of obtaining mirror duals to theories given by affine $\hat{A}_N$ quivers. Circular D3 branes which wrap around all the NS5 branes are selfdual, so it is straightforward to read off the data of the mirror quiver. Later in \secref{sec:DhatDouble} we shall consider a framed $\hat{A}_3$ quiver and its mirror.

\subsection{Parameter Space of Vacua for $A_L$ Quivers}
We need to introduce one more ingredient -- the space of mass parameters of supersymmetric vacua $\CL$ for quiver gauge theories in question. However, in order to define $\CL$ we need to deform the setup twofold (see \cite{Gaiotto:2013bwa} for details): First, we compactify the theory on  $\mathbb{R}^2\times S^1$, and, second, we turn on another mass deformation which breaks the supersymmetry down to $\CN=2$. Both modifications are absolutely necessary in order to transform $\CL$ into a complex symplectic manifold with symplectic form
\begin{equation}
\Omega = \sum_i dp_m^i\wedge dm_i + \sum_a dp_t^a \wedge dt_a\,, 
\end{equation}
where the conjugate momenta $p_{t,m}$ to (now complexified) coordinates $m_i$ and $t_a$ are defined through the following generating function
\begin{equation}
p_m^i = \frac{\partial \CW(s,m,t)}{\partial m_i}\,,\quad p_t^a = \frac{\partial \CW(s,m,t)}{\partial t_a}\,.
\label{eq:pmuptaurat}
\end{equation}
The generating function $\CW(s,m,t)$ is nothing but the twisted effective superpotential which describes the massive vacua of the $\CN=2$ theory. The twisted superpotential can be derived straightforwardly from the UV description of the theory by integrating out all chiral multiplets \cite{Nekrasov:2009uh}.  As is explained in \cite{Gaiotto:2013bwa}, $\CW$ serves as a generating function on the parameter space of vacua, which represents itself as a symplectic Lagrangian submanifold $\CL\subset\CM$ inside the complex vector space of all coordinates (masses and FI terms) and the corresponding conjugate momenta.

The $\CN=2^*$ deformation is implemented by the canonical embedding of the $\CN=2$ supersymmetry algebra inside the $\CN=4$ supersymmetry algebra, namely, the $U(1)$ R-symmetry generator of the $\CN=2$ subalgebra is given by the sum of two Cartan generators of $SU(2)_{\text{Higgs}}\times SU(2)_{\text{Coulomb}}$ $\CN=4$ algebra R-symmetry $\mathfrak{j}_R = \mathfrak{j}^3_{\text{Higgs}}+\mathfrak{j}^3_{\text{Coulomb}}$. The orthogonal Cartan generator $\mathfrak{j}_\epsilon = \mathfrak{j}^3_{\text{Higgs}}-\mathfrak{j}^3_{\text{Coulomb}}$ commutes with the $\CN=2$ subalgebra and generates $U(1)_\epsilon$ flavor symmetry with $\epsilon$ being the corresponding twisted mass. 

Finally, the circle compactification provides us with complex mass parameters which are obtained by combining real masses and FI terms with the corresponding flavor Wilson lines. Due to the periodicity along the compact direction it is convenient to replace tuple $(m_i,t_a,\epsilon)$ by its trigonometric version
\begin{equation}
\mu_i = e^{2\pi R m_i}\,,\quad \tau_a= e^{2\pi R t_a}\,,\quad \eta = e^{4\pi R \epsilon}\,,
\label{eq:Lcoord}
\end{equation}
where the numerical factors in the exponential are conventions. Analogously to \eqref{eq:pmuptaurat} we introduce exponentiated momenta 
\begin{equation}
p_\mu^i = e^{2\pi R p_m^i}\,,\quad p_\tau^a= e^{2\pi R p_t^a}\,, \quad p_\eta=e^{4\pi R \frac{\partial \CW}{\partial \epsilon}}\,,
\end{equation}
where in the end we have introduced the momentum conjugate to $\epsilon$, which can also be treated an independent coordinate.

The twisted superpotential $\CW(s,m,t,\epsilon)$ is to be minimized with respect to the adjoint scalar $s$ of the $\mathcal{N}=2$, 3d vector superfield, which can also be exponentiated
\begin{equation}
\sigma_i = e^{2\pi R s_i}\,.
\label{eq:sigmasdef}
\end{equation}
In the same symplectic fashion we introduce canonical momenta which are conjugate to $s$
The condition for supersymmetric vacua is thus precisely the extrema of $\CW$
\begin{equation}
p_\sigma^i:=\exp 2\pi R\frac{\partial \CW}{\partial s_i}=1\,.
\label{eq:psigmadef}
\end{equation}
Therefore algebraically vacua moduli space $\CL$ is a Lagrangian submanifold in $\CM$ given by specifying the conjugate momenta to the full set of variables: $s,m_i,t^a,\epsilon$.

The mirror symmetry in three dimensions interchanges FI terms and masses, hence it should also interchange the corresponding conjugate momenta. In particular it implies that $p_\mu^i$ of one model should coincide with $p_\tau^a$ of the mirror dual model up to (possibly) some identifications of the mass deformations on both sides. On top of that the mirror symmetry flips the sign of $\epsilon$ since it negates the action of the $U(1)_\epsilon$ generator we introduced above. 

\subsection{The Partition Function on $S^3$}
 Localization methods have emerged as a powerful toolbox for computing various observables exactly in QFTs with enough supersymmetry\cite{Pestun:2007rz, Kapustin:2009kz, Hama:2011ea, Festuccia:2011ws}. The study of localization for $\CN \geq 2$ quiver gauge theories on $S^3$ was initiated in \cite{Kapustin:2009kz} and in recent years such computations have been carried out extensively for various three-manifolds including the squashed sphere $S^3_b$ \cite{Hama:2011ea}. In the $b\to 0$ limit the squashed sphere partition function simplifies dramatically, namely it becomes the exponential of the twisted effective superpotential $\mathcal{Z}_{S^3_b}\sim e^{-i\CW}$. Therefore we recover the classical parameter space of the mass deformations $\CL$ in this limit. On the other hand, $b\sim 1$ value corresponds to an intrinsic quantum regime\footnote{The radius of the sphere coincides with $b$ in this limit}.

The computations of the squashed sphere partition function are slightly cumbersome due to the presence of special functions constructed  from double infinite products. However, those functions reduce to exponentials for the round sphere when $b=1$. The partition function on round sphere is therefore a particularly convenient object for studying dualities in 3d quiver gauge theories.  Explicit computations of $S^3$ partition functions as tools to check three dimensional mirror symmetry for $\CN=4$  quiver gauge theories was discussed in \cite{Kapustin:2010xq}. This approach was also taken in \cite{Dey:2011pt,Dey:2013nf} where mirror symmetry for a large class of affine $D$-type quivers was discussed.

Given an $\CN = 4$ quiver gauge theory, the rules for writing down the $S^3$ partition function may be summarized in the following fashion. Localization ensures that the partition function of the theory reduces to a matrix integral over the Cartan of the gauge group. Since $S^3$ does not have any instantons, any such partition function may be schematically represented as 
\begin{equation}
\CZ=\int  \frac{d^k s}{|\mathcal{W}|} \prod_{\alpha} \alpha(s) \exp{S_{\text{cl}}[s]}\, \CZ_{\text{1-loop}}[s]\,,
\end{equation}
where $s$ is the real adjoint scalar that sits inside a 3d $\mathcal{N}=2$ vector multiplet. One can use a constant gauge transformation to make $s$ lie in the Cartan subalgebra of the gauge group.  In the above formula $\prod_{\alpha} \alpha(s)$ is the Vandermonte determinant where the product is over all roots of the gauge group. This factor appears in the measure as a result of gauge fixing the matrix model such that $s$ lies in the Cartan of the gauge group. $|\mathcal{W}|$ represents the order of the Weyl group - the $\frac{1}{|\mathcal{W}|}$ factor is needed to account for the residual gauge symmetry after $s$ is gauge-fixed to lie in the Cartan subalgebra.

The contribution of vectors and hypers in the $\CN = 4$ theory to the above partition are as follows. For every $U(1)$ factor in the gauge group, one obtains the following classical contribution
\begin{equation}
S^{\text{FI}}_{\text{cl}}= 2\pi i \eta\, \text{Tr}(s)\,,
\end{equation}
where $\eta$ is a FI parameter. Each $\mathcal{N}=4$ vector multiplet contributes with
\begin{equation}
Z^v_{\text{1-loop}}=\prod_{\alpha} \frac{\sinh^2{\pi \alpha(s)}}{\pi \alpha(s)}\,,
\end{equation}
where the product extends over all the roots of the Lie algebra of G. In fact, this is precisely the contribution of an $\mathcal{N}=2$ vector multiplet since contribution of the adjoint chiral which is part of the $\mathcal{N}=4$ vector multiplet is trivial \cite{Kapustin:2010xq}.

Finally, each $\mathcal{N}=4$ hypermultiplet contributes with
\begin{equation}
\CZ^h_{\text{1-loop}}=\prod_{\rho} \frac{1}{\cosh{\pi \rho(s+m)}}\,,
\end{equation}
where the product extends over all the weights of the representation R of the gauge group G and $m$ is a real mass parameter.

Note that the Vandermonte factor in the measure exactly cancels with the denominator of the 1-loop contribution of the vector multiplet for each factor in the gauge group and we can therefore ignore this contribution in the matrix integral.

Now let us illustrate how $S^3$ partition function may be useful in studying mirror symmetry of $\mathcal{N}=4$ quiver gauge theories. Consider the linear quiver pair which we have already discussed in this section: the A-model quiver with labels  $(1,0)(2,2)(1,0)$ (see \figref{fig:quiver121f2HW}) and its mirror, the B-model which has labels $(2,4)$.
The  A-model partition function is given by
\begin{equation}
\begin{split}
&\CZ_A=\int \prod^2_{\alpha=1} d s_{\alpha}  \frac{d^2 s_0}{2!} \frac{\prod^2_{\alpha=1} e^{2\pi i s_{\alpha} \eta_{\alpha}} \prod^2_{i=1} e^{2\pi i s^i_0 \eta_0} \sinh^2{\pi(s^1_0-s^2_0)}}{\prod^2_{i=1}\cosh{\pi(s_1-s^i_0)}\prod^2_{a=1} \cosh{\pi(s^i_0+m_a)}\cosh{\pi(s_2-s^i_0)}} \\
&= e^{-2\pi i m_1(t_1+t_2)} e^{2\pi i m_2(t_3+t_4)}\frac{1}{2\sinh{\pi(t_1-t_2)}\sinh{\pi(t_3-t_4)}\sinh^2{\pi(m_1-m_2)}} \\
\times & \Big[\frac{(e^{2\pi i t_4(m_1-m_2)}-e^{2\pi i t_1(m_1-m_2)})(e^{2\pi i t_3(m_1-m_2)}-e^{2\pi i t_2(m_1-m_2)})}{\sinh{\pi(t_2-t_3)}\sinh{\pi(t_1-t_4)}} \\
&-\frac{(e^{2\pi i t_3(m_1-m_2)}-e^{2\pi i t_1(m_1-m_2)})(e^{2\pi i t_4(m_1-m_2)}-e^{2\pi i t_2(m_1-m_2)})}{\sinh{\pi(t_1-t_3)}\sinh{\pi(t_2-t_4)}}\Big]\,,
\end{split}
\label{eq:ZAex}
\end{equation}
where $m_1$ and $m_4$ are the masses of the fundamental hypermultiplets in the middle node. We have also defined $\eta_1=t_1-t_2, \eta_0=t_2-t_3, \eta_2=t_3-t_4$ which are the Abelian coupling constants for the three gauge groups in the quiver. 

The partition function of the B-model $(2,4)$ is 
\begin{equation}
\begin{split}
&\CZ_B=\int \frac{d^2s}{2!} \frac{\prod^2_{i=1} e^{2\pi i s^i (\widetilde{t}_1-\widetilde{t}_2)}  \sinh^2{\pi(s^1-s^2)}}{\prod^2_{i=1}\prod^4_{a=1} \cosh{\pi(s^i+M_a)}} \\
&= \frac{1}{2\sinh{\pi(M_1-M_2)}\sinh{\pi(M_3-M_4)}\sinh^2{\pi(\widetilde{t}_1-\widetilde{t}_2)}}\\
& \times \Big[\frac{(e^{2\pi i M_1(\widetilde{t}_1-\widetilde{t}_2)}-e^{2\pi i M_4(\widetilde{t}_1-\widetilde{t}_2)})(e^{2\pi i M_2(\widetilde{t}_1-\widetilde{t}_2)}-e^{2\pi i M_3 (\widetilde{t}_1-\widetilde{t}_2)})}{\sinh{\pi(M_1-M_4)}\sinh{\pi(M_2-M_3)}} \\
&-\frac{(e^{2\pi i M_1(\widetilde{t}_1-\widetilde{t}_2)}-e^{2\pi i M_3(\widetilde{t}_1-\widetilde{t}_2)})(e^{2\pi i M_2(\widetilde{t}_1-\widetilde{t}_2)}-e^{2\pi i M_4 (\widetilde{t}_1-\widetilde{t}_2)})}{\sinh{\pi(M_1-M_3)}\sinh{\pi(M_2-M_4)}}\Big]\,.
\end{split}
\label{eq:ZBex}
\end{equation}
For convenience we impose the following constraints on mass parameters and FI parameters: $\widetilde{t}_2+\widetilde{t}_1=0$ and $M_1+M_2+M_3+M_4=0$.

From the formulae for $\CZ_A$ and $\CZ_B$, it is evident that they are equivalent under the mirror map
\begin{equation}
t_i \leftrightarrow M_i\,,\quad  m_a \leftrightarrow \widetilde{t}_a\,,
\label{eq:MirrorMarLinEx}
\end{equation}
up to a phase factor $e^{2\pi i m_1(t_1+t_2)} e^{-2\pi i m_2(t_3+t_4)}$ which vanishes when one imposes the constraints $m_1+m_2=0, \; t_1+t_2+t_3+t_4=0$.

In the following sections, we shall make extensive use of the $S^3$ partition function to obtain various quiver gauge theories from linear quiver pairs using the technique of abelian/non-abelian gauging. 

\subsection{The Hilbert Series of the Coulomb Branch}
In this section we review a general formula for the Hilbert series of the Coulomb branch of a $3d$ $\cN=4$ theory discussed in \cite{Cremonesi:2013lqa}.

It is convenient to use the $3d$ $\CN=2$ formalism, in which the $\CN=4$ vector multiplet decomposes into a $\CN=2$ vector multiplet and a chiral multiplet  $\Phi$ in the adjoint representation of the gauge group. On a generic point of the Coulomb moduli space, the triplet of scalars in the $\CN=4$ vector multiplets acquires a vacuum expectation value, and the gauge fields that remain massless are abelian and can be dualized to scalar fields.  The classical Coulomb branch is parametrized by the collection of such massless scalar fields.  The Coulomb branch, however, receives many quantum corrections.  The asymptotic hyperk\"ahler metric in the weak coupling region of the Coulomb branch can be computed at one loop. Yet this method does not provide a suitable description for the strongly coupled region.

It is shown in \cite{Borokhov:2002ib} that there is a description of the quantum Coulomb branch that bypasses the dualization of free abelian vector multiplets.  This realization involves 't Hooft monopole operators, which are local disorder operators that can be defined directly in the infrared CFT.  The magnetic charges $\vec m$ of the monopole operators are labelled by the weight lattice $\Gamma^*_{G^\vee}$ of the GNO (Langlands) dual gauge group $G^\vee$ \cite{Goddard:1976qe}.  The GNO monopole charges $\vec m$ breaks the gauge group $G$ to a residual gauge group $H_{\vec m}$, which is the commutant of $\vec m$ inside $G$. The components of the complex scalar $\phi$ that are moduli of the BPS monopole configuration reside in the Lie algebra of the group $H_{\vec m}$ and are left unbroken by the monopole flux \cite{Cremonesi:2013lqa}. The monopoles can be dressed with the scalar components $\phi$ of the chiral multiplet $\Phi$ that preserves some amount of supersymmetry. The residual gauge symmetry in the monopole background contains continuous part $H_{\vec m}$ and a discrete part, namely Weyl group $W_{G^\vee}$ of $G^\vee$; they act on $\vec m$ and $\phi$.  The gauge invariant operators are labelled by $\vec m \in \Gamma^*_{G^\vee}/W_{G^\vee}$, \ie~ a Weyl chamber of weight lattice $\Gamma^*_{G^\vee}$.  Such operators are dressed by all possible products of $\phi$ which are invariant under the action of the residual group $H_{\vec m}$.

The Hilbert series is the generating function of the chiral ring that counts gauge invariant BPS operators parametrizing the Coulomb branch, graded according to their dimension and quantum numbers under global symmetries.  From the above discussion, the general formula for the Coulomb branch Hilbert series reads
\bea\label{Hilbert_series}
H_G(t,z)=\sum_{\vec m\,\in\, \Gamma^*_{G^\vee}/W_{G^\vee}} t^{\Delta(\vec m)} P_G(t,\vec m) z^{J(\vec m)} \;,
\eea
where the notation is explained as follows:
\begin{itemize}
\item The sum is taken over a Weyl Chamber of the weight lattice $\Gamma^*_{G^\vee}$,
\item The function $P_G(t,m)$ counts Casimir gauge invariants of the residual gauge group $H_{\vec m}$ made with the adjoint $\phi$, according to their dimension; it is given by
\bea\label{classical_dressing}
P_G(t; \vec m)=\prod_{i=1}^r \frac{1}{1-t^{d_i(\vec m)}} \;,
\eea
where $d_i(\vec m)$, $i=1,\dots,{\rm rank}\; H_{\vec m}$ are the degrees of the Casimir invariants of the residual gauge group $H_{\vec m}$ left unbroken by the GNO magnetic flux $\vec m$.
\item The factor $t^{\Delta(m)}$ takes into account quantum dimensions of monopole operators which is given by \cite{Borokhov:2002cg,Gaiotto:2008ak,Benna:2009xd,Bashkirov:2010kz}
\bea\label{dimension_formula}
\Delta(\vec m)=-\sum_{\vec \alpha \in \Delta_+(G)} |\alpha(\vec m)| + \frac{1}{2}\sum_{i=1}^n\sum_{\vec \rho_i \in R_i}|\vec \rho_i(\vec m)|\;,
\eea
where the first sum over positive roots $\vec \alpha\in\Delta_+(G)$ of $G$ is the contribution of $\cN=4$ vector multiplets and the second sum over the weights $\vec \rho_i$ of the matter field representation $R_i$ under the gauge group is the contribution of the $i$-th $\cN=4$ hypermultiplet.
\item For a non-simple connected group $G$, there is a non-trivial topological symmetry under which the monopole operators are charged.  $J(\vec m)$ denotes the topological charge of the monopole operator of GNO charges $\vec m$, and $z$ is the fugacity associated with the topological charge.
\end{itemize}
We refer to \eref{Hilbert_series} as the {\it monopole formula} of the Coulomb branch Hilbert series.

\section{Gauging Quivers: A Basic Example}\label{Sec:GaugingBasic}
Let us describe a simple example which illustrates the main idea of the gauging method. In this section we shall only discuss Abelian gauging by which we shall mean gauging a single, or several U(1) factors. In later sections we shall address non-Abelian gauging, which will result in a nontrivial deformation of the Coulomb branch of the mirror model.

\subsection{Parameter Space Approach}
In this work we shall be repeatedly using the embedding of parameter spaces of various quiver gauge theories $\CL$ into larger parameter spaces of some linear $A_L$ quivers whose mirrors can be easily constructed. Using the results of \cite{Gaiotto:2013bwa} we can approach the desired parameter space by taking some singular limit of the $A_L$ mirror pair. Each $\mathcal{N}=2$ $A_L$ quiver theory has a fairly large parameter space of masses $\mu^{(I)}_i$ ($I$ corresponds to $I$-th gauge node) and FI couplings $\tau_a$. Similar set of parameters exists on the mirror side $\mu^{\vee\,(I)}_i$ and $\tau^\vee_a$. For the $A_L$ quivers the mirror map simply interchanges the masses and the FI terms. For other quivers a more complicated mapping is expected.

Let us look at one of the simplest examples of quivers which are not linear, say the $\hat{D}_4$ quiver. Its mirror is known and is described as $Sp(1)$ (or $SU(2)$) gauge theory with eight fundamental half-hypermultiplets with $SO(8)$ global symmetry. Below we shall describe how to reproduce this result using the gauging method starting from another mirror pair of linear quivers.

\begin{figure}[H]
\begin{center}
\includegraphics[scale=0.5]{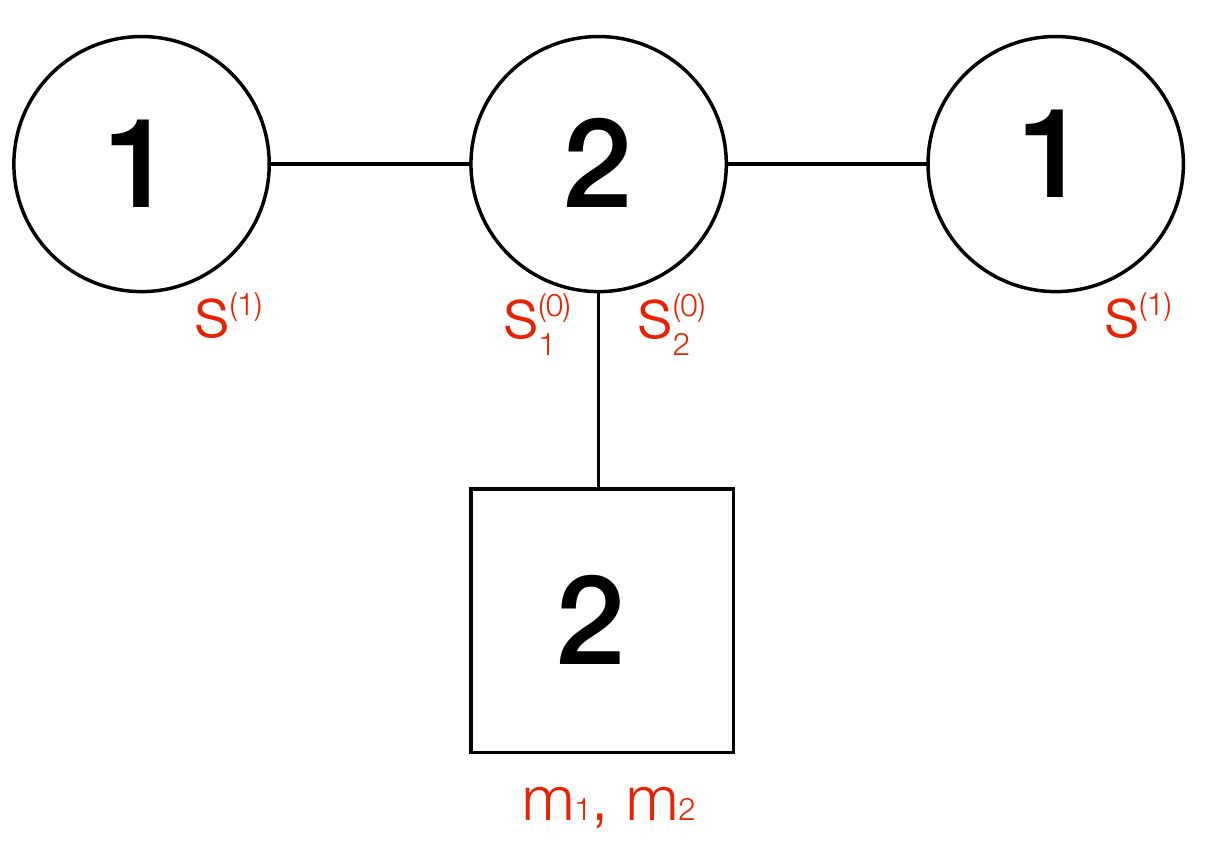}
\caption{$A_3$ quiver with labels $(1,0)_1(2,2)_0(1,2)_2$ together with all Coulomb branch parameters and masses.}
\label{fig:quiver102211det}
\end{center}
\end{figure}

Coincidentally, the proper mirror pair consists of the two theories which we have already described in the previous section: A-models with labels $(1,0)(2,2)(1,0)$ (see \figref{fig:quiver121f2HW}, \ref{fig:quiver102211det}) and B-model, which is $U(2)$ theory with 4 flavors. 
We have concluded that these two theories are mirror dual to each other by applying the S-duality to their brane descriptions. Now we shall look at these two models more carefully by studying their supersymmetric vacua.

The vacua of the A quiver are governed by the following Bethe-type equations\footnote{We shall be using two terms -- ``Bethe equations'' and ``SUSY vacua equations'' in the paper interchangeably.} 
\begin{align}
\label{eq:AmodelMassEx}
\frac{\tau_2}{\tau_1} \prod_{a=1}^2\frac{\eta\sigma^{(1)}-\sigma^{(0)}_a}{\eta\sigma^{(0)}_a-\sigma^{(1)}}&=1\,,\notag\\
\frac{\tau_4}{\tau_3} \prod_{a=1}^2\frac{\eta\sigma^{(2)}-\sigma^{(0)}_a}{\eta\sigma^{(0)}_a-\sigma^{(2)}}&=1\,,\notag\\
\frac{\tau_3}{\tau_2} \prod_{I=1}^2\frac{\eta\sigma^{(0)}_1-\sigma^{(I)}}{\eta\sigma^{(I)}-\sigma^{(0)}_1} \cdot \prod_{a=1}^2\frac{\eta\sigma^{(0)}_1-\mu^{(0)}_a}{\eta\mu^{(0)}_a-\sigma^{(0)}_1} \cdot \frac{\eta^{-1}\sigma^{(0)}_1-\eta\sigma^{(0)}_2}{\eta^{-1}\sigma^{(0)}_2-\eta\sigma^{(0)}_1}&=1\,,\\
\frac{\tau_3}{\tau_2} \prod_{I=1}^2\frac{\eta\sigma^{(0)}_2-\sigma^{(I)}}{\eta\sigma^{(I)}-\sigma^{(0)}_2} \cdot \prod_{a=1}^2\frac{\eta\sigma^{(0)}_2-\mu^{(0)}_a}{\eta\mu^{(0)}_a-\sigma^{(0)}_2} \cdot \frac{\eta^{-1}\sigma^{(0)}_2-\eta\sigma^{(0)}_1}{\eta^{-1}\sigma^{(0)}_1-\eta\sigma^{(0)}_2}&=1\,.\notag
\end{align}
These four equations are to be solved with respect to four A-model Coulomb branch parameters: $\sigma^{(1)},\sigma^{(0)}_1,\sigma^{(0)}_2$ and $\sigma^{(2)}$, where upper indices designate the corresponding gauge groups in the quiver (see \figref{fig:quiver102211det}).
Recall that FI terms $t_1$ through $t_4$ denote the coordinates of the NS5 branes along the $x^3$ direction (see \figref{fig:quiver121f2HW}) and their differences $t_2-t_1, t_3-t_2, t_4-t_3$ give the FI couplings. Note also that \eqref{eq:AmodelMassEx} contains all these variables in a trigonometric form, see \eqref{eq:Lcoord} and \eqref{eq:sigmasdef}. The first and the second equations of \eqref{eq:AmodelMassEx} arise from minimizing the effective twisted superpotential of the theory with respect to $s^{(1)}$ and $s^{(2)}$ respectively. These two Coulomb coordinates only appear in the bifundamental hypermultiplets (chiral parts inside those hypers contribute to the numerators, anti-chiral parts give the denominators) which connect nodes (1) and (0) and nodes (2) and (0); that is why the corresponding Bethe equations are fairly simple. At the middle node (0), however, there are more contributions. First, there are two variables $s^{(0)}_1$ and $s^{(0)}_2$ for each Cartan generator of $U(2)$, so there is a contribution from the adjoint field, and second, in addition to the bifundamental fields there are two more fundamental hypers with masses $m^{(0)}_1$ and $m^{(0)}_2$. Note that $\CN=2^*$ mass $\epsilon$ enters differently in the expressions for chiral and vector fields due to the special R-charge assignments: chiral fields have charge $1$ and vectors fields have charge $-2$. We refer the reader to \cite{Gaiotto:2013bwa} for more details.

In order to fully describe the parameter space $\CL$ in addition to writing Bethe equations \eqref{eq:psigmadef}, which can be viewed as
\begin{equation}
p_{\sigma\,i}^{(I)}=1\,,\quad i=1,\dots N_I\,,\quad I=1,\dots, L\,,
\end{equation}
we specify the momenta conjugate to the mass parameters and FI parameters. For the case in hand we have for the middle node\footnote{For this example we omit the $(0)$ superscript for the masses.} 
\begin{equation}
p_\mu^a=\tau_1\tau_2\prod_{j=1}^2\frac{\eta\mu_a-\sigma^{(0)}_j}{\eta\sigma^{(0)}_j-\mu_a}\,,
\label{eq:Pmuexample}
\end{equation}
together with the corresponding formulae for $p_\tau$. 

\begin{figure}[H]
\begin{center}
\includegraphics[scale=0.3]{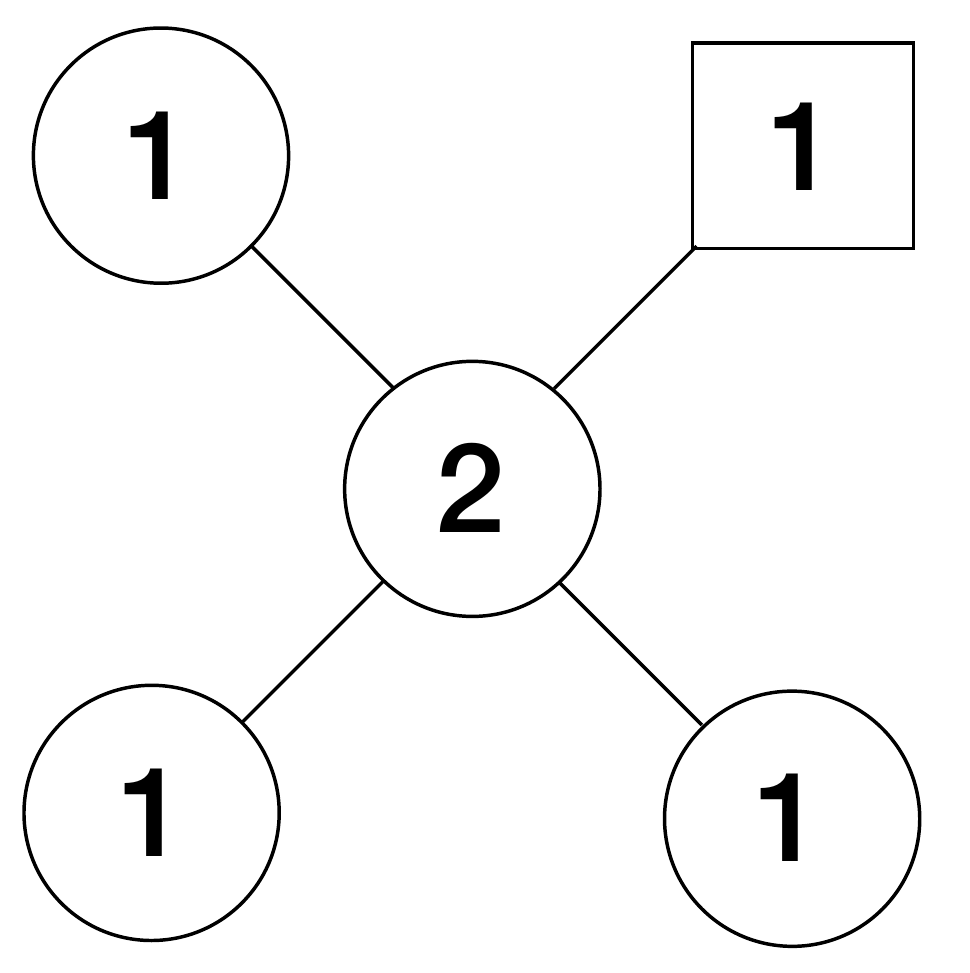}\quad\quad\quad\quad\quad \includegraphics[scale=0.4]{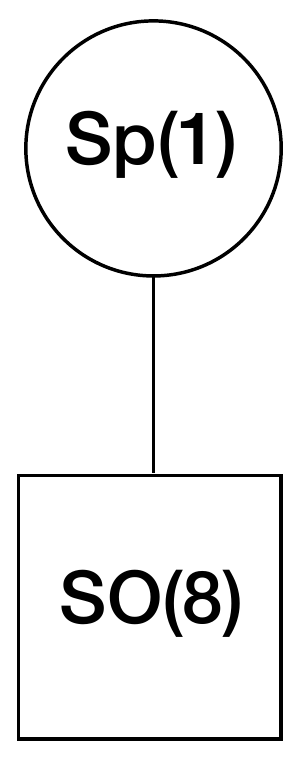}
\caption{$D_4$ quiver and its mirror dual obtained from the mirror pair of linear quivers with labels $(1,0)(2,2)(1,2)$ and $(2,4)$.}
\label{fig:quiverD4}
\end{center}
\end{figure}

Already at this point we can make an interesting observation. Let us treat one of the equations (say for $m_1$) in \eqref{eq:Pmuexample} as a new Bethe equation (we shall now formally relabel $m_1$ into $\sigma^{(3)}$) as if the momentum $p_\mu^{1}$ is fixed to some constant value. 
One can now rewrite this equation as follows
\begin{equation}
\frac{\tau_1\tau_2}{p_\mu^3} \prod_{j=1}^2\frac{\eta\sigma^{(3)}-\sigma^{(0)}_j}{\eta\sigma^{(0)}_j-\sigma^{(3)}} =1\,.
\label{eq:pmufix1}
\end{equation}
We now add this equation to \eqref{eq:AmodelMassEx} in order to form a new set of Bethe equations with respect to five variables: four Coulomb coordinates of the $A_3$ quiver we have started with and $\sigma^{(3)}$. One recognizes in this set of five equations the vacua equations for $D_4$ quiver with $U(2)$ gauge group in the middle (label $(0)$) node, three $U(1)$ gauge nodes labeled by $(1)-(3)$, and one global $U(1)$ symmetry \figref{fig:quiverD4}.

From the point of view of the parameter space of the mirror theory fixing the momentum in \eqref{eq:pmufix1}, which is equivalent to losing mass parameter $m_1$ (because we now need to solve the new set of equations with respect to it) corresponds to (up to some rescaling, which we shall fix promptly) eliminating one of the FI parameters. Indeed, according to the mirror map in the linear quivers FI parameters and masses get interchanged, as well as their conjugate momenta; so removing $m_1$ on the A-side corresponds to eliminating, say, $\tau^\vee_1$ on the B-side. An exact expression for $m_1$ as a solution of the A model Bethe equations may be quite cumbersome, and requires the knowledge of the solution of some high-degree polynomial equations. However, in order to identify the content of the B model after fixing $p_\mu^1$ we can use the expression for the momentum conjugate to $\tau^\vee_1$. Recall that in our case the mirror quiver is $A_1$ with labels $(2,4)$, so it has two FI parameters: $\tau^\vee_1$ and $\tau^\vee_2$ and two Coulomb parameters $\sigma^\vee_1$ and $\sigma^\vee_2$. According to \cite{Gaiotto:2013bwa} we have 
\begin{equation}
p^{\vee\,1}_\tau = \frac{1}{\sigma^\vee_1\sigma^\vee_2}\,.
\label{eq:ptau1}
\end{equation}
We recall that under the mirror map $\mu^\vee_a=\tau_a$. Our prescription now requires to us to fix $p^{\vee\,1}_\tau=1$, therefore we impose a constraint for $\sigma^\vee_1$ and $\sigma^\vee_2$, namely that one variable is inversely proportional to another, or, in terms of rational coordinates, $s^\vee_1 = -s^\vee_2$. This constraint yields us $SU(2)=Sp(1)$ gauge theory with four flavors. Therefore we have shown how the  parameter spaces of the two mirror quivers in \figref{fig:quiverD4} can be embedded inside the parameter spaces of two linear quivers.

The statement will become obvious upon proper identification of the momentum \eqref{eq:Pmuexample} with the twist parameters of the A quiver. It works as follows
\begin{equation}
\frac{\tau_1\tau_2}{p_\mu^1}=\frac{\tau_2}{\tau_1}\,,
\end{equation}
therefore $p_\mu^1=\tau_1^2$. In order to match it with $p^{\vee\,1}_\tau=1$, which we just used to derive the $Sp(1)$ theory we need to assume $\tau^2_1=1$ or $t_1=0$, which in terms of the brane construction of \figref{fig:quiver121f2HW} fixes the location of the location of the leftmost NS fivebrane to the origin in the $x^3$ direction.

In a moment we shall demonstrate how the procedure we have just performed (also known as \textit{gauging} of an Abelian symmetry, or merely  \textit{Abelian gauging}) can also be carried out at the level of  partition functions of the A and B models \eqref{eq:ZAex}, \eqref{eq:ZBex}. This computation will turn out to be very effective in deriving mirror pairs via Abelian gauging as well as obtaining exact relationships between the mass parameters/FI parameters of the dual theories (so called mirror maps).

The dimensions of the Coulomb and Higgs branches of the A-model quiver before the gauging \figref{fig:quiver102211det} were $4$ and $2$ respectively (correspondingly these numbers give the dimensions of the Higgs and Coulomb branches of the B-model before the gauging). 
After gauging these dimensions on the A-side (see \figref{fig:quiverD4}) have become $5$ and $1$ respectively, which agrees with the dimensions of the Higgs and Coulomb branches of the $Sp(1)$ theory with four flavors. Therefore we can see that by gauging a $U(1)$ subgroup of the $U(2)$ node on the left in \figref{fig:quiver121} we increase the dimension of the Coulomb branch by one and decrease the dimension of the Higgs branch by one.

As we have already mentioned before, the mirror map (exact correspondence of the mass/FI parameters on both sides of the duality) for linear quivers is very simple. Indeed, one simply interchanges the roles of twisted masses and FI parameters. However, after the gauging has been implemented, the mirror map will change as well. In order to derive the exact form of this map, as well as verify the proposed mirror pair using exact localization methods, we appeal to the computations of partition function on a three-sphere.

\subsection{Partition Function Approach}
For gauging the $U(2)$ flavor symmetry we have to treat $m_1$ and $m_2$ as independent parameters, without imposing any constraints. Note that from \eqref{eq:ZAex} and \eqref{eq:ZBex} we have
\begin{equation}
\CZ_A =e^{-2\pi i m_1(t_1+t_2)} e^{2\pi i m_2(t_3+t_4)} \CZ_B\,.
\end{equation}
Now let us implement the gauging by $m_1 (\widetilde{t}_1)\to -s_3$. Since we are gauging a single $U(1)$ flavor symmetry, the partition function of \textit{gauged A-model} (which we denote by $\widetilde{\CZ}_A$) may be simply obtained by multiplying $\CZ_A$ with the appropriate FI contribution $e^{-2\pi i {\eta}_3 s_3}$ and integrating over $s_3$. Therefore,
\begin{equation}
\begin{split}
\widetilde{\CZ}^{(1)}_A&=\int ds_3 e^{-2\pi i {\eta}_3 s_3} \CZ_A\\
&= \int \prod^2_{\alpha=1} d s_{\alpha}  ds_3 \frac{d^2 s_0}{2!} \frac{e^{-2\pi i {\eta}_3 s_3}\prod^2_{\alpha=1} e^{2\pi i s_{\alpha} \eta_{\alpha}} \prod^2_{i=1} e^{2\pi i s^i_0 \eta_0} \sinh^2{\pi(s^1_0-s^2_0)}}{\prod^2_{i=1}\cosh{\pi(s_1-s^i_0)}\cosh{\pi(s^i_0-s_3)}\cosh{\pi(s_2-s^i_0)}\cosh{\pi(s^i_0+m_2)}} \\
&=e^{2\pi i m_2(t_3+t_4)} \int ds_3 e^{-2\pi i s_3(t_1+t_2)} e^{-2\pi i {\eta}_3 s_3} \CZ_B \\
&=e^{2\pi i m_2(t_3+t_4)} \int ds_3 e^{-2\pi i s_3(t_1+t_2)} e^{-2\pi i {\eta}_3 s_3} \frac{d^2s}{2!} \frac{\prod^2_{i=1} e^{2\pi i s^i \eta}  \sinh^2{\pi(s^1-s^2)}}{\prod^2_{i=1}\prod^4_{a=1} \cosh{\pi(s^i+M_a)}}\,.
\end{split}
\end{equation}
The partition function $\widetilde{\CZ}^{(1)}_A$ corresponds to the  gauged A-model quiver which, from the second line, is a $U(1)^3 \times U(2)$ gauge theory with one fundamental hyper (\figref{fig:quiverD4}). To determine the mirror of this quiver, one needs to consider the formula on the third line, which essentially rewrites the partition function of the gauged A-model quiver in terms of the partition function of the original B-model. 
One may then identify the dual theory  by computing $\widetilde{\CZ}^{(1)}_A$ using the third/fourth line of the above equation.\\
\begin{equation}
\begin{split}
&\widetilde{\CZ}^{(1)}_A =e^{2\pi i m_2(t_3+t_4)}  \int ds_3 \frac{d^2s}{2!} \frac{ \sinh^2{\pi(s^1-s^2)}}{\prod^2_{i=1}\prod^4_{a=1} \cosh{\pi(s^i+M_a)}} \prod^2_{i=1} e^{2\pi i s^i (-m_2 +s_3)} e^{2\pi i s_3(-t_1-t_2 - {\eta}_3)}\\
&=e^{2\pi i m_2(t_3+t_4)}   \int \frac{d^2s}{2!} \frac{ \sinh^2{\pi(s^1-s^2)} \delta\left(s^1+s^2 -t_1-t_2-{\eta}_3\right) }{\prod^2_{i=1}\prod^4_{a=1} \cosh{\pi(s^i+M_a)}} e^{-2\pi i m_2(s^1+s^2)}\\
&=  e^{2\pi i m_2(-t_1-t_2-{\eta}_3 +t_3 +t_4)} \int \frac{d^2s}{2!} \frac{ \sinh^2{\pi(s^1-s^2)} \delta\left(s^1+s^2 \right) }{\prod^2_{i=1}\prod^4_{a=1} \cosh{\pi(s^i+(t_1+t_2+{\eta}_3)/2+M_a)}}\,.
\end{split}
\end{equation}

Therefore, up to the prefactor (a pure phase) indicated above, one can easily identify the above as the $S^3$ partition function of a $SU(2)$ gauge theory with 4 hypers. Note that the masses of the fundamental hypers are shifted as $M_a \to M_a + (t_1+t_2+{\eta}_3)/2$ as we gauge the flavor in the A-model. The mirror 
map for the mirror dual pair in  \figref{fig:quiverD4} is therefore
\begin{equation}
M^{(1)}_a = M_a +(t_1+t_2+{\eta}_3)/2 = t_a + (t_1+t_2+{\eta}_3)/2
\end{equation}

The mirror map is very similar to that of the original linear quiver pairs - each mass just gets shifted by the same factor. For the new mirror pair, 
the A-model has four independent FI parameters, namely $t_a$s ($a=1,2,3,4$) with one constraint and $\eta_3$ -which matches the four independent masses in the B-model. Note that the masses of the B-model no longer satisfy the constraint of zero sum - in fact $\sum_a M^{(1)}_a =2(t_1+t_2+{\eta}_3) \neq 0$.

\begin{figure}[H]
\begin{center}
\includegraphics[scale=0.3]{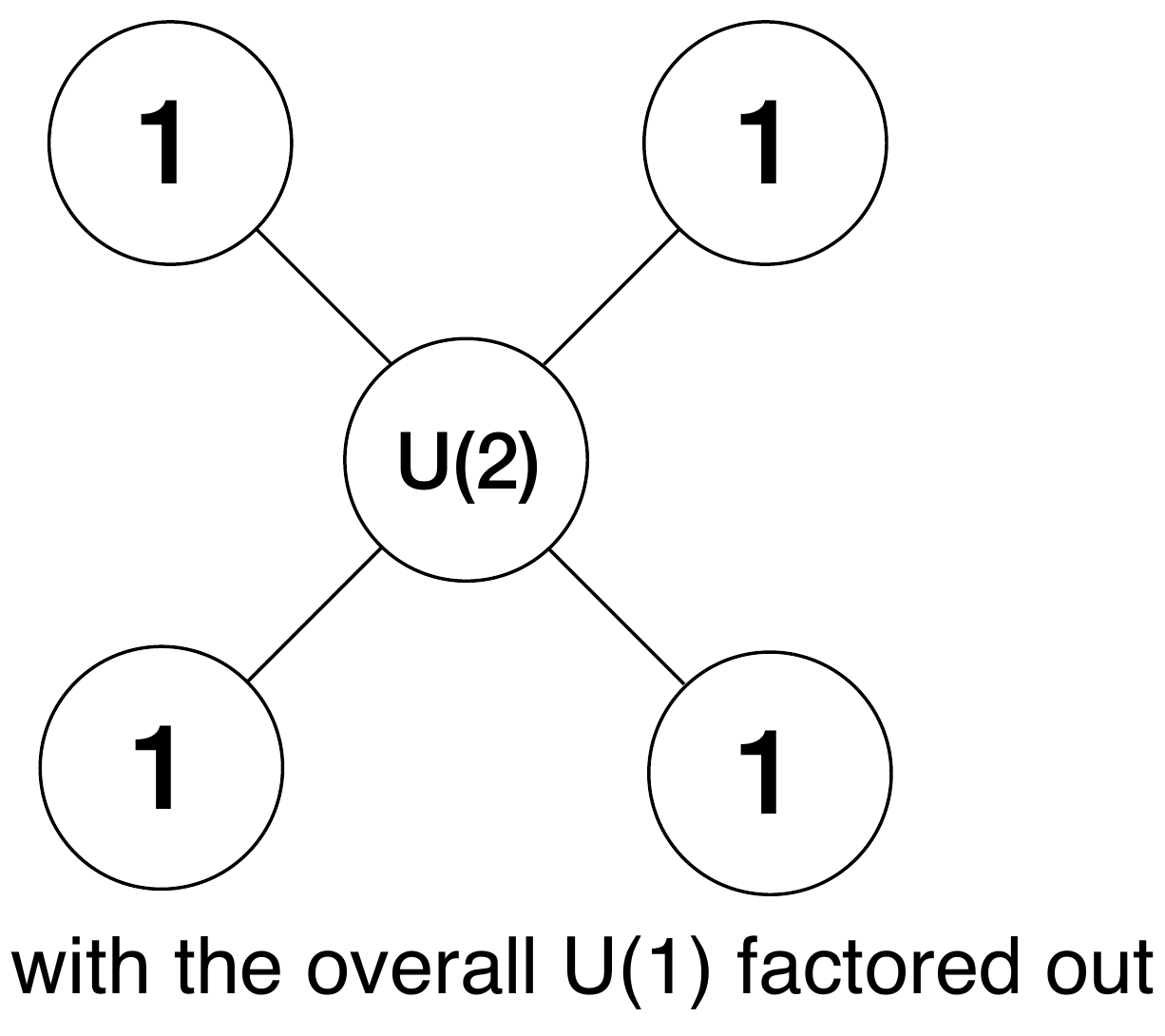}
\caption{$\hat{D}_4$ with $U(2)$ gauge group in the middle node and overall $U(1)$ factored out. It may be considered as a result of two Abelian gaugings which were implemented on the middle node of the linear $(2,0)(2,2)(2,0)$ quiver.}
\label{fig:quiver222SU(2)}
\end{center}
\end{figure}

 One can gauge the remaining $U(1)$ flavor in exactly the same way. In this case, one obtains,
\begin{equation}
\begin{split}
\widetilde{\CZ}^{(2)}_A =& \delta \left(t_1+t_2+{\eta}_3 +{\eta}_4-t_3 -t_4\right)\int \frac{d^2s}{2!} \frac{ \sinh^2{\pi(s^1-s^2)} \delta\left(s^1+s^2 \right) }{\prod^2_{i=1}\prod^4_{a=1} \cosh{\pi(s^i+(t_1+t_2+{\eta}_3)/2+t_a)}}\\
=&\delta \left(\eta_1+\eta_2+{\eta}_3 +{\eta}_4+2\eta_0\right)\int \frac{d^2s}{2!} \frac{ \sinh^2{\pi(s^1-s^2)} \delta\left(s^1+s^2 \right) }{\prod^2_{i=1}\prod^4_{a=1} \cosh{\pi(s^i+(t_1+t_2+{\eta}_3)/2+t_a)}}\,.
\end{split}
\end{equation}
The B-model is therefore the same, but the A-model will be $\hat{D}_4$ quiver with an overall $U(1)$ removed as imposed by the delta function constraint in the previous equation. 
\begin{equation}
\eta_1+\eta_2+2\eta_0+{\eta}_3 +{\eta}_4=0\,. \label{constraintFI-1}
\end{equation}
Note that the form of the constraint is of the form $\sum_i \eta_i l_i=0$, where the sum runs over the nodes of the $\hat{D}_4$ quiver while $l_i$ denotes the Dynkin label of the $i$-th node.\\
One therefore has a quiver with the gauge group $U(1)^4 \times U(2) // U(1)$. On the B-model side, this simply amounts to imposing the constraint \eqref{constraintFI-1}. The mirror map formally remains the same, subject to this extra constraint, see \figref{fig:quiver222SU(2)}.

One may wonder however if there is a preferred choice in imposing extra constraint \eqref{constraintFI-1}. For instance, one could try to consider $\hat{D}_4$ quiver with $SU(2)$ node in the middle instead of the $U(2)$. As it turns out, this choice, as well as all the other $U(1)$ quotients, except for $S(U(1)^4\times U(2))$ or the configuration presented in \figref{fig:quiverD4} where an overall $U(1)$ factor is decoupled, does not give a correct mirror description. Below we demonstrate this fact by computing the Hilbert series on the Coulomb branches \cite{Cremonesi:2013lqa} of the corresponding $\hat{D}_4$ quivers.

\subsection{Checking mirror symmetry: The Hilbert series of $\hat{D}_4$ quiver}
The Hilbert series of $\hat{D}_4$ quiver in \figref{fig:quiver222SU(2)} can be obtained via formula \eref{Hilbert_series}; the result is as follows:
\bea \label{HD4hat}
H[\widehat{D}_4](t, \vec x) &= \sum_{m_1,\ldots, m_4 \in \BZ} \; \sum_{n_1 \geq n_2 = 0} t^{\Delta(\vec m, \vec n)}\left(z_0^{n_1+n_2}  \prod_{i=1}^4 {z_i^{m_i}}  \right)\times  \nn \\
& \qquad  \left[ P_{U(1)}(t) \right]^4 (1-t) P_{U(2)} (t, \vec n)\,.
\eea
Let us explain each part of the above formula as follows.  The dimension formula of monopole operators is given by
\bea
\Delta(\vec m, \vec n) = - |n_1-n_2|+ \frac{1}{2}\sum_{i=1}^4 \sum_{j=1}^2 |m_i - n_j| ~,
\eea
where $m_1,\dots, m_4$ are monopole charges associated with each $U(1)$ group, and $n_1, n_2$ are monopole charges associated with $U(2)$ in the center of the quiver. The product in the brackets in the first line of \eref{HD4hat} corresponds to the refinement of various global charges: each fugacity $z_i$ keeps track of the charge $m_i$ for each $U(1)$ and the fugacity $z_0$ keeps track of the topological charge $n_1+n_2$ of $U(2)$.  Functions $P_{U(1)}(t)$ and $P_{U(2)} (t, \vec n)$ are contributions from the Casimir invariants of $U(1)$ and $U(2)$ gauge groups given by \eref{classical_dressing}:
\bea
P_{U(1)}(t) = \frac{1}{1-t}~, \qquad 
P_{U(2)}(t, \vec n) = \begin{cases} \frac{1}{(1-t)(1-t^2)}~, & \qquad n_1 = n_2 \\ \frac{1}{(1-t)^2}~, & \qquad n_1 \neq  n_2~. \end{cases}
\eea

An overall $U(1)$ in the quiver \figref{fig:quiver222SU(2)} is factored out from the $U(2)$ middle node via the following steps:
\ben
\item fixing the charge $n_2$ associated with $U(2)$ gauge group to zero, as stated in the second summation; 
\item multiplying the factor $(1-t)$ in front of $P_{U(2)} (t, \vec n)$, and 
\item by imposing the condition 
\bea
z_0^2 z_1 z_2 z_3 z_4 =1~.
\eea  
\een
Note that this procedure of gauge fixing is not unique.  One can instead, for example, take any of the $U(1)$ nodes in \figref{fig:quiver222SU(2)} to be a flavour node (see \eg, section 3.4 of \cite{Cremonesi:2013lqa}) and obtain the same answer.

In order to make the $SU(2)$ associated to each leg manifest and to fix the overall $U(1)$, we write
\bea \label{fixfuga}
z_0 = x_1 x_2 x_3 x_4, \qquad z_i= x_i^{-2}~,
\eea
where $x_i$ (with $i=1,2,3,4$) are the $SU(2)$ fugacities corresponding to each leg.

Since $SU(2)$ gauge theory with $4$ flavors has an $SO(8)$ flavour symmetry, it is expected that the Hilbert series should be written in terms of characters of $SO(8)$ representations.  In order to do so, we may use the following fugacity map\footnote{This is the same as (4.5) of \cite{Hanany:2010qu}}
\bea
y_1 = x_1 x_2, \quad y_2 = x_2^2, \quad y_3 =x_3 x_2, \quad y_4 = x_4 x_2~,
\eea
where $y_1, y_2, y_3, y_4$ are the $SO(8)$ fugacities and $x_1, x_2, x_3, x_4$ are the $SU(2)^4$ fugacities.  To make a connection with the fugacities $z_0, z_1, \ldots, z_4$, we have
\bea
z_0 = y_1 y_2^{-1} y_3 y_4~, \quad z_1 = y_1^2 y_2^{-1}, \quad z_2 = y_2, \quad z_3 = y_2^{-1} y_3^2, \quad z_4= y_2^{-1} y_4^2~.
\eea

In terms of $y_1, \ldots, y_4$, the power series of \eref{HD4hat} in $t$ is given by
\bea \label{SO8exp}
H[\widehat{D}_4](t, \vec y)  = \sum_{k=0}^\infty [0,k,0,0]_{\vec y} t^{k}~,
\eea
where $[0,k,0,0]_{\vec y}$ denotes the character of representation $[0,k,0,0]$ of $SO(8)$ written in terms of $y_1, \ldots, y_4$.  
Henceforth, we use a square bracket $[ \ldots]_{\vec y}$ to denote the character of our representation written in terms of the variables in the subscript, which is $\vec y$ in this case.\footnote{The characters can be computed using Weyl's character formula or using LiE online service in the following link: \url{http://young.sp2mi.univ-poitiers.fr/cgi-bin/form-prep/marc/LiE_form.act?action=character&type=D&rank=4&highest_rank=8}.}  To avoid the cumbersome notation, we drop the subscript when there is no potential confusion.
 
Setting $y_i=1$, which amounts to taking the dimension of the representations in \eref{SO8exp}, we obtain the unrefined Hilbert series
\bea \label{unrefD4}
H[\widehat{D}_4](t, \{ y_i =1 \}) &= \sum_{k=0}^\infty \dim~[0,k,0,0] t^k \nn \\
&= \sum_{k=0}^\infty \frac{(k+1)(k+2)^3(k+3)^3(k+4)(2k+5)}{4320} t^k \nn \\
&=  \frac{(1+t)(1+17 t +48 t^2 +17 t^3+t^4)}{(1-t)^{10}} \nn \\
&=1 + 28 t + 300 t^2 + 1925 t^3 + 8918 t^4+\ldots~.
\eea

\paragraph{A remark on gauge fixing.}
We emphasize that the gauge fixing procedure described above is {\it different} from taking the middle node of \figref{fig:quiver222SU(2)} to be $SU(2)$.  Let us compare \eref{HD4hat} with the Hilbert series of the same quiver with the central node taken to be $SU(2)$.  The latter is given by
\bea \label{HSSU2central}
\widetilde{H}(t, \vec x)&= \sum_{m_1,\ldots, m_4 \in \BZ} \; \sum_{n = 0}^\infty t^{\widetilde{\Delta}(\vec m, \vec n)}\left(  \prod_{i=1}^4 {x_i^{-2m_i}}  \right)\times  \nn \\
& \qquad  \left[ P_{U(1)}(t) \right]^4 (1-t) P_{U(2)} (t, n, -n) ~,
\eea
where
\bea
\widetilde{\Delta}(\vec m, n) = - |2n|+ \frac{1}{2}\sum_{i=1}^4 \sum_{j=1}^2 (|m_i - n|+|m_i + n|) ~.
\eea
Indeed, the summand of \eref{HSSU2central} is equal to that of \eref{HD4hat} with $n_1=n_2=-n$.  However, after taking the summations into account, we see that this is not compatible with the gauge fixing described above, where we fixed $n_2=0$ rather than taking $n_2$ to be $-n_1$.  For reference, we present a few terms of \eref{HSSU2central}:
\bea
\widetilde{H}(t, \{ x_i =1\}) &= 1 + 12 t + 156 t^2 + 949 t^3 + 4486 t^4+\ldots~.
\eea
This is different from \eqref{unrefD4}.

In the following section, while discussing a balanced or any generic flavorless 3d quiver, we will implicitly assume that an overall $U(1)$ vector multiplet decouples from the theory. Note that in the classical analysis of the parameter space of vacua $\CL$ there is a notion of the ``center of mass'' for twisted masses and FI terms, which is naturally associated with the translational symmetry of the system of D5 and NS5 branes. Therefore one can gauge the entire global symmetry of the quiver on the level of $\CL$, except for a single $U(1)$ factor. This is the reason why in the A model quiver in \figref{fig:quiverD4} one global $U(1)$ symmetry is present. However, as we have shown above, the Hilbert series for this quiver and for balanced flavorless $\hat{D}_4$ from \figref{fig:quiver222SU(2)} are identical. Classically the statement boils down to the fact that a linear rank-$L$ quiver has $L+1$ NS5 branes and therefore $L+1$ NS5 positions. However, the FI parameters appearing in the $\mathcal{N}=4$ Lagrangian correspond to the differences $\eta_i=t_{i+1}-t_i$. In the remainder of the paper we shall be using this observation -- by specifying a proper submanifold in $\CL$ and obtaining a quiver with a single global $U(1)$ symmetry we will automatically arrive at the corresponding flavorless quiver and its mirror description. 

The rest of the paper consists of the analysis of 3d mirror pairs involving quivers of different shapes: generic $\hat{D}_N$ and star-shaped quivers in the next section, exceptional $E_{6,7,8}$ quivers and their extensions in \secref{Sec:Emirrors}, and framed $A_1$ $Sp(N_c)$ quivers in \secref{Sec:Spmirrors}.

\section{$\hat{D}_N$ and Star-shaped Quivers}\label{Sec:Dquivers}
In this section, we analyze various examples of the framed $\hat{D}_N$ quivers shown in \tabref{Tab:StarD} and closely related star-shaped quivers using the Abelian gauging technique. 

A straightforward generalization of the $\hat{D}_4$ quiver from \figref{fig:quiver222SU(2)} is a star-shaped quiver with more than four nodes with $U(1)$ gauge groups on those nodes.  
Later we shall also discuss star quivers with longer legs.

\subsection{Star-shaped Quivers via Abelian Gauging}\label{star-quiver}
Let us consider an obvious generalization of the example from \figref{fig:quiverD4}. 

\begin{figure}[H]
\begin{center}
\includegraphics[scale=0.3]{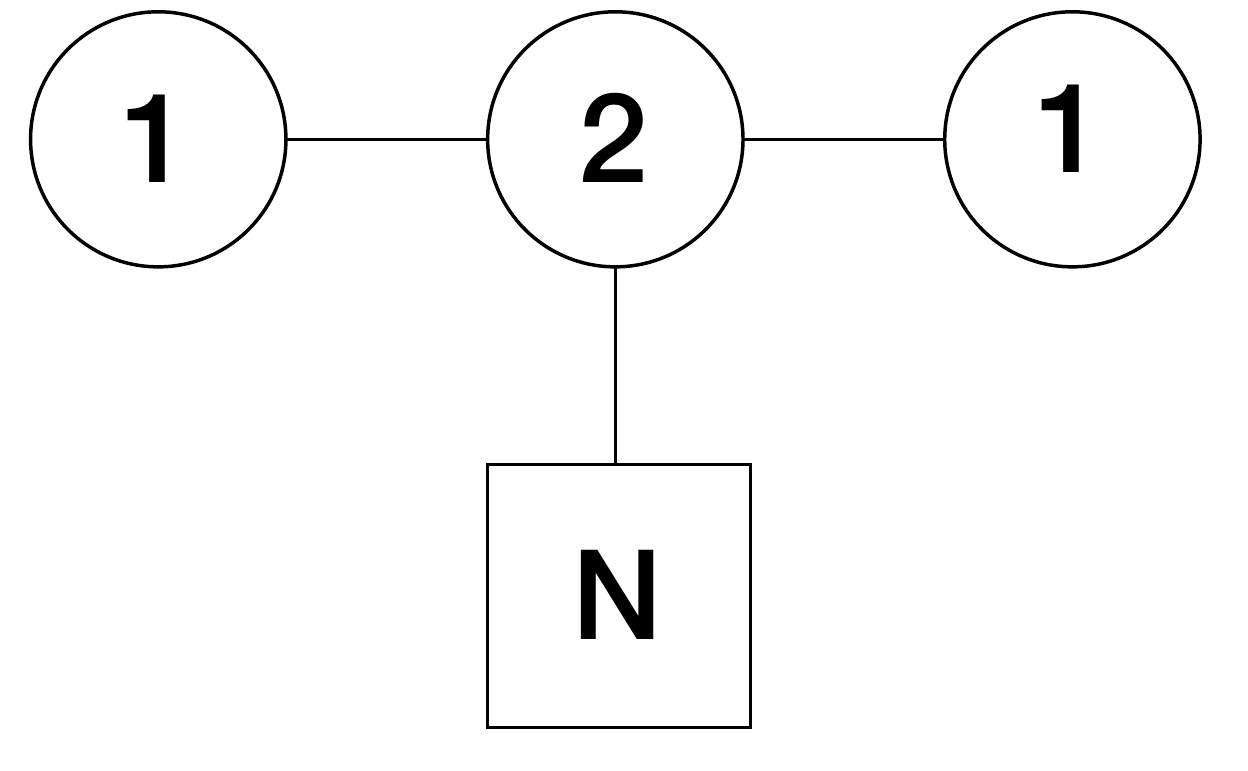}\quad\quad\quad \includegraphics[scale=0.3]{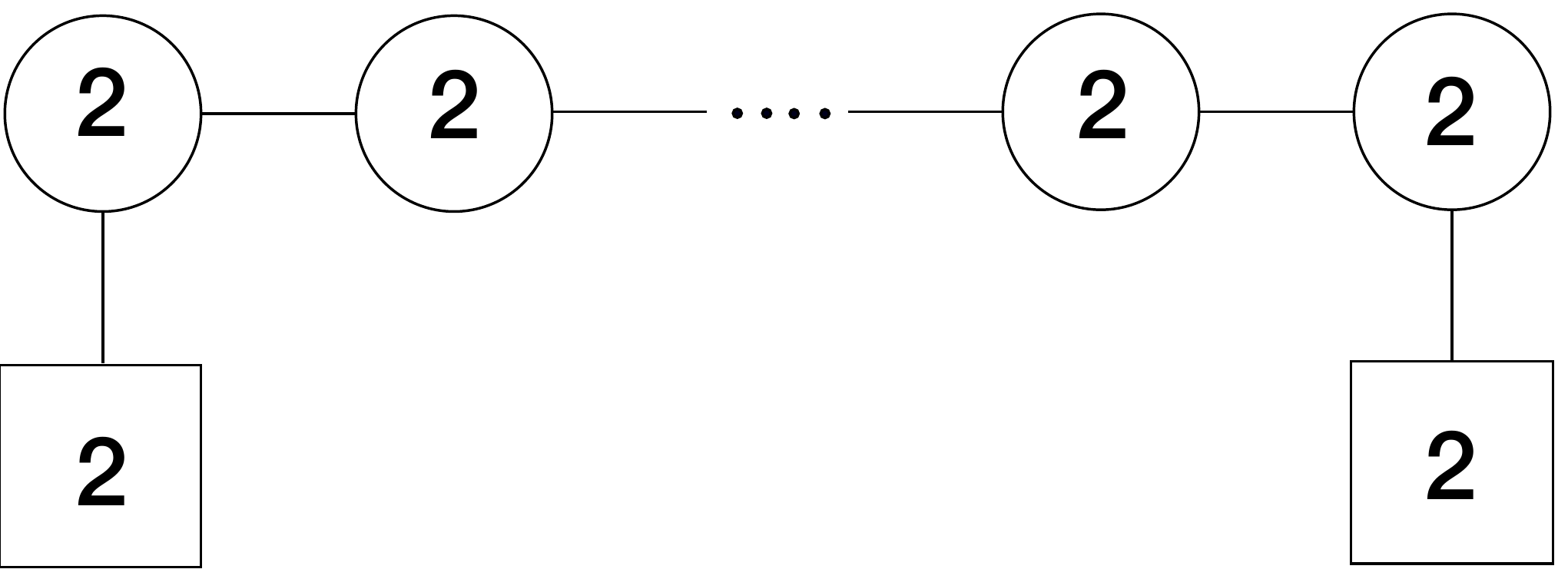}
\caption{Linear quivers which are mirror dual to each other. For the quiver on the right there are $N-1$ $U(2)$ gauge groups in the chain.}
\label{fig:quiver121}
\end{center}
\end{figure}

Consider a mirror pair of linear quivers (\figref{fig:quiver121}) where the  A-model is $(1,0)(2,N)(1,0)$ quiver and the B-model is
$(2,2)_1(2,0)_2\cdots(2,0)_{N-2}(2,2)_{N-1}$ -- the subscript denotes that there are $N-1$ nodes in the latter quiver . 

\begin{figure}[H]
\begin{center}
\includegraphics[scale=0.3]{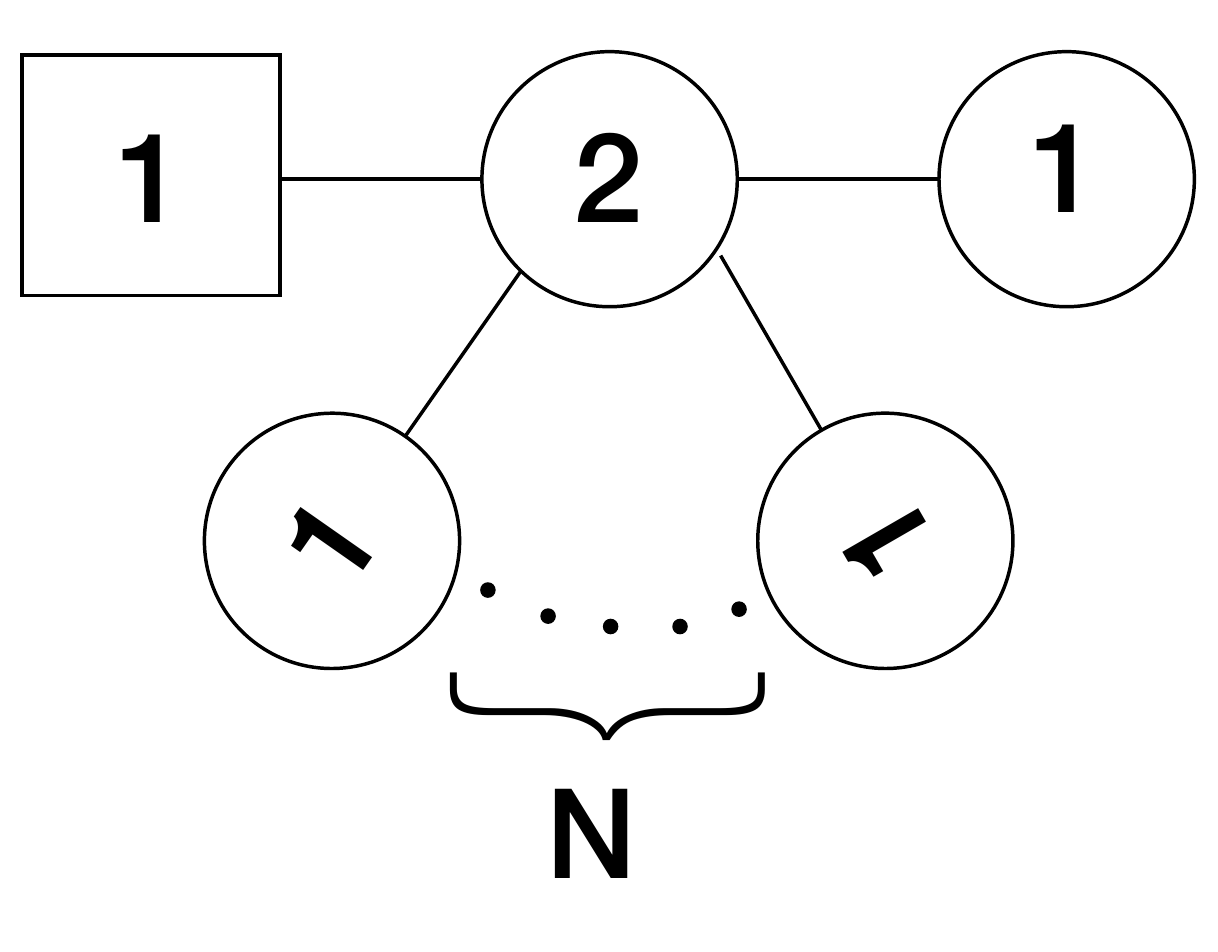}\quad\quad\quad \includegraphics[scale=0.35]{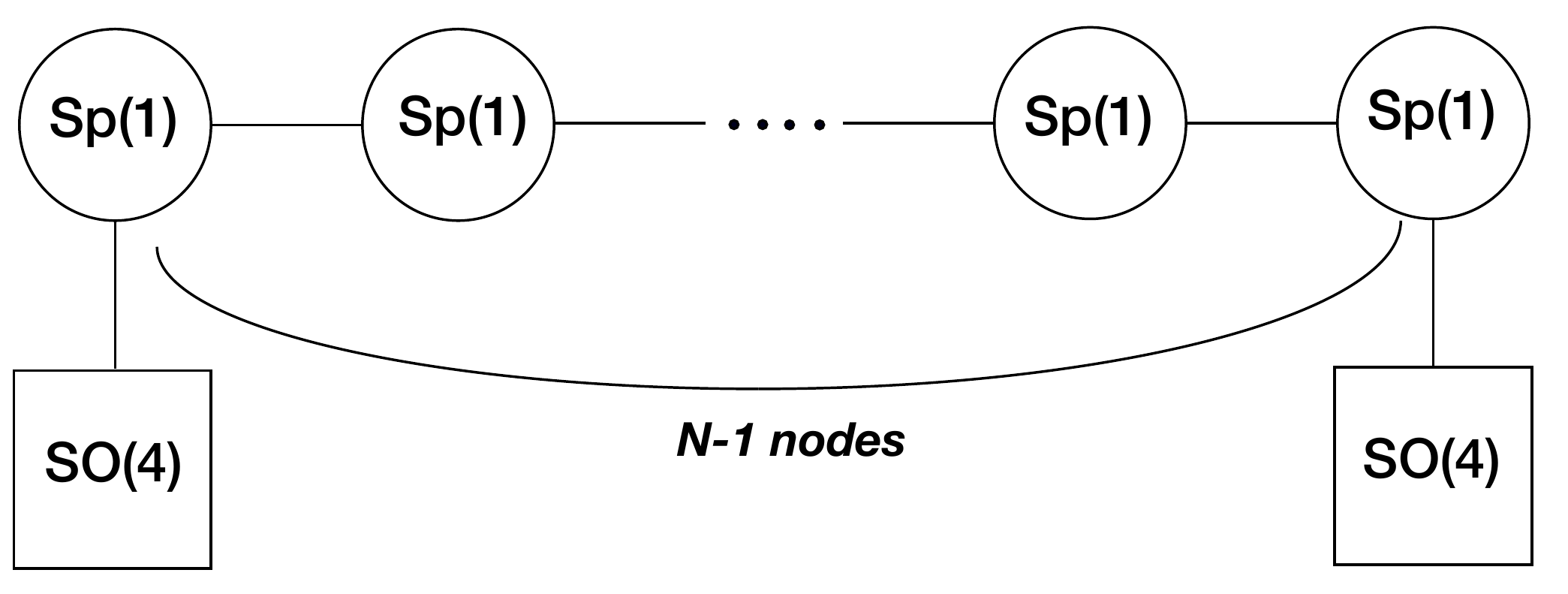}
\caption{New mirror pair obtained from linear quivers \figref{fig:quiver121}. For the quiver on the right there are $N-1$ $Sp(1)$ gauge groups in the chain.}
\label{fig:quiver121new}
\end{center}
\end{figure}

We can now see how the `gauging' trick works, namely, it splits a flavor node $M_i$ in an $A_L$ quiver with $U(N_i)$ gauge group on its $i$-th node into a maximum of $M_i$ $U(1)$ gauge group factors which leads us to more generic constructions of mirrors, as shown for example in \figref{fig:quiver121new}. 

Now, let us demonstrate how the gauging procedure may be performed at the level of partition functions.
For a generic $N$, the partition function on a round 3-sphere for the A-model is
\begin{equation}
\begin{split}
&\CZ_A=\int \prod^2_{\alpha=1} d s_{\alpha}  \frac{d^2 s_0}{2!} \frac{\prod^2_{\alpha=1}  e^{2\pi i s_{\alpha} \eta_{\alpha}} \prod^2_{i=1} e^{2\pi i s^i_0 \eta_0} \sinh^2{\pi(s^1_0-s^2_0)}}{\prod^2_{i=1}\cosh{\pi(s_1-s^i_0)}\prod^N_{a=1} \cosh{\pi(s^i_0+M_a)}\cosh{\pi(s_2-s^i_0)}} \\
&= -\int  \frac{d^2 s_0}{2!} \frac{\prod^2_{i=1} e^{2\pi i s^i_0(t_2-t_3)} \left(e^{2\pi i s^1_0(t_1-t_2)}-e^{2\pi i s^2_0(t_1-t_2)}\right)\left(e^{2\pi i s^1_0(t_3-t_4)}-e^{2\pi i s^2_0(t_3-t_4)}\right)}{4\sinh{\pi(t_1-t_2)}\sinh{\pi(t_3-t_4)}\prod^N_{a=1} \prod^2_{i=1}\cosh{\pi(s^i_0+M_a)}}
\end{split}
\end{equation}
where we have defined $\eta_1=t_1-t_2, \eta_0=t_2-t_3, \eta_2=t_3-t_4$, with the constraint $t_1+t_2+t_3+t_4=0$. The masses obey the constraint $\sum^N_{a=1}M_a=0$. To obtain the second line, we have simply integrated out the two boundary $U(1)$ nodes in the integral.

The partition function for the B-model $(2,2)_1(2,0)_2 \ldots (2,0)_{N-2} (2,2)_{N-1}$ is
\begin{equation}
\begin{split}
&\CZ_B=\int \prod^{N-1}_{\alpha=1} \frac{d^2 s_{\alpha}}{2!}  \frac{\prod^{N-1}_{\alpha=1} \prod^2_{i=1} e^{2\pi i s^i_{\alpha} \widetilde{\eta}_{\alpha}}  \sinh^2{\pi(s^1_{\alpha}-s^2_{\alpha})}}{\prod^2_{i=1}\prod^2_{a=1}\cosh{\pi(s^i_1+{m}_a)}\cosh{\pi(s^i_{N-1}+\bar{m}_a)}\prod^{N-2}_{\alpha=1} \prod_{i,j}\cosh{\pi(s^i_{\alpha}-s^j_{\alpha+1})}} 
\end{split}
\end{equation}
where we again set $\widetilde{\eta}_{\alpha}=\widetilde{t}_{\alpha} -\widetilde{t}_{\alpha +1}$ with the constraint $\sum^{N}_{\alpha=1} \widetilde{t}_{\alpha}=0$. As shown in the case for $N=2$, one can show that $\CZ_A$ and $\CZ_B$ merely differ by a phase. Using Cauchy determinant formula and the associated tool-box for manipulating $S^3$ partition functions, as explained for example in \cite{Dey:2013nf}, we obtain,
\begin{equation}
\begin{split}
\CZ_B(m_a,\bar{m}_a;\widetilde{t}_{\alpha})=&e^{2\pi i\widetilde{t}_1(t_1+t_2)-2\pi i\widetilde{t}_N(t_3+t_4)}\CZ_A (M_{\alpha}; t_i)\\
\implies \CZ_A(M_{\alpha}; t_i)= &e^{-2\pi i\widetilde{t}_1(t_1+t_2)+2\pi i\widetilde{t}_N(t_3+t_4)}\CZ_B (m_a,\bar{m}_a;\widetilde{t}_{\alpha})
\end{split}
\end{equation}
The mirror map is simply given by 
\begin{equation}
\begin{split}
&M_{\alpha} \leftrightarrow \widetilde{t}_{\alpha}\\
 &t_1,t_2 \leftrightarrow m_1,m_2\\
 &t_3,t_4 \leftrightarrow \bar{m}_1,\bar{m}_2.
\end{split}
\end{equation}
One can now gauge the Cartan of the $U(N)$ flavor symmetry labeled by $M_{\alpha} (=\widetilde{t}_{\alpha})$ in $N$ steps starting with $M_1(=\widetilde{t}_1)$. As before, in the gauging procedure, we treat the $M_{\alpha} (=\widetilde{t}_{\alpha})$s as independent complex parameters without any constraint. Therefore, gauging the first $U(1)$ in the A-model, which in the dual theory corresponds to one of the nodes with fundamental matter, we have
\begin{equation}
\begin{split}
&\widetilde{\CZ}^{(1)}_A(M_{2},\ldots.M_N; t_i,\eta_3)= e^{2\pi i\widetilde{t}_N(t_3+t_4-t_1-t_2-\eta_3)} \int \prod^{N-1}_{\alpha=1} \frac{d^2 s_{\alpha}}{2!} \prod^2_{i=1} \int \prod^{N-1}_{\alpha=2} e^{2\pi i s^i_{\alpha}\left( \widetilde{t}_{\alpha}-\widetilde{t}_{\alpha+1}\right)}  \\
&\times \frac{\delta(s^1_1+s^2_1)\prod^{N-1}_{\alpha=1} \sinh^2{\pi(s^1_{\alpha}-s^2_{\alpha})}}{\prod^2_{i=1}\prod^2_{a=1}\cosh{\pi(s^i_1+{t}_a+\frac{t_1+t_2+\eta_3}{2})}\cosh{\pi(s^i_{N-1}+{t}_{2+a} +\frac{t_1+t_2+\eta_3}{2} )}\prod^{N-2}_{\alpha=1} \prod_{i,j}\cosh{\pi(s^i_{\alpha}-s^j_{\alpha+1})}}
\end{split}
\end{equation}
where $\eta_3$ is the FI parameter corresponding to the gauged $U(1)$. The dual theory can be immediately read off from the above partition function - it is the same quiver as the B-model in \figref{fig:quiver121} with the first $U(2)$ replaced by a $Sp(1)$. The mirror map is also obvious from the above formula- the A-model has $(N-1)$ mass parameters matching the number of remaining $\widetilde{t}_{\alpha}$ parameters for the B-model. In addition, the 4 independent mass parameters of the B-model are related to the 4 independent FI parameters ($t_i$s with one constraint and $\eta_3$) in the A-model in the following fashion: 
\begin{equation}
\begin{split}
& \widetilde{t}_{\alpha} = M_{\alpha}\; (\alpha=2,\ldots,N)\\
&m_a={t}_a+\frac{t_1+t_2+\eta_3}{2}\; (a=1,2)\\
&\bar{m}_a={t}_{2+a} +\frac{t_1+t_2+\eta_3}{2}\; (a=1,2)
\end{split}
\end{equation}
Carrying on with gauging the second $U(1)$, one gets
\begin{equation}
\begin{split}
&\widetilde{\CZ}^{(2)}_A(M_{3},\ldots.M_N; t_i,\eta_3,\eta_4)= e^{2\pi i\widetilde{t}_N(t_3+t_4-t_1-t_2-\eta_3-\eta_4)} \int \prod^{N-1}_{\alpha=1} \frac{d^2 s_{\alpha}}{2!} \prod^2_{i=1} \int \prod^{N-1}_{\alpha=3} e^{2\pi i s^i_{\alpha}\left( \widetilde{t}_{\alpha}-\widetilde{t}_{\alpha+1}\right)}  \\
&\times \frac{\delta(s^1_1+s^2_1)\delta(s^1_2+s^2_2)\prod^{N-1}_{\alpha=1} \sinh^2{\pi(s^1_{\alpha}-s^2_{\alpha})}}{\prod_{a,i}\cosh{\pi(s^i_1+{t}_a+\frac{t_1+t_2+\eta_3}{2})}\cosh{\pi(s^i_{N-1}+{t}_{2+a} +\frac{t_1+t_2+\eta_3+\eta_4}{2} )}\prod^{N-2}_{\alpha=2} \prod_{i,j}\cosh{\pi(s^i_{\alpha}-s^j_{\alpha+1})}}\\
&\times \frac{1}{\prod_{i,j}\cosh{\pi(s^i_{1}-s^j_{2}-\frac{\eta_4}{2})}}
\end{split}
\end{equation}
The mirror theory is now given by the B-model in \figref{fig:quiver121} with the first two $U(2)$s replaced by $Sp(1)$s and the mirror map in this case can be read off as follows:
\begin{equation}
\begin{split}
& \widetilde{t}_{\alpha} = M_{\alpha}\; (\alpha=3,\ldots,N)\\
&m_a={t}_a+\frac{t_1+t_2+\eta_3}{2}\; (a=1,2)\\
&\bar{m}_a={t}_{2+a} +\frac{t_1+t_2+\eta_3+\eta_4}{2}\; (a=1,2)\\
&m^{bif}_1 =\frac{\eta_4}{2}
\end{split}
\end{equation}
Note that there is a non-zero mass for the hypermultiplet in the bifundamental representation of $Sp(1)\times Sp(1)$  in addition to the four fundamental masses. The number of FI parameters in the A-model therefore agrees with the number of mass parameters of the B-model.

Gauging $(N-1)$ of the $N$ $U(1)$s in a manner outlined above, one obtains the desired A-model of \figref{fig:quiver121new} . The resultant partition function
\begin{equation}
\begin{split}
&\widetilde{\CZ}^{(N-1)}_A(M_N; t_i,\eta_3,\eta_4,\ldots,\eta_{N+1})= \int \frac{d^2 s_{\alpha}}{2!} \frac{e^{2\pi i\widetilde{t}_N(t_3+t_4-t_1-t_2-\eta_3-\ldots -\eta_{N+1})}}{\prod^{N-2}_{\alpha=1} \prod_{i,j}\cosh{\pi(s^i_{\alpha}-s^j_{\alpha+1}-m^{bif}_{\alpha})}} \\
&\times\frac{\prod^N_{\alpha=1}\delta(s^1_{\alpha}+s^2_{\alpha})\prod^{N-1}_{\alpha=1} \sinh^2{\pi(s^1_{\alpha}-s^2_{\alpha})}}{\prod_{a,i}\cosh{\pi(s^i_1+{t}_a+\frac{t_1+t_2+\eta_3}{2})}\cosh{\pi(s^i_{N-1}+{t}_{2+a} +\frac{t_1+t_2+\eta_3+\eta_4+\ldots+ \eta_{N+1}}{2} )}}
\end{split}
\end{equation}
Therefore, up to a field-independent phase, one obtains the expected dual theory of  \figref{fig:quiver121new}. The masses of the B-model are related to the FI-parameters of the A-model in the following way,
\begin{equation}
\begin{split}
&m_a={t}_a+\frac{t_1+t_2+\eta_3}{2}\;\; (a=1,2)\\
&\bar{m}_a={t}_{2+a} +\frac{t_1+t_2+\eta_3+\eta_4+\ldots+ \eta_{N+1}}{2}\;\; (a=1,2)\\
&m^{bif}_{\alpha} =\frac{\eta_{\alpha+3}}{2} \;\; (\alpha=1,2,\ldots,N-2)
\end{split}
\end{equation}
One can further gauge the remaining mass $M_N$. The partition function of the gauged theory is
\begin{equation}
\begin{split}
&\widetilde{\CZ}^{(N)}_A(t_i,\eta_3,\eta_4,\ldots,\eta_{N+1})= \int \frac{d^2 s_{\alpha}}{2!} \frac{\delta(-t_3-t_4+t_1+t_2+\eta_3+\ldots +\eta_{N+1}+\eta_{N+2})}{\prod^{N-2}_{\alpha=1} \prod_{i,j}\cosh{\pi(s^i_{\alpha}-s^j_{\alpha+1})}} \\
&\times\frac{\prod^N_{\alpha=1}\delta(s^1_{\alpha}+s^2_{\alpha})\prod^{N-1}_{\alpha=1} \sinh^2{\pi(s^1_{\alpha}-s^2_{\alpha})}}{\prod_{a,i}\cosh{\pi(s^i_1+{t}_a+\frac{t_1+t_2+\eta_3}{2})}\cosh{\pi(s^i_{N-1}+{t}_{2+a} +\frac{t_1+t_2+\eta_3+\eta_4+\ldots+ \eta_{N+1}}{2} )}}
\end{split}
\end{equation}
The mirror map remains unchanged. As we saw in the case for $N=2$, gauging the final $U(1)$ imposes an extra constraint on the FI parameters of the A-model, namely
\begin{equation}
\eta_1+\eta_2+\eta_3+\ldots +\eta_{N+1}+\eta_{N+2}+2\eta_0=0\,.
\end{equation}
The A-model is therefore a Star-shaped quiver with the gauge group $U(1)^{N+2}\times U(2) // U(1)$ with $U(2)$ being the central node.

\begin{figure}[H]
\begin{center}
\includegraphics[scale=0.4]{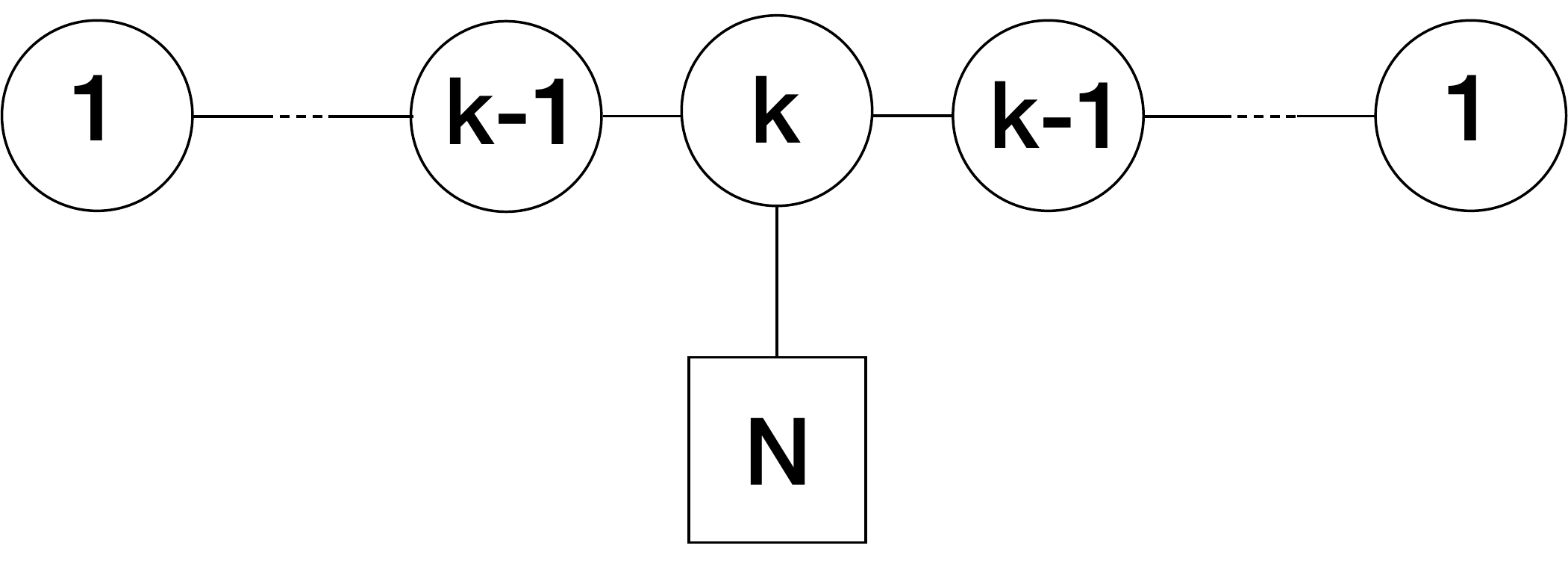}
\caption{Linear $A_{2k-1}$ quiver with two $T[U(k-1)]$ with framing at the middle node which is used to derive the mirror dual for star shaped quivers.}
\label{fig:quiveNk2tails}
\end{center}
\end{figure}

At this point it is easy to propose a higher-rank generalization of the mirror pair from \figref{fig:quiver121new}. If we start off with the following linear quiver 
\begin{equation}
(k,k)_1(k,0)_2\dots (k,0)_{N-2}\dots(k,k)_{N-1}\,,
\label{eq:QuivLinBalk}
\end{equation}
then its mirror will be a $U(k)$ theory with $N$ flavors and two $T[U(k-1)]$ tails attached to it (see \figref{fig:quiveNk2tails}). 
We can now gauge the maximal torus of the $U(N)$ flavor symmetry on the middle node. This gauging imposes a simple constraint on the Cartan of each $U(k)$ gauge group in the dual quiver \eqref{eq:QuivLinBalk} (and a constraint on the FI parameters if the Cartan of $U(N)$ is fully gauged) which amounts to removing a $U(1)$ subgroup from each $U(k)$ gauge group. This leads us to the mirror pair presented in the first line of \tabref{Tab:StarD}.

\subsection{Flavorless $\hat{D}_{N}$ Quivers}
\begin{figure}[H]
\begin{center}
\includegraphics[scale=0.6]{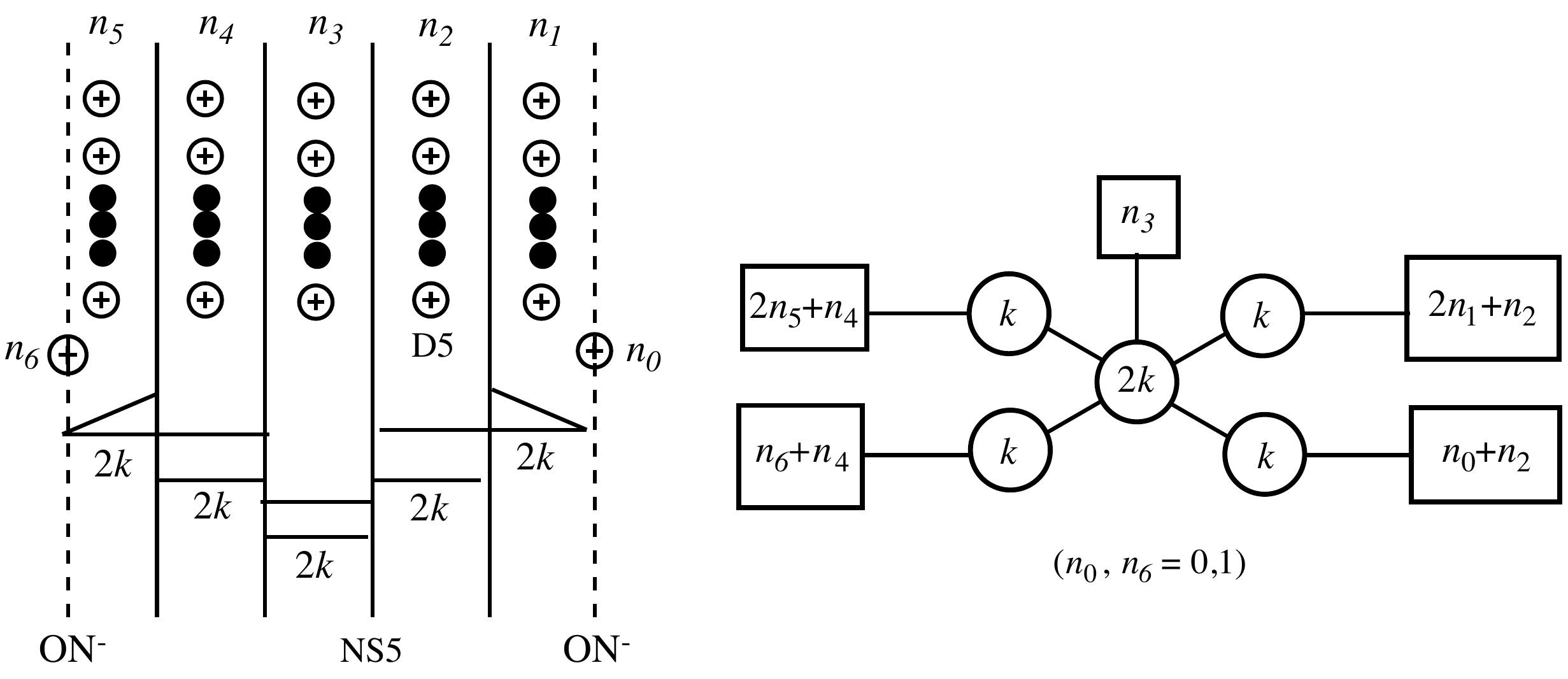}
\caption{Brane construction of a generic flavored $\hat{D}_{4}$ quiver. The labels indicate that there are $2k$ D3 branes in each interval, \ie~ $k$ copies of the D3 branes drawn in the diagram.  The numbers $n_1, \ldots, n_5$ label the numbers of D5-branes at each interval, and the numbers $n_0, n_6=0,1$ labels the numbers of D5-branes stuck on each $ON^-$ plane.}
\label{fig:BraneflavoredD4}
\end{center}
\end{figure}

The $\hat{D}_{N}$ quiver and its mirror may be obtained by starting from the same pair of linear quivers as we used to obtain the Star-shaped quiver and its mirror dual in \secref{star-quiver}. 
The brane constructions of such quivers using $ON^-$ planes are depicted in \figref{fig:BraneflavoredD4}.  In \cite{Lindstrom:1999pz} several of them are considered and interesting global symmetries of the quivers are discussed.  

Start with the following mirror pair of linear quivers (see \figref{fig:quiver121}, with $N \to N-2$) \\
\textbf{A-model}: $(2,2)_1(2,0)_2\ldots (2,0)_{N-4}(2,2)_{N-3}$  \\
\textbf{B-model}: $(1,0)(2,N-2)(1,0)$. \\
 
 From our computation in the previous section one can readily see that (note that what we called B-model there is the A-model in the present case)
\begin{equation}
\begin{split}
\CZ_A(m_a,\bar{m}_a;\widetilde{t}_{\alpha})=e^{2\pi i\widetilde{t}_1(t_1+t_2)-2\pi i\widetilde{t}_{N-2}(t_3+t_4)}\CZ_B (M_{\alpha}; t_i)\,.
\end{split}
\end{equation}
The mirror map is the same as before and can be read off from the above equation.
\begin{equation}
\begin{split}
&M_{\alpha} \leftrightarrow \widetilde{t}_{\alpha}\\
 &t_1,t_2 \leftrightarrow m_1,m_2\\
 &t_3,t_4 \leftrightarrow \bar{m}_1,\bar{m}_2.
\end{split}
\end{equation}
In the example of the Star-shaped quiver, we gauged the $U(N)$ flavor symmetry of the linear quiver $(1,0)(2,N)(1,0)$ as $U(1)^N$ to obtain the $SU(2)\times U(1)^N$ Star-shaped quiver. To obtain the $\hat{D}_{N}$ quiver, we gauge each of the two $U(2)$ flavor symmetries of the linear quiver $(2,2)_1(2,0)_2\ldots (2,0)(2,2)_{N-3}$ as $U(1)^2$. The partition function of the gauged theory is given as
\begin{equation}
\begin{split}
\widetilde{\CZ}_A(\zeta_j;\widetilde{t}_{\alpha})=&\int  \prod^2_{\alpha=1} d s_{\alpha}  \frac{d^2 s_0}{2!} \prod^4_{j=1}d t_j e^{2\pi i \zeta_j t_j} e^{2\pi i\widetilde{t}_1(t_1+t_2)-2\pi i\widetilde{t}_{N-2}(t_3+t_4)} e^{2\pi i s_{1} (t_1-t_2)} e^{2\pi i s_{2} (t_3-t_4)}\\
&\times \frac{ \prod^2_{i=1} e^{2\pi i s^i_0 (t_2-t_3)} \sinh^2{\pi(s^1_0-s^2_0)}}{\prod^2_{i=1}\cosh{\pi(s_1-s^i_0)}\prod^{N-2}_{a=1} \cosh{\pi(s^i_0+\widetilde{t}_a)}\cosh{\pi(s_2-s^i_0)}} \\
\end{split}
\end{equation}
Integrating over the $t_j$s, we get
\begin{equation}
\begin{split}
\widetilde{\CZ}_A(\zeta_j;\widetilde{t}_{\alpha})=&\int  \prod^2_{\alpha=1} d s_{\alpha}  \frac{d^2 s_0}{2!} \delta(s_1+\zeta_1+\widetilde{t}_1)\delta(s_2-\zeta_4+\widetilde{t}_{N-2})\delta(s^1_0+s^2_0-s_1+\widetilde{t}_1+\zeta_2)\\
&\times \frac{ \delta(s^1_0+s^2_0-s_2+\widetilde{t}_{N-2}-\zeta_3) \sinh^2{\pi(s^1_0-s^2_0)}}{\prod^2_{i=1}\cosh{\pi(s_1-s^i_0)}\prod^{N-2}_{a=1} \cosh{\pi(s^i_0+\widetilde{t}_a)}\cosh{\pi(s_2-s^i_0)}} \\
=& \int    \frac{d^2 s_0}{2!}  \delta(s^1_0+s^2_0+\zeta_1+2\widetilde{t}_{1}+\zeta_2) \delta(s^1_0+s^2_0-\zeta_4+2\widetilde{t}_{N-2}-\zeta_3)\\
&\times \frac{\sinh^2{\pi(s^1_0-s^2_0)}}{\prod^2_{i=1}\cosh{\pi(s^i_0+\zeta_1+\widetilde{t}_1)}\prod^{N-2}_{a=1} \cosh{\pi(s^i_0+\widetilde{t}_a)}\cosh{\pi(s^i_0-\zeta_4+\widetilde{t}_{N-2})}}
\end{split}
\end{equation}
Finally, shifting the integration variables appropriately, we obtain the final form of the partition function for the flavorless $\hat{D}_{N}$ quiver.
\begin{equation}
\begin{split}
&\widetilde{\CZ}_A(\zeta_j;\widetilde{t}_{\alpha})= \delta(\zeta_1+2\widetilde{t}_{1}+\zeta_2+\zeta_4-2\widetilde{t}_{N-2}+\zeta_3) \int    \frac{d^2 s_0}{2!} \delta(s^1_0+s^2_0)\sinh^2{\pi(s^1_0-s^2_0)}\\
&\times \frac{1}{\prod^2_{i=1}\cosh{\pi(s^i_0+\frac{(\zeta_1-\zeta_2)}{2})}\prod^{N-2}_{a=1} \cosh{\pi(s^i_0-\frac{\zeta_1+\zeta_2}{2}-\widetilde{t}_1+\widetilde{t}_a)}\cosh{\pi(s^i_0+\frac{\zeta_3-\zeta_4}{2})}}
\end{split}
\end{equation}
The delta function indicates that there exists one constraint involving FI parameters of the $\hat{D}_{N}$ quiver, which is equivalent to saying that an overall $U(1)$ factor decouples from the gauge group. Note that this should be taken as part of the definition of the flavorless $\hat{D}_{N}$ quiver. Explicitly, the constraint can be written as,
\begin{equation}
\zeta_1+\zeta_2+\zeta_3+\zeta_4+2\widetilde{t}_{1}-2\widetilde{t}_{N-2}=0 \Leftrightarrow \sum^4_{i=1}\zeta_i +2\sum^{N-3}_{j=1}\widetilde{\eta}_j=0\,.
\end{equation}
The constraint is again of the form $\sum_i \eta_i l_i=0$ - where the sum runs over all the nodes of the quiver and $l_i$ denotes the Dynkin label of the $i$-th node. Taken with the other constraint $\sum_a t_a=0$, this tells us that there are exactly $N$ independent FI parameters - $N+2$ parameters with 2 constraints.\\
The mirror dual of the $\hat{D}_{N}$ quiver so defined, can now be read off from the partition function of the theory -- it is a $Sp(1)$ gauge theory with $N$ fundamental hypers. The mirror map relating the $N$ masses to the FI parameters of the $\hat{D}_{N}$ quiver is
\begin{equation}
\begin{split}
& M_1=\frac{\zeta_2-\zeta_1}{2} \\
& M_a=\frac{\zeta_1+\zeta_2}{2}+\widetilde{t}_1-\widetilde{t}_a \; \;(a=1,2,3,\ldots,N-2)\\
& M_N=\frac{\zeta_4-\zeta_3}{2}
\end{split}
\end{equation}
Therefore the Abelian gauging technique allows one to derive the $\hat{D}_{N}$ quiver and its mirror from a pair of mirror dual linear quivers.

\subsection{$\hat{D}_{N}$ Quivers with Single Framing}\label{sec:DhatSing}
Let us start analyzing quivers with framing. For framed $\hat{D}_N$ quivers there are two obvious cases which require separate analysis: a framed node on the bifurcated edge of the quiver or a framed internal node. We look at both cases below.

\subsubsection{Framing on an internal node}
\begin{figure}[H]
\begin{center}
\includegraphics[scale=0.35]{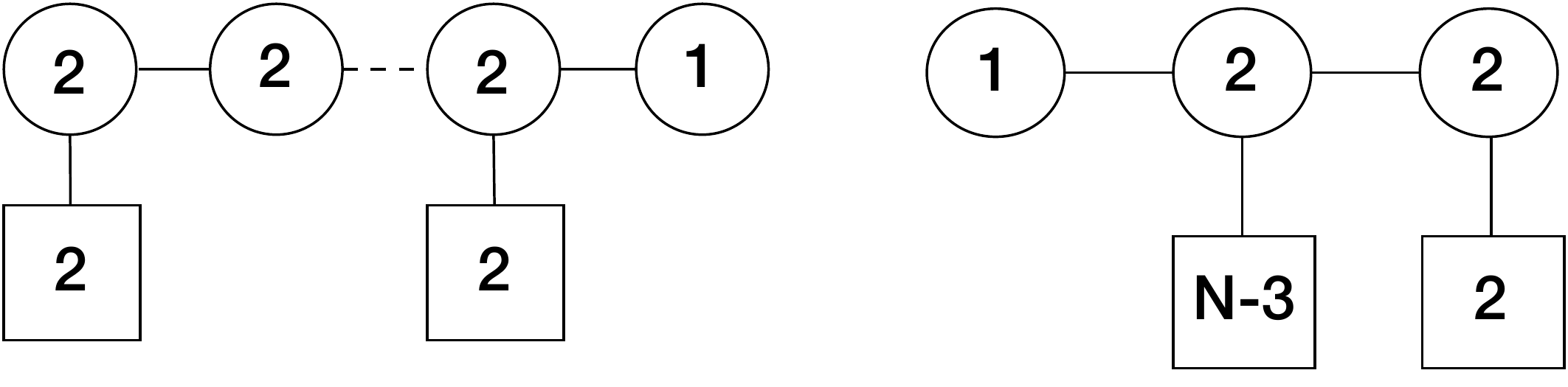}
\caption{A-model $(2,2)_{1}(2,0)_2\ldots(2,0)_{N-4}(2,2)_{N-3}(1,0)_{N-2}$ and B-model $(1,0)(2,N-3)(2,2)$.}
\label{fig:MirrPair4311}
\end{center}
\end{figure}

\begin{figure}[H]
\begin{center}
\includegraphics[scale=0.3]{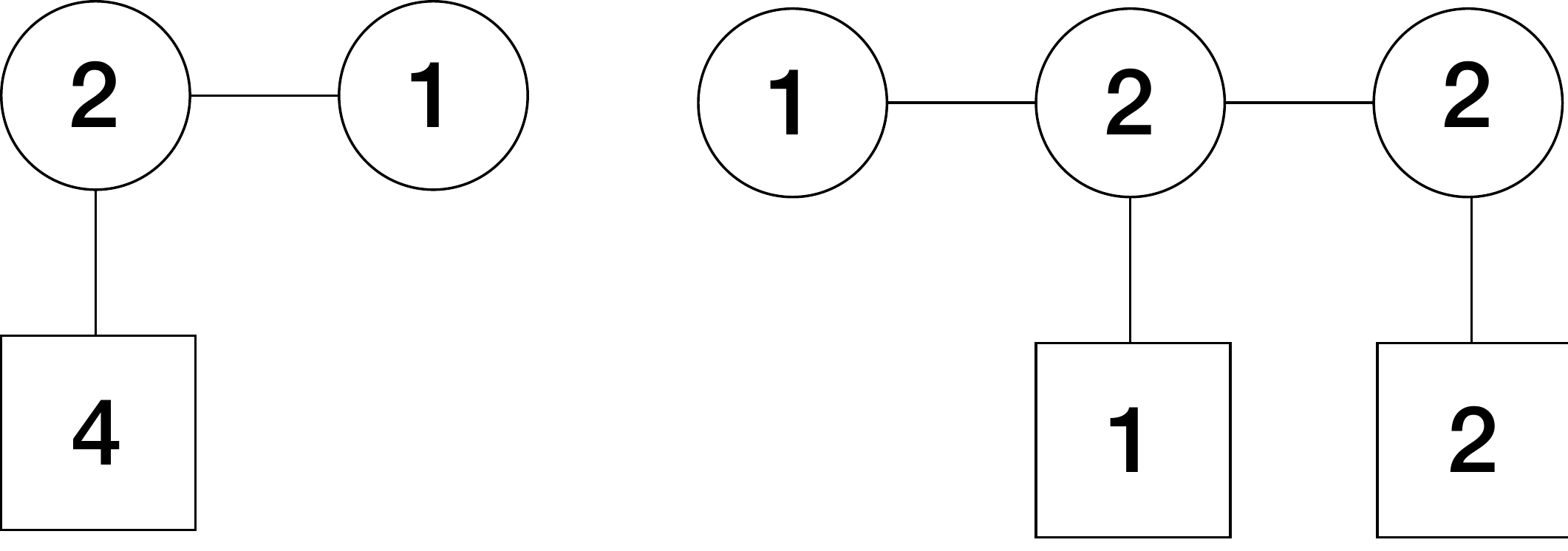}
\caption{A-model $(2,4)(1,0)$ and B-model $(1,0)(2,1)(2,2)$.}
\label{fig:MirrPair431}
\end{center}
\end{figure}

In order to obtain a generic mirror pair in this class by Abelian gauging, we start from the following linear quivers (\figref{fig:MirrPair4311})
There is however a special case when $N=4$ -- the A-model quiver looks slightly different (see \figref{fig:MirrPair431}) in this case.
The partition functions of two linear quivers from \eqref{fig:MirrPair4311} are related in the following way
\begin{equation}
\begin{split}
\CZ_A(m_a;\widetilde{t}_{\alpha})=e^{2\pi i\widetilde{t}_1(m_1+m_2-m_3)-2\pi i\widetilde{t}_{N-2}(m_3-m_4)-2\pi i\widetilde{t}_{N-1}m_4}\CZ_B (M_{\alpha}; t_a) \,,
\end{split}
\end{equation}
where $a=1,2,3,4; \;\alpha=1,2,\ldots, N-1$.  The mirror map  can be read off from the above equation.
\begin{equation}
\begin{split}
&M_{\alpha} = \widetilde{t}_{\alpha}\,,\\
&t_a = m_a\,.
\end{split}
\end{equation}
In order to obtain the appropriate $\hat{D}_{N}$ quiver, one has to gauge the Cartan of the $U(2)_1$ flavor symmetry (parametrized by $m_1,m_2$) and partially gauge the Cartan of $U(2)_{N-3}$ (parametrized by $m_3,m_4$). One can carry out the gauging one $U(1)$ at a time and at each step one obtains a new family of mirror pairs.\\

\subsubsection*{Step 1: Gauging $m_1$}
\begin{figure}[H]
\begin{center}
\includegraphics[scale=0.4]{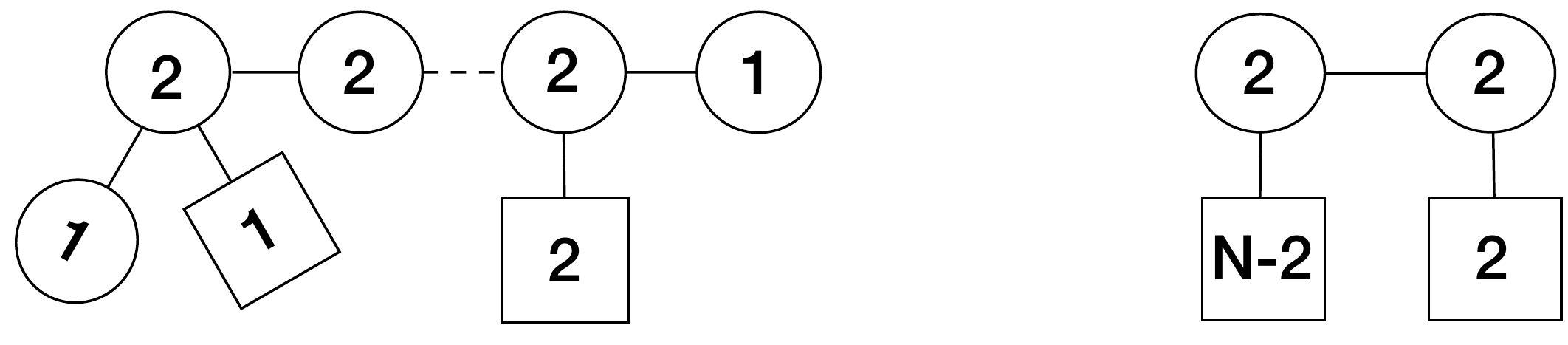}
\caption{A-model $(1,0)(2,1)_{1}(2,0)_2\ldots(2,2)_{N-3}(1,0)_{N-2}$ and B-model $(2,N-2)(2,2)$.}
\label{fig:MirrPair431g}
\end{center}
\end{figure}

The mirror pair obtained by gauging $m_1$ is given in \figref{fig:MirrPair431g}. 
On the A side we gauged $U(1)\subset U(2)$ of the global symmetry on the first node, thereby enlarging the A-quiver by $(1,0)$ node. On the B-side the $U(1)$ factor got ``ungauged'' and become a global symmetry on the second node which resulted in the increase of the rank of the corresponding global symmetry group.

The partition functions of the two theories are related in the following fashion
\begin{equation}
\begin{split}
\widetilde{\CZ}^{(1)}_A(m_a;\widetilde{t}_{\alpha},\zeta_1)=e^{2\pi i m_2 (\zeta_1+2\widetilde{t}_1)}e^{-2\pi i\widetilde{t}_1 m_3-2\pi i\widetilde{t}_{N-2}(m_3-m_4)-2\pi i\widetilde{t}_{N-1}m_4}\widetilde{\CZ}^{(2)}_B (M_{\beta}; t_a) 
\end{split}
\end{equation}
where $a=2,3,4$, $\alpha=1,2,\ldots,N-1$ and $\beta=1,2,\ldots, N$. Here, $\zeta_1$ corresponds to the FI parameter of the newly introduced $U(1)$ node.\\
The mirror map can be read off from the above equation as before.
\begin{equation}
\begin{split}
&M_{\alpha} = \widetilde{t}_{\alpha}\\
&M_N=\zeta_1 + \widetilde{t}_1\\
&t_a = m_a
\end{split}
\end{equation}

\subsubsection*{Step 2: Gauging $m_2$}
\begin{figure}[H]
\begin{center}
\includegraphics[scale=0.4]{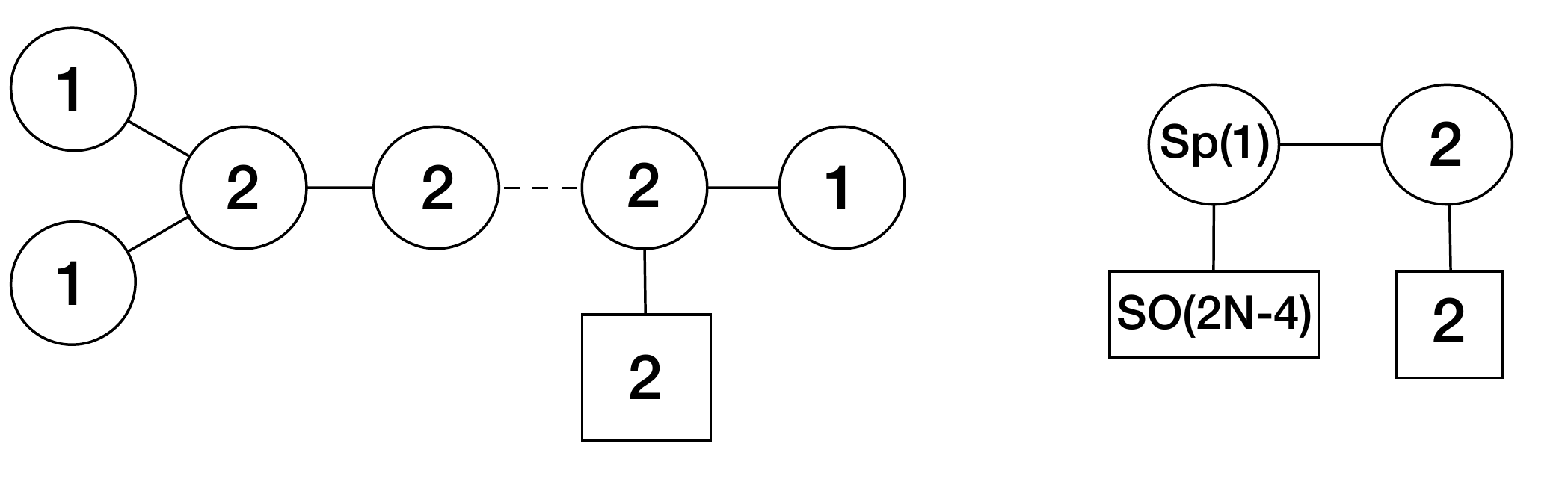}
\caption{A-model $(1,0)^2(2,0)_{1}(2,0)_2\ldots (2,0)_{N-4}(2,2)_{N-3}(1,0)_{N-2}$ and B-model $(Sp(1),N-2)(2,2)$.}
\label{fig:MirrPair431gg}
\end{center}
\end{figure}

Next we gauge the remaining global $U(1)$ on the second node of the A-quiver (\figref{fig:MirrPair431gg}) as a result of which the associated $U(2)$ gauge group is partially ``ungauged" to an $Sp(1)$.

The partition functions of the two theories are related in the following fashion
\begin{equation}
\begin{split}
\widetilde{\CZ}^{(2)}_A(m_a;\widetilde{t}_{\alpha},\zeta_1,\zeta_2)=e^{2\pi i(-\widetilde{t}_{1}+\widetilde{t}_{N-2})m_3} e^{2\pi i(\zeta_1+\zeta_2+2\widetilde{t}_1-\widetilde{t}_{N-2}-\widetilde{t}_{N-1})m_4}\widetilde{\CZ}^{(2)}_B (M_{\beta}; t_a) 
\end{split}
\end{equation}
where $a=3,4$, $\alpha=1,2,\ldots,N-1$ and $\beta=1,2,\ldots, N$. Here, $\zeta_2$ corresponds to the FI parameter of the newly introduced $U(1)$ node.\\
The mirror map can be read off from the above equation as before.
\begin{equation}
\begin{split}
&M_{\alpha} = \frac{\zeta_1+\zeta_2}{2}+\widetilde{t}_{1}-\widetilde{t}_{\alpha},\; a=1,2,\ldots,N-1 \\
&M_N=\frac{\zeta_2-\zeta_1}{2}\\
&t_3 = m_3, \; t_4=m_4\,.
\end{split}
\end{equation}

\subsubsection*{Step 3: Gauging $m_3$}
\begin{figure}[H]
\begin{center}
\includegraphics[scale=0.4]{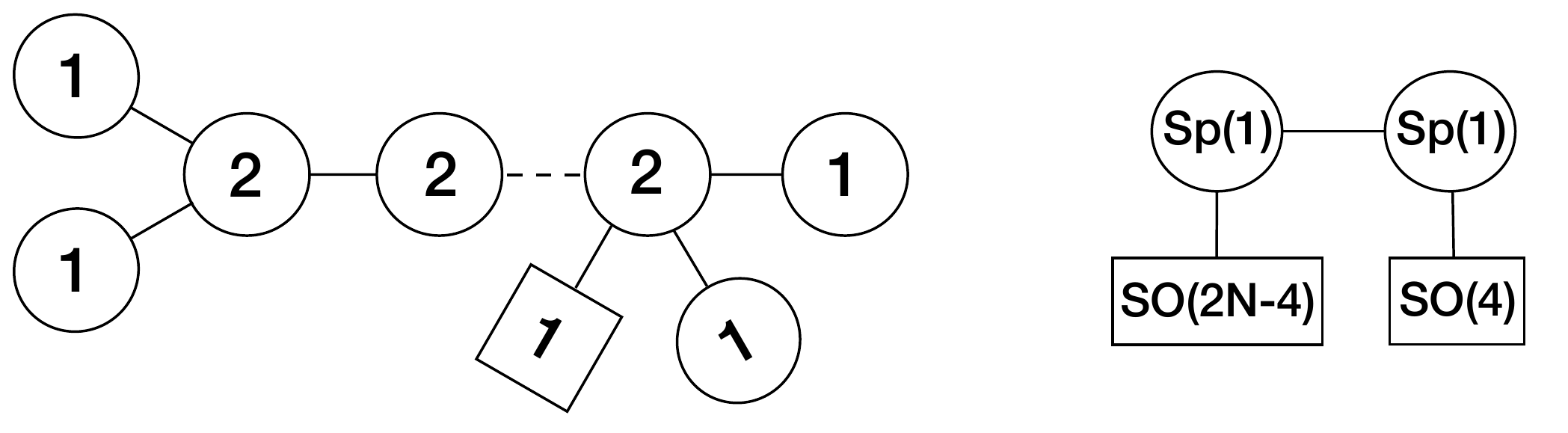}
\caption{A-model $\hat{D}_{N}$ quiver with labels $(1,0)^2(2,0)_{1}(2,0)_2\ldots(2,1)_{N-3}(1,0)^2$ and B-model $(Sp(1),N-2)(Sp(1),2)$.}
\label{fig:MirrPair431ggg}
\end{center}
\end{figure}

\begin{figure}[H]
\begin{center}
\includegraphics[scale=0.6]{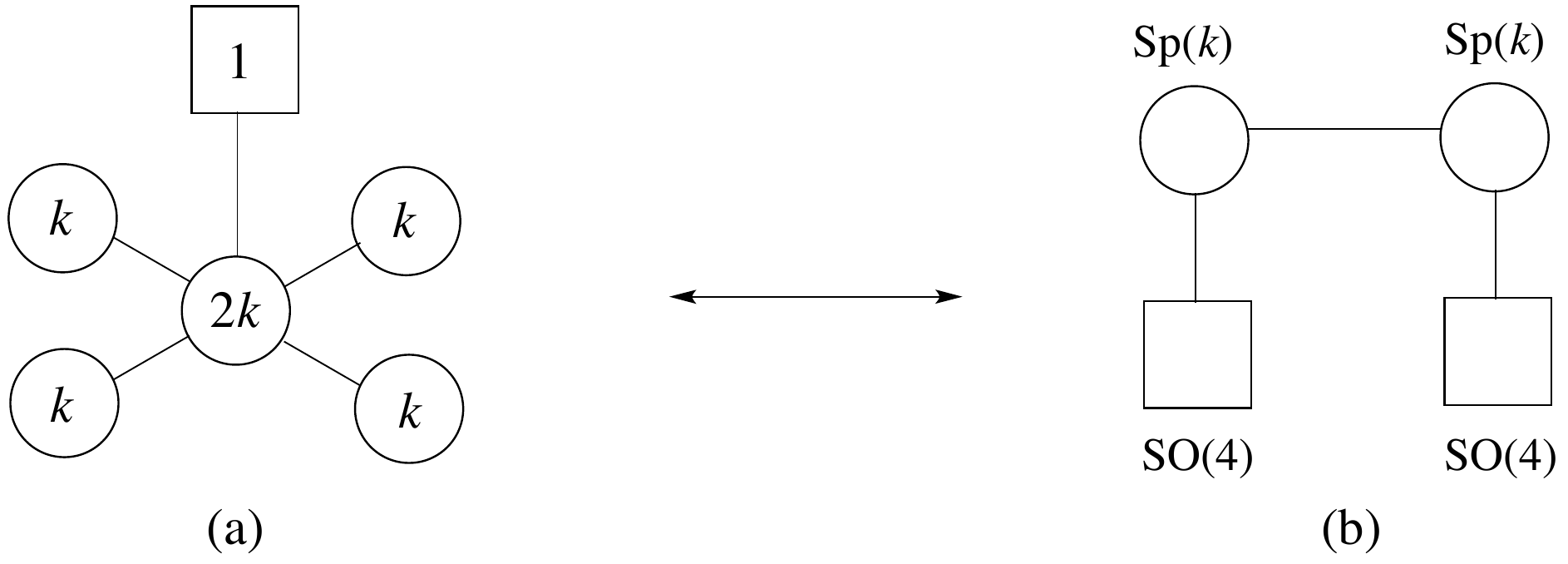}
\caption{Framed $\hat{D}_4$ quiver on the left and its mirror on the right. The Higgs branch of quiver (a) is the moduli space of $k$ $SU(2)$ instantons on $\BC^2/\hat{D}_4$ and that of quiver (b) corresponds to $k$ $SO(8)$ instantons on $\BC^2/\BZ_2$.}
\label{fig:QuiverflavoredD41}
\end{center}
\end{figure}

Finally we gauge $U(1)\subset U(2)$ on the second node from the right of the A-quiver to obtain the desired framed $\hat{D}_N$ quiver and the Sp-SO-type quiver on the mirror side (\figref{fig:MirrPair431ggg}).
For $N=4$, the mirror pair specified above is exactly the one in \figref{fig:QuiverflavoredD41}.
The partition functions of the two theories are related in the following fashion
\begin{equation}
\begin{split}
\widetilde{\CZ}^{(3)}_A(m_4;\widetilde{t}_{\alpha},\zeta_1,\zeta_2,\zeta_3)= e^{2\pi i(\zeta_1+\zeta_2+\zeta_3+\widetilde{t}_1-\widetilde{t}_{N-1})m_4}\widetilde{\CZ}^{(3)}_B (M_{\beta}, m_{bif}; t_4) 
\end{split}
\end{equation}
where $\alpha=1,2,\ldots,N-1$ and $\beta=1,2,\ldots, N$. Here, $\zeta_3$ corresponds to the FI parameter of the newly introduced $U(1)$ node. $m_{bif}$ is a non-zero mass parameter for the $Sp(1) \times Sp(1)$ bi-fundamental hyper. \\
The mirror map can be read off from the above equation as before.
\begin{equation}
\begin{split}
&M_{\alpha} = \frac{\zeta_1+\zeta_2}{2}+\widetilde{t}_{1}-\widetilde{t}_{\alpha},\; \alpha=1,2,\ldots,N-3\\
&M_{N-2}=\frac{\zeta_2-\zeta_1}{2}\\
&M_{N-1}=\frac{\zeta_1+\zeta_2+\zeta_3}{2} +\frac{\widetilde{t}_{1}-\widetilde{t}_{N-2}}{2}\\
&M_{N}=\frac{\zeta_1+\zeta_2+\zeta_3}{2} + \frac{\widetilde{t}_{1}+\widetilde{t}_{N-2}-2\widetilde{t}_{N-1}}{2}\\
&m_{bif}=\frac{\widetilde{t}_{1}-\widetilde{t}_{N-2}-\zeta_3}{2}\\
&t_4=m_4
\end{split}
\end{equation}
Note that the number of independent FI parameters for the A-model is $N+1$, namely $N-1$ parameters $\{\widetilde{t}_{\alpha}\}$  with one constraint and three $\zeta_i$. This exactly matches with the number of mass parameters of the B-model, namely $N$ fundamental masses and one bi-fundamental mass.

Obviously, one can go ahead and gauge the remaining $U(1)$ as well. As we saw in the previous section, this does not change the mirror map in any way but imposes a constraint on the FI parameters of the A-model which is tantamount to having an overall $U(1)$ factor decouple from the gauge group. More explicitly, the mirror pair in this case is the following\\
\textbf{A-model}: $(1,0)^2(2,0)_{1}(2,0)_2\ldots(2,1)_{N-3}(1,0)^3// U(1)$\\
\textbf{B-model}: $(Sp(1),N-2)(Sp(1),2)$.

A generic $\hat{D}_N$ quiver with a framing in this class (see \figref{fig:QuiverflavoredD41} for $N=4$) and gauge groups of arbitrary rank as well as its mirror dual can be derived by starting from the following linear quivers:\\
\textbf{A-model}: $(2k,2k)_1(2k,0)_{2}(2k,0)_3\ldots(2k,k+1)_{N-3}(k,0)$\\
\textbf{B-model}: $(1,0)(2,0)\ldots(2k-1,0)(2k,N-3)(2k,2)(2k-2)\ldots(4,0)(2,0)$\\
To obtain the appropriate $\hat{D}_N$ quiver, one needs to completely gauge the $U(2k)$ flavor group as $U(k)\times U(k)$, while for the $U(k+1)$ flavor group, a $U(k)$ subgroup should be gauged.\\

\subsubsection{Framing on boundary nodes}
\begin{figure}[H]
\begin{center}
\includegraphics[scale=0.6]{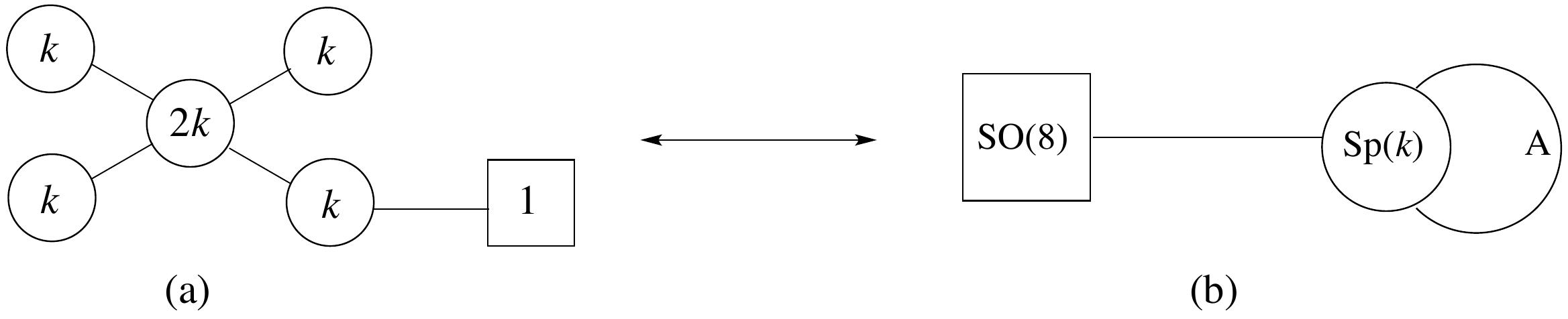}
\caption{Single framed $\hat{D}_4$ quiver and its mirror. The Higgs branch of quiver (a) is the moduli space of $k$ $U(1)$ instantons on $\BC^2/\hat{D}_4$ and that of quiver (b) corresponds to $k$ $SO(8)$ instantons on $\BC^2$. On the right quiver `A' stands for the (reducible) antisymmetric representation of $Sp(k)$.}
\label{fig:QuiverflavoredD42}
\end{center}
\end{figure}

\begin{figure}[H]
\begin{center}
\includegraphics[scale=0.3]{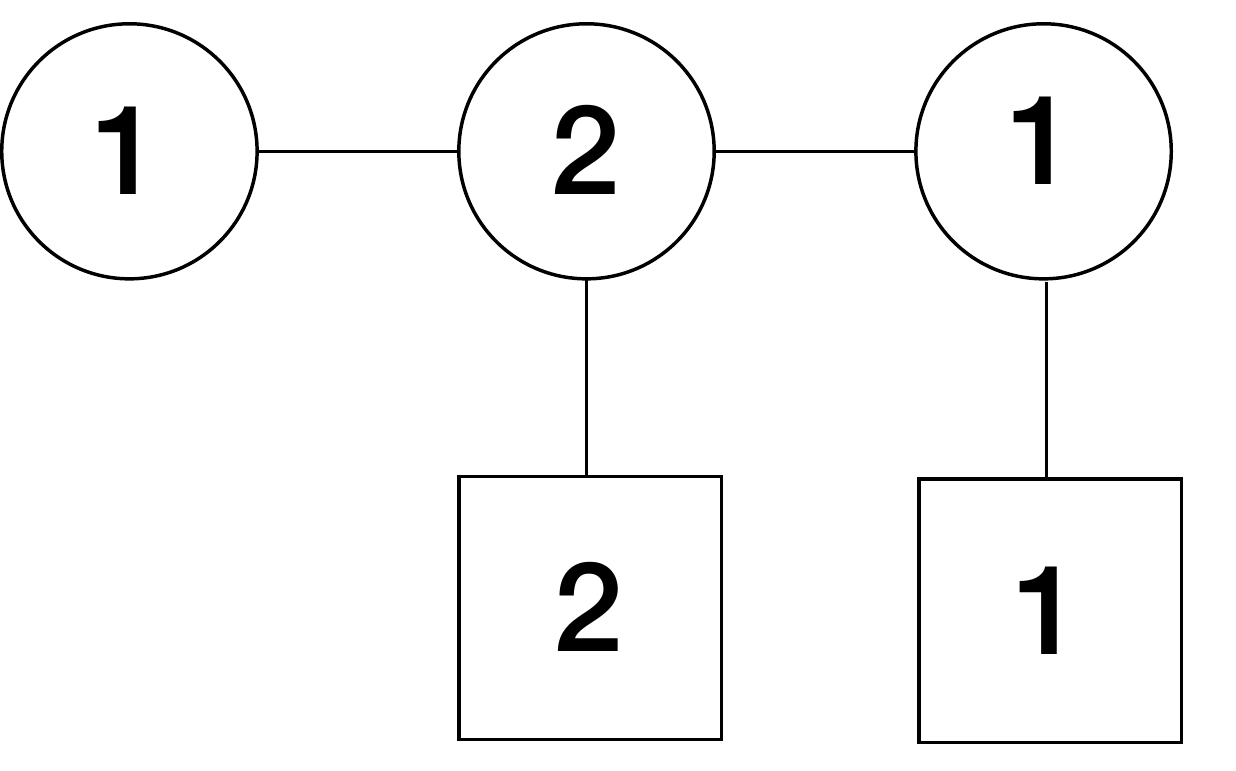}\qquad\qquad\qquad\qquad\includegraphics[scale=0.3]{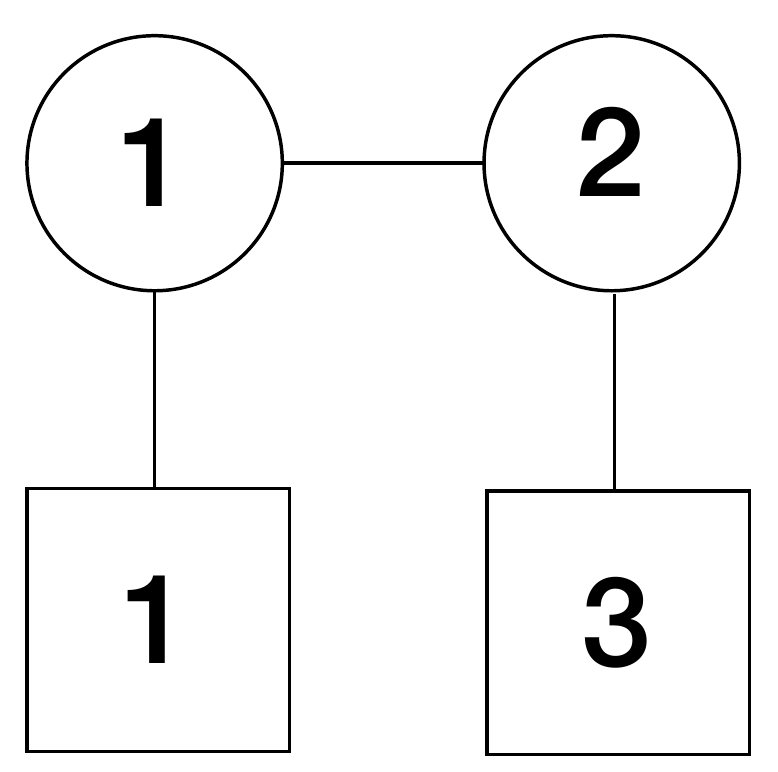}
\caption{Two mirror dual linear quivers which give rise to mirror pair in \figref{fig:QuiverflavoredD42}.}
\label{fig:QuiverflavoredD42Lin}
\end{center}
\end{figure}

Let us proceed with $\hat{D}_{4}$ quiver, this time with a single hypermultiplet at a boundary node (like quiver (a) in \figref{fig:QuiverflavoredD42}). 
In order to derive a mirror for this quiver by Abelian gauging, we start from the following linear quivers --  A-model with labels $(1,0)(2,2)(1,1)$ and B-model with labels $(1,1)(2,3)$ (see \figref{fig:QuiverflavoredD42Lin}).

We perform the abelian gauging trick on the $U(2)$ flavor symmetry on the second node of the left quiver in \figref{fig:QuiverflavoredD42Lin} in order to map it onto the A model quiver in \figref{fig:QuiverflavoredD42}. Let us see what happens with its mirror. The dimensions of the Coulomb and Higgs branches on the A-model quiver are $4$ and $3$ respectively. After the gauging is done they become $6$ and $1$ correspondingly. In the classical parameter space description, this happens because two momenta $p_\mu^{a}$ on the A-model quiver get fixed, and two more vectormultiplets are introduced. Thus on the B side the dimension of the Coulomb branch has to drop by two. 
We can see that this indeed happens if we multiply two Bethe equations for the two gauge nodes of the B quiver. Because of the mirror constraints on $p_\tau^{\vee\,j}$ the value of $\sigma^{(2)}$ is fixed by $p_\tau^{\vee\,3}$. The second mirror constraint provides the projection of $U(2)\to Sp(1)$ via $\sigma^{\vee\,(1)}_1 \sigma^{\vee\,(1)}_2=1$, which can be implemented by adjusting the momenta and twists. Notice that classical analysis does not explicitly show the contribution of the singlet multiplet A in \figref{fig:QuiverflavoredD42}.

In order to see how abelian gauging leads to the mirror pair we want and in particular how antisymmetric matter (a singlet when $k=1$) appears in the mirror we perform gauging of the linear quivers using the partition function approach. For the two linear quivers from \figref{fig:QuiverflavoredD42Lin} the partition functions are
\begin{equation}
\begin{split}
\CZ_A(m_a;{t}_{\alpha})=& \int \prod^2_{\alpha=1} d s_{\alpha}  \frac{d^2 s_0}{2!} \frac{\prod^2_{\alpha=1} e^{2\pi i s_{\alpha} \eta_{\alpha}} \prod^2_{i=1} e^{2\pi i s^i_0 \eta_0} \sinh^2{\pi(s^1_0-s^2_0)}}{\prod^2_{i=1}\cosh{\pi(s_1-s^i_0)}\prod^2_{a=1} \cosh{\pi(s^i_0+m_a)}\cosh{\pi(s_2-s^i_0)}} \\
&\times \frac{1}{\cosh{\pi(s_2+m_{3})}}\,,\\
\CZ_B (M_{\alpha}; \widetilde{t}_a) =&\int  d s_{1}  \frac{d^2 s_0}{2!} \frac{e^{2\pi i s_{1} \widetilde{\eta}_{1}} \prod^2_{i=1} e^{2\pi i s^i_0 \widetilde{\eta}_0} \sinh^2{\pi(s^1_0-s^2_0)}}{\prod^2_{i=1}\cosh{\pi(s^i_0-s_1)}\prod^3_{a=1} \cosh{\pi(s^i_0+M_a)}\cosh{\pi(s_1+M_{4})}}\,,   
\end{split}
\end{equation}
where $a=1,2,3,4; \;\alpha=1,2,\ldots, N-1$. In the A-model, the FI parameters are defined as $\eta_1=t_1-t_2, \eta_0= t_2-t_3, \eta_2=t_3-t_4$, while for the B-model, these are $\widetilde{\eta}_0=\widetilde{t}_1-\widetilde{t}_2, \widetilde{\eta}_1=\widetilde{t}_2-\widetilde{t}_3$.\\

The mirror symmetry implies that $\CZ_A(m_a;{t}_{\alpha})=\CZ_B (M_{\alpha}; \widetilde{t}_a)$ up to some overall phase (which we shall ignore in this example and the subsequent ones) provided the parameters are related as follows:
\begin{equation}
\begin{split}
&M_{\alpha} = {t}_{\alpha}\\
&\widetilde{t}_a  = m_a\,.
\end{split}
\end{equation}
Now, we gauge the $U(2)$ flavor symmetry of the A-model as a $U(1) \times U(1)$, which gives a $\hat{D}_4$ quiver with a single flavor on one of the boundary nodes. The partition function of this theory is 
\begin{equation}
\begin{split}
\widetilde{\CZ}_A(\zeta_1,\zeta_2, m_3;{t}_{\alpha})=&\int dm_1 dm_2 e^{2\pi i m_1 \zeta_1} e^{2\pi i m_2 \zeta_2} \CZ_A (m_a;{t}_{\alpha})\\
=&\int dm_1 dm_2 e^{2\pi i m_1 \zeta_1} e^{2\pi i m_2 \zeta_2} \CZ_B (t_{\alpha}; m_a)\,,
\end{split}
\end{equation}
where the second equality follows from the mirror symmetry of the linear quivers. From the second equality, completing the integration over $m_1$ and $m_2$ we have
\begin{equation}
\begin{split}
\widetilde{\CZ}_A(\zeta_1,\zeta_2, m_3;{t}_{\alpha})=& \int  d s_{1}  \frac{d^2 s_0}{2!} \frac{\delta(s^1_0+s^2_0+\zeta_1)\delta(-s^1_0-s^2_0+s_1+\zeta_2) \sinh^2{\pi(s^1_0-s^2_0)}}{\prod_i \prod^3_{\alpha=1}\cosh{\pi(s^i_0+t_{\alpha})}} \\
&\times \frac{e^{-2\pi i m_3 s_1}}{\prod_i \cosh{\pi(s^i_0-s_1)} \cosh{\pi(s_1+t_4)}}\,.
\end{split}
\end{equation}
Finally, integrating over $s_1$ using the delta function and shifting the remaining integration variables appropriately, we have 
\begin{equation}
\begin{split}
\widetilde{\CZ}_A(\zeta_1,\zeta_2, m_3;{t}_{\alpha})=& \int  \frac{d^2 s_0}{2!} \frac{\delta(s^1_0+s^2_0)\sinh^2{\pi(s^1_0-s^2_0)}}{\prod_i \prod^3_{\alpha=1}\cosh{\pi(s^i_0-t_{\alpha}-\frac{\zeta_1}{2})}\prod_i\cosh{\pi(s^i_0 +\zeta_1+\zeta_2)}} \\
&\times \frac{e^{2\pi i m_3 (\zeta_1+\zeta_2)}}{\cosh{\pi(\zeta_1+\zeta_2+t_4)}}\\
&=\widetilde{\CZ}_B(M_i, M_{singlet})\,.
\end{split}
\end{equation}
The dual theory can be immediately read off -- the first line is identified as the partition function of a $Sp(1)$ gauge theory with 4 flavors while the second line is the partition function of a single free hyper (up to a phase). The mirror map of this mirror pair is
\begin{equation}
\begin{split}
&M_a=t_a +\frac{\zeta_1}{2} \; (a=1,2,3)\\
&M_4=\zeta_1+\zeta_2\\
&M_{singlet}=t_4 + \zeta_1+\zeta_2\,.
\end{split}
\end{equation}
Note that the number of parameters exactly matches on both sides. For the A-model, we have 5 independent parameters - $\{t_1,t_2,t_3,t_4\}$ with one constraint and $\{\zeta_1,\zeta_2\}$. This is matched by the 5 mass parameters for the B-model.\\

In order to obtain a generic mirror pair in this class (for rank of the quiver $N>4$) by gauging, we start from the following linear quivers:\\
\textbf{A-model}: $(1,0)(2,1)_{1}(2,0)_2\ldots(2,0)_{N-4}(2,1)_{N-3}(1,1)_{N-2}$ and \\
\textbf{B-model}: $(2,N-1)(1,1)$.\\

One needs to gauge the $U(1)$ flavor symmetries of the nodes $(2,1)_1$ and $(2,1)_{N-3}$ to obtain the appropriately framed $\hat{D}_N$ quiver. Proceeding as before, the mirror is found to be a $Sp(1)$ gauge theory with $N$ fundamental hypers and one singlet hyper. The mirror map in this case is an obvious generalization of the $\hat{D}_4$ case.
\begin{equation}
\begin{split}
&M_a=t_a +\frac{\zeta_1}{2} \; (a=1,2,\ldots,N-1)\\
&M_N=\zeta_1+\zeta_2\\
&M_{singlet}=t_N + \zeta_1+\zeta_2\,.
\end{split}
\end{equation}

\subsection{$\hat{D}_{N}$ Quivers with Double Framing}\label{sec:DhatDouble}
Next we analyze $\hat{D}_{N}$ quivers with two hypermultiplets. Framing on one or more internal nodes can be treated in a fashion analogous to the example of single framing on an internal node discussed above. However, framing on the boundary nodes may be done in two possible ways -- one can either have double framing on a single boundary node, or one can have two framed boundary nodes at different locations in the quiver. We will treat each of these cases individually.

\subsubsection{Framed $\hat{D}_3$ quiver}
\begin{figure}[H]
\begin{center}
\includegraphics[scale=0.5]{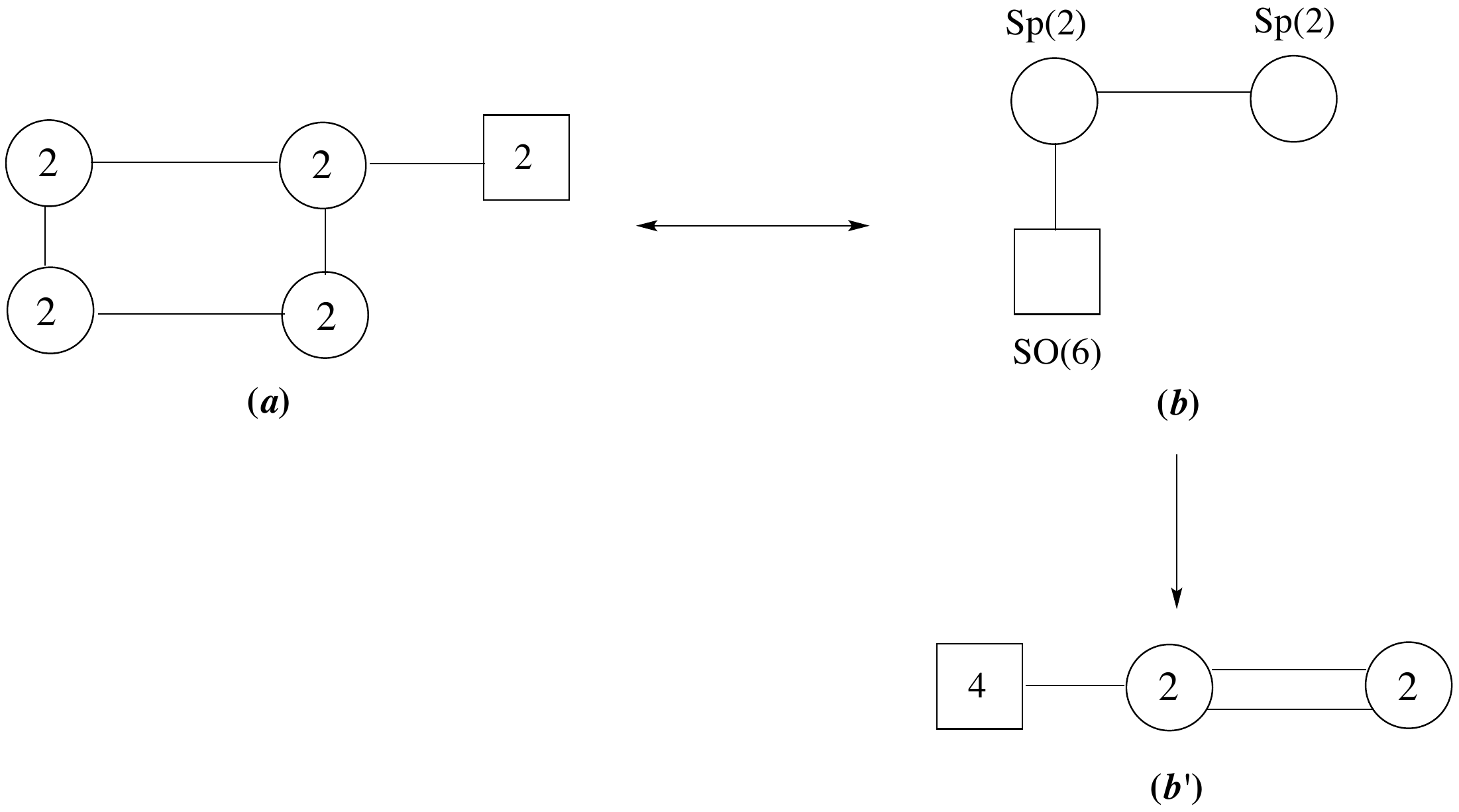}
\caption{Framed $\hat{D}_3$ quiver (a), its Hanany-Witten mirror (b), and its ``good'' mirror $(b')$.  According to \cite{Dey:2013fea}, the Higgs branches of quivers $(a)$, $(b)$ and $(b')$ are as follows. $(a)$: the moduli space of $2$ $SU(2)$ instantons on $\BC^2/\hat{D}_3 \simeq \BC^2/\BZ_4$. $(b)$: the moduli space of $2$ $SO(6)$ instantons on $\BC^2/\BZ_2$.   $(b')$: the moduli space of $2$ $SU(4)$ instantons on $\BC^2/\BZ_2$. }
\label{fig:QuiverflavoredD3}
\end{center}
\end{figure}

Let us start with a warm-up example of a framed $\hat{D}_3$ quiver (see $(a)$ in \figref{fig:QuiverflavoredD3}). In some sense it can be treated as an $SO(6)$ toy-example of our next quiver theory presented in \figref{fig:QuiverflavoredD44}. The Hanany-Witten mirror $(b)$ in \figref{fig:QuiverflavoredD3} appears to be bad. However, due to the isomorphism $\hat{D}_3\simeq \hat{A}_3$, we can apply known rules for the circular quiver $\hat{A}_3$ and obtain ``good'' quiver as its mirror -- $(b')$ in the figure. Therefore we expect the two quivers -- $(b)$ and $(b')$ to have the same infrared physics. 
One can readily check the proposed mirror symmetry between the framed $\hat{D}_3$ quiver and the ``good'' quiver $(b')$ in \figref{fig:QuiverflavoredD3} using Hilbert series.  For convenience, we introduce the notation
\bea
\tau = t^{1/2}
\eea
and use this to write the Higgs branch Hilbert series.
The Hilbert series for the Higgs branch of quiver (a) of \figref{fig:QuiverflavoredD3} can conveniently be computed using the localization method (see \eg~ \cite{Dey:2013fea}). It reads 
\bea
H^{H}_{(a)}(\tau, y) &=1+([0]+[2])\tau^2+ (5 [0]+2 [2]+[4])\tau^4+(7 [0]+9 [2]+2 [4]+[6])\tau^6 \nn \\
& \qquad +(19 [0]+17 [2]+10 [4]+2 [6]+[8])\tau^8 +\ldots~.
\eea
Setting $y=1$, we obtain
\bea
H^{H}_{(a)} (\tau, y=1) &= \frac{1}{\left(1-\tau^2\right)^8 \left(1+\tau^2\right)^4 \left(1+\tau^2+\tau^4\right)^3} (1+3 \tau^2+8 \tau^4+20 \tau^6+41 \tau^8+61 \tau^{10} \nn \\
& \qquad +78 \tau^{12}+84 \tau^{14}+78 \tau^{16} + \text{palindrome} + \tau^{28}) \nn \\
&= 1+4 \tau^2+16 \tau^4+51 \tau^6+143 \tau^8+350 \tau^{10}+\ldots~,
\eea
where `palindrome' denotes the repetitions of the coefficients that have been written before in the reverse order.

The Coulomb branch Hilbert series of diagram (a) in \figref{fig:QuiverflavoredD3} is given by
\bea
H^C_{(a)}(t, \vec a) &=  \sum_{\alpha=1}^4 \sum_{m_{\alpha,1} =-\infty}^\infty \sum_{m_{\alpha,2}=-\infty}^{m_{\alpha,1}} t^{\Delta(\vec m)} P_{U(2)} (t, m_{\alpha,1}, m_{\alpha,2}) \prod_{\alpha=1}^4 \prod_{i=1}^2 a_{\alpha, i}^{m_{\alpha,i}}~.
\eea
where $m_{\alpha,1}, m_{\alpha,2}$ are the monopole charges associated with the $\alpha$-th $U(2)$ gauge group, where $\alpha=1,\ldots, 4$.  Here $\Delta(\vec m, \vec n)$ is the dimension of the monopole operators:
\bea
\Delta(\vec m) = \frac{1}{2} \left[ 2|m_{1,1}|+2 |m_{1,2}| + \sum_{\alpha=1}^4 \sum_{i,j=1}^2 |m_{\alpha,i} - m_{\alpha,j} | \right] - \sum_{\alpha=1}^4 |m_{\alpha,1} - m_{\alpha,2} |~.
\eea
For simplicity, we set $a_{\alpha,i}=1$ and obtain
\bea \label{CoulombD3a}
H^C_{(a)}(t, \{ a_{\alpha,i}=1 \} )=  1+18 t+221 t^2+1898 t^3+12663 t^4+\ldots~.
\eea

The dimensions for the monopole operators in quiver $(b)$ of \figref{fig:QuiverflavoredD3} are given by
\bea
\Delta(m_1, m_2; n_1, n_2)&= \frac{1}{2} \left[ 3\sum_{i=1}^2 (|n_i|+\left|-n_i \right|) + \sum_{s_1,s_2=0}^1 \sum_{i,j=1}^2 | (-1)^{s_1} m_i +(-1)^{s_2} n_j|  \right] \nn \\
& \qquad - (2m_1 + |m_1-m_2| +|m_1+m_2|)  \nn \\
& \qquad - (2n_1 + |n_1-n_2| +|n_1+n_2|) ~,
\eea 
where $m_1, m_2$ and $n_1, n_2$ are the monopole charges for the two $Sp(2)$ gauge groups.

Observe that $\Delta(2, 0; 0, 0)=0$; hence the theory contains a monopole operator of charge zero.  The quiver is a ``bad'' theory in the sense of \cite{Gaiotto:2008ak}.

Then we compute the Hilbert series of the Higgs branch of $(b')$ of \figref{fig:QuiverflavoredD3}
\bea
H^{H}_{(b')}(\tau, x, \vec y) &= 1+([0; 1, 0, 1] + [2; 0, 0, 0]) \tau^2 + (2 [0; 0, 0, 0] + [0; 0, 2, 0] + [0; 1, 0, 1] + [0; 2, 0, 2] \nn \\
& \qquad + 2 [2; 1, 0, 1] + 2 [4; 0, 0, 0]) \tau^4+ + \ldots~.
\eea
Note that we cannot factorize $\BC^2/\BZ_2$ from this Hilbert series.  Setting $x= y_i=1$, we obtain
\bea
H^{H}_{(b')} (\tau,x=1, \{ y_i =1 \}) &= \frac{1}{(1 - \tau^2)^{16} (1 + \tau^2)^8}(1+10 \tau^2+97 \tau^4+498 \tau^6+1917 \tau^8+4990 \tau^{10} \nn \\
& \qquad +10065 \tau^{12}+14784 \tau^{14}+17144 \tau^{16}+14784 \tau^{18} \nn \\
& \qquad + \text{palindrome}+ \tau^{32}) \nn \\
&= 1+18 \tau^2+221 \tau^4+1898 \tau^6+12663 \tau^8+\ldots~.
\eea
Note that this is in agreement with \eref{CoulombD3a}. 

Finally, the Coulomb branch Hilbert series of diagram $(b')$ in \figref{fig:QuiverflavoredD3} is given by
\bea
H^C_{(a)}(t, \vec a) &=  \sum_{\alpha=1}^2 \sum_{m_{\alpha,1} =-\infty}^\infty \sum_{m_{\alpha,2}=-\infty}^{m_{\alpha,1}} t^{\Delta(\vec m)} P_{U(2)} (t, m_{\alpha,1}, m_{\alpha,2}) \prod_{\alpha=1}^2 \prod_{i=1}^2 a_{\alpha, i}^{m_{\alpha,i}}~.
\eea
where $m_{\alpha,1}, m_{\alpha,2}$ are the monopole charges associated with the $\alpha$-th $U(2)$ gauge group, where $\alpha=1,2$.  Here $\Delta(\vec m, \vec n)$ is the dimension of the monopole operators:
\bea
\Delta(\vec m) = \frac{1}{2} \left[ 4|m_{1,1}|+4 |m_{1,2}| + 2 \sum_{i,j=1}^2 |m_{1,i} - m_{2,j} | \right] - \sum_{\alpha=1}^2 |m_{\alpha,1} - m_{\alpha,2} |~.
\eea
For simplicity, we set $a_{\alpha,i}=1$ and obtain
\bea \label{CoulombD3b'}
H^C_{(b')}(t, \{ a_{\alpha,i}=1 \} ) &=  \frac{1}{\left(1-t^2\right)^8 \left(1+t^2\right)^4 \left(1+t^2+t^4\right)^3} (1+3 t^2+8 t^4+20 t^6+41 t^8+61 t^{10} \nn \\
& \qquad +78 t^{12}+84 t^{14}+78 t^{16} + \text{palindrome} + t^{28}) \nn \\
&= H^H_{(a)}(t, \{ a_{\alpha,i}=1 \} ) 
\eea
This is equal to the Higgs branch Hilbert series of quiver $(a)$.

\subsubsection{Framing at a single node of $\hat{D}_N$ quivers}
\begin{figure}[H]
\begin{center}
\includegraphics[scale=0.6]{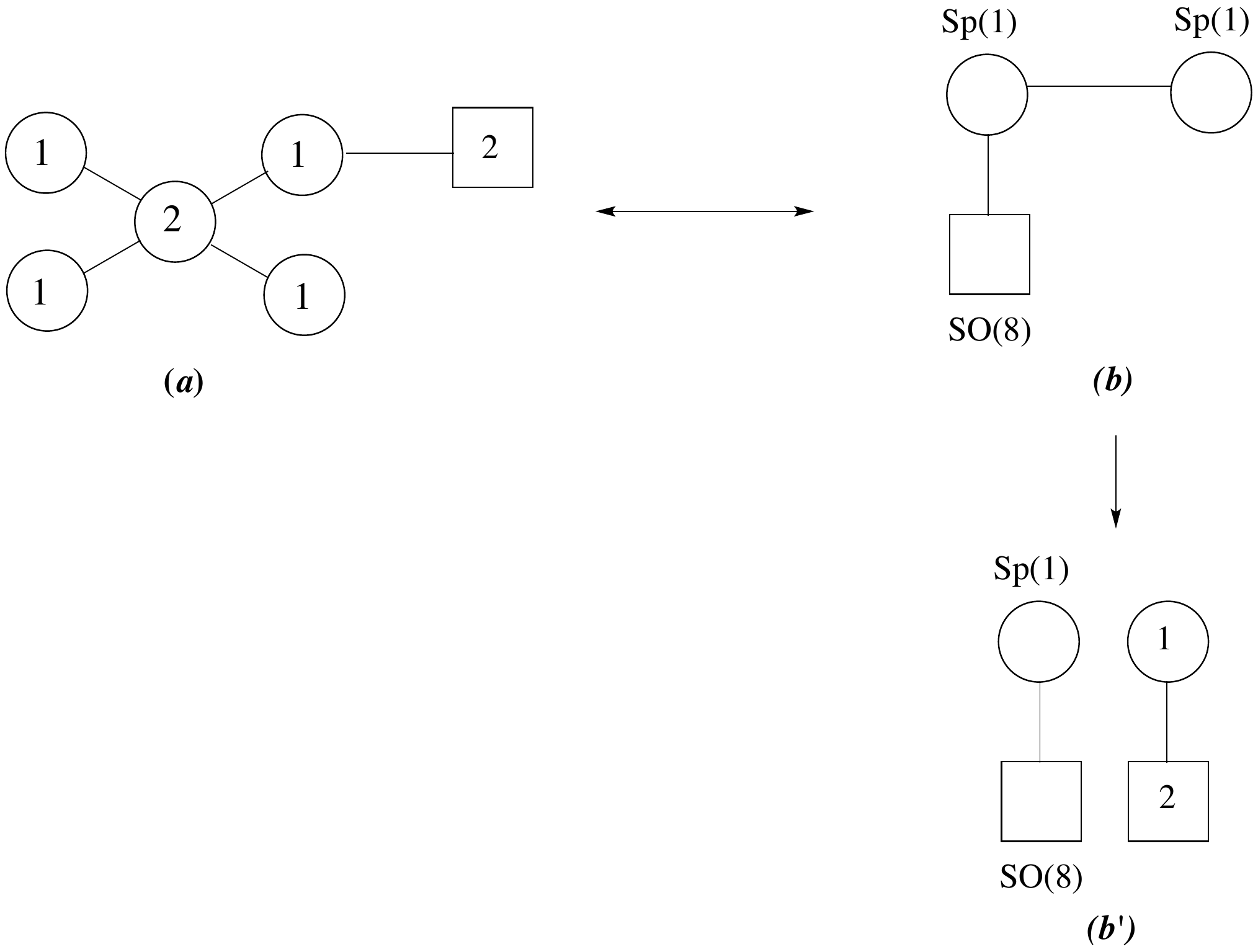}
\caption{Doubly Framed $\hat{D}_4$ quiver $(a)$, its Hanany-Witten mirror $(b)$, and its ``good'' mirror $(b')$. The Higgs branches of $(a)$, $(b)$ and $(b')$ are as follows. $(a)$: the moduli space of $1$ $SU(2)$ instanton on $\BC^2/\hat{D}_4$.  $(b)$: the moduli space of $1$ $SO(8)$ instanton on $\BC^2/\BZ_2$; see \cite{Dey:2013fea}.  $(b')$:  the reduced instanton moduli space of $1$ $SO(8)$ instanton on $\BC^2$ times $\BC^2/\BZ_2$.  The factorisation of the Higgs branch of quiver $(b')$ is discussed in \cite{Dey:2013fea}.}
\label{fig:QuiverflavoredD44}
\end{center}
\end{figure}

Let us first consider the $\hat{D}_4$ quiver with two hypermultiplets on one of its external nodes, \figref{fig:QuiverflavoredD44} (a). We can realize this quiver theory using branes (see \figref{fig:BraneflavoredD4}) and S-duality or the Hanany-Witten realisation \cite{Porrati:1996xi} to generate the mirror quiver (b) in \figref{fig:QuiverflavoredD44}.
We cannot be completely satisfied with the $(b)$ picture in \figref{fig:QuiverflavoredD44} since the quiver is ``bad'' on the unframed $Sp(1)$ node. Inability to find a ``good'' quiver by formally applying the S-duality is not an uncommon phenomenon while working with quivers involving $Sp$ and $SO$ gauge groups \cite{Gaiotto:2008ak}. Therefore we expect to be able to find another ``good'' quiver theory which flows in the infrared to the same SCFT as theory $(b)$ flows to. We will show in this section that this is the quiver $(b')$ in \figref{fig:QuiverflavoredD44} from a straightforward application of Abelian gauging using sphere partition functions.

\begin{figure}[H]
\begin{center}
\includegraphics[scale=0.3]{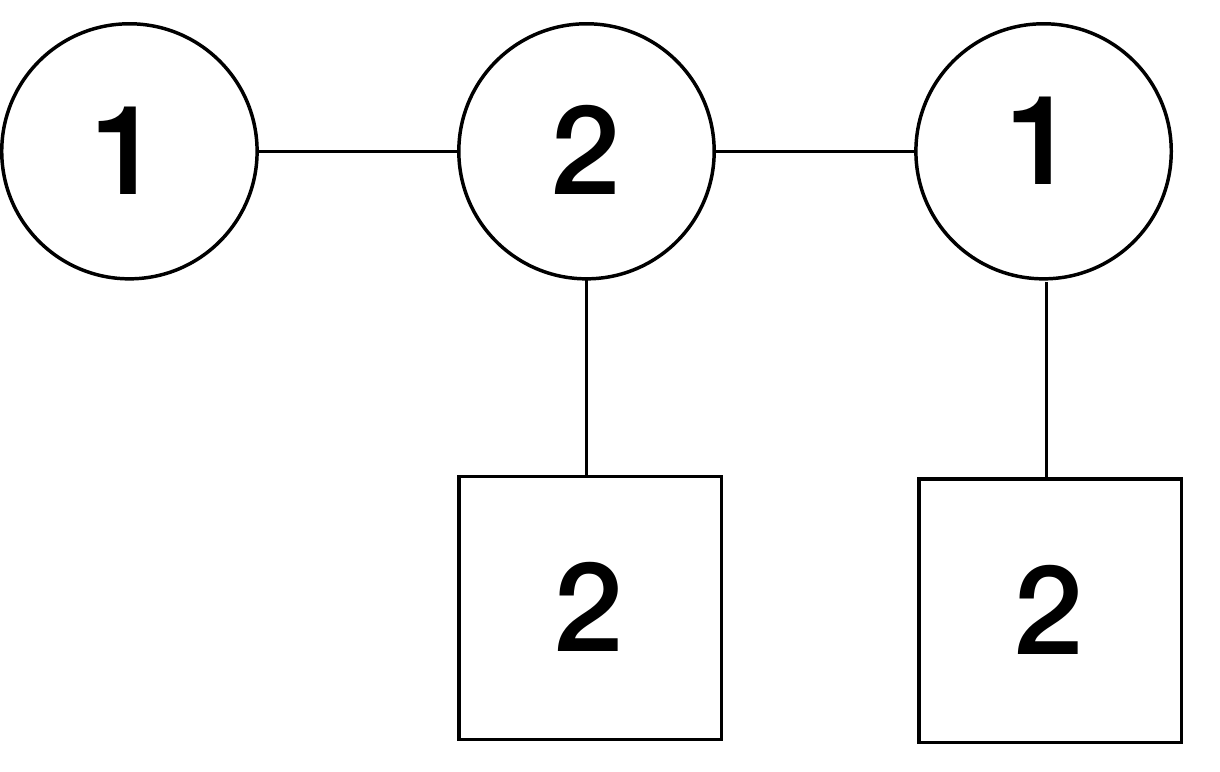}\qquad\qquad\qquad\includegraphics[scale=0.3]{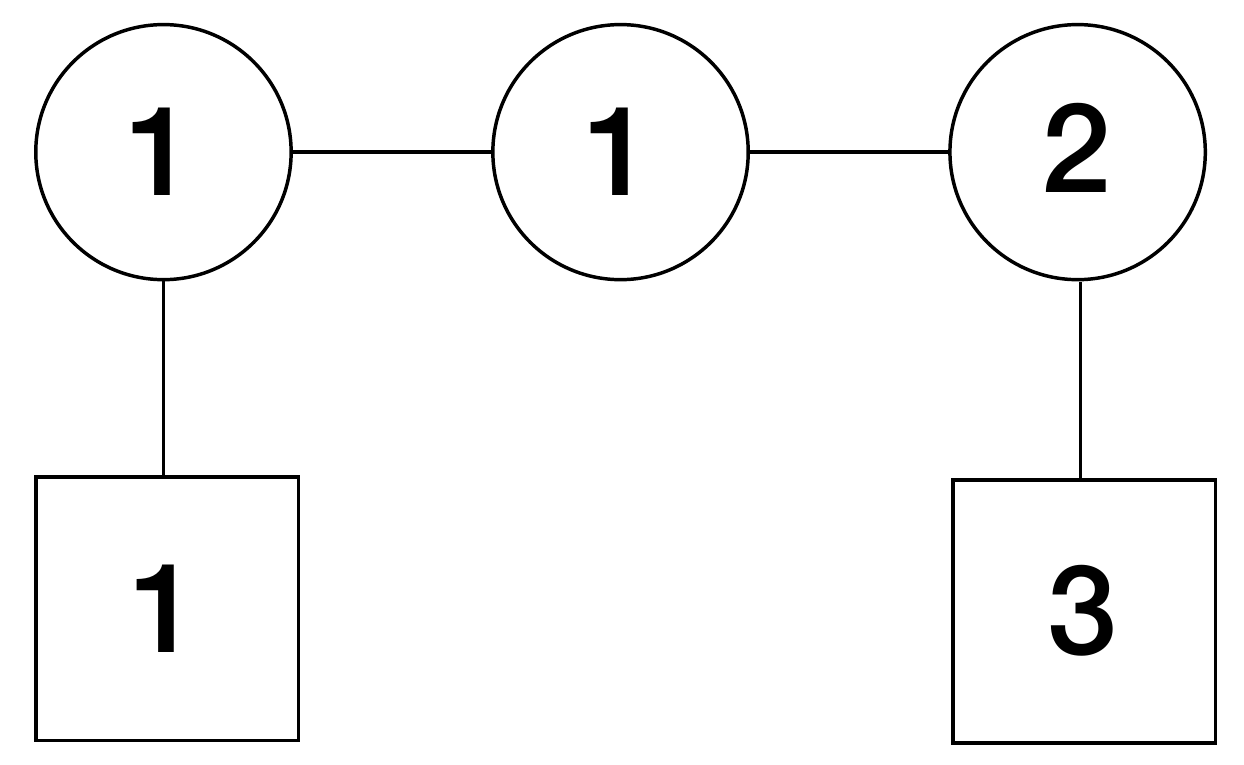}\\
\includegraphics[scale=0.55]{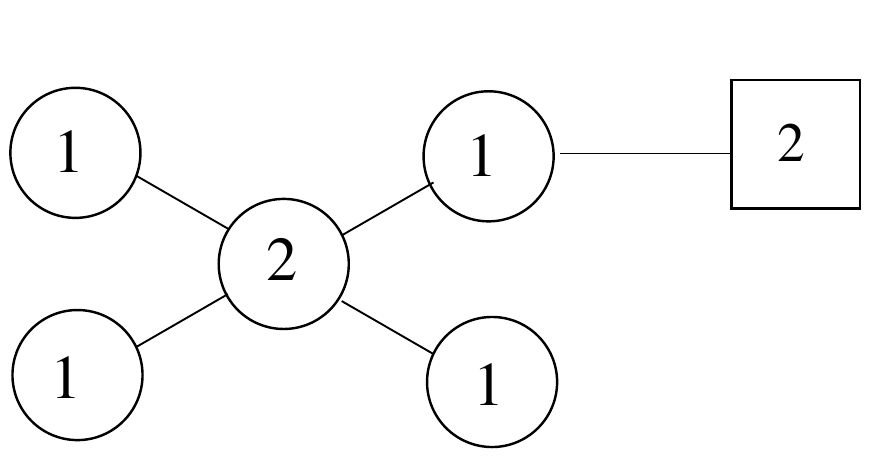}\qquad\qquad\qquad\includegraphics[scale=0.3]{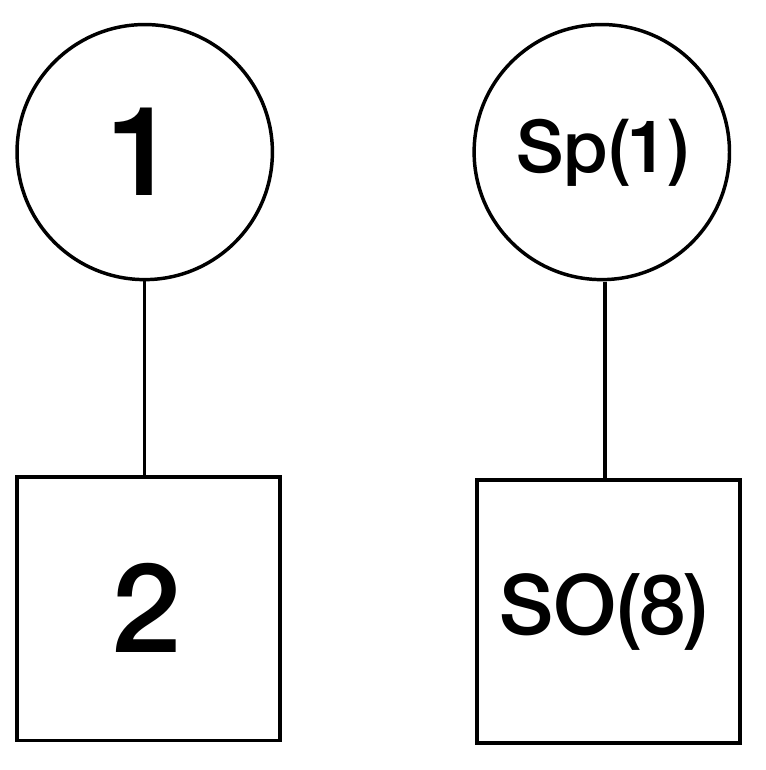}
\caption{Mirror linear quivers (top row) and the new mirror pair after the gauging trick (bottom row).}\label{fig:QuiverflavoredD44Lin}
\end{center}
\end{figure}

\subsubsection*{Partition function approach}
Consider the partition functions for the linear quivers shown in the top row of \figref{fig:QuiverflavoredD44Lin}. They take the following form
\begin{equation}
\begin{split}
\CZ_A(m_a;{t}_{\alpha})=& \int \prod^2_{\alpha=1} d s_{\alpha}  \frac{d^2 s_0}{2!} \frac{\prod^2_{\alpha=1} e^{2\pi i s_{\alpha} \eta_{\alpha}} \prod^2_{i=1} e^{2\pi i s^i_0 \eta_0} \sinh^2{\pi(s^1_0-s^2_0)}}{\prod^2_{i=1}\cosh{\pi(s_1-s^i_0)}\prod^2_{a=1} \cosh{\pi(s^i_0+m_a)}\cosh{\pi(s_2-s^i_0)}} \\
&\times \frac{1}{\prod^2_{a=1}\cosh{\pi(s_2+m_{2+a})}}\\
\CZ_B (M_{\alpha}; \widetilde{t}_a) =&\int \prod^2_{\alpha=1} d s_{\alpha}  \frac{d^2 s_0}{2!} \frac{\prod^2_{\alpha=1} e^{2\pi i s_{\alpha} \widetilde{\eta}_{\alpha}} \prod^2_{i=1} e^{2\pi i s^i_0 \widetilde{\eta}_0} \sinh^2{\pi(s^1_0-s^2_0)}}{\prod^2_{i=1}\cosh{\pi(s^i_0-s_1)}\prod^3_{a=1} \cosh{\pi(s^i_0+M_a)}\cosh{\pi(s_1-s_{2})}} \\
&\times \frac{1}{\cosh{\pi(s_2+M_{4})}}   
\end{split}
\end{equation}
where $a=1,2,3,4; \;\alpha=1,2,\ldots, 4$. In the A-model, the FI parameters are defined as $\eta_1=t_1-t_2, \eta_0= t_2-t_3, \eta_2=t_3-t_4$, while for the B-model, these are $\widetilde{\eta}_0=\widetilde{t}_1-\widetilde{t}_2, \widetilde{\eta}_1=\widetilde{t}_2-\widetilde{t}_3, \widetilde{\eta}_2=\widetilde{t}_3-\widetilde{t}_4$. Mirror Symmetry implies that $\CZ_A(m_a;{t}_{\alpha})=\CZ_B (M_{\alpha}; \widetilde{t}_a)$ up to some overall phase provided the parameters are related as follows:
\begin{equation}
\begin{split}
&M_{\alpha} = {t}_{\alpha}\\
&\widetilde{t}_a  = m_a\,.
\end{split}
\end{equation}

Now we gauge the left $U(2)$ flavor symmetry in the top left quiver in \figref{fig:QuiverflavoredD44Lin} as a $U(1) \times U(1)$, which gives a $\hat{D}_4$ quiver with two hypers on a single boundary node (lower left quiver in \figref{fig:QuiverflavoredD44Lin}). The partition function of this theory is 
\begin{equation}
\begin{split}
\widetilde{\CZ}_A(\zeta_1,\zeta_2, m_3,m_4;{t}_{\alpha})=&\int dm_1 dm_2 e^{2\pi i m_1 \zeta_1} e^{2\pi i m_2 \zeta_2} Z_A (m_a;{t}_{\alpha})\\
=&\int dm_1 dm_2 e^{2\pi i m_1 \zeta_1} e^{2\pi i m_2 \zeta_2} Z_B ({t}_{\alpha}; m_a)\,,
\end{split}
\end{equation}
where the second equality follows from the mirror symmetry of the linear quivers. From the second equality, completing the integration over $m_1$ and $m_2$ we have
\begin{equation}
\begin{split}
\widetilde{\CZ}_A(\zeta_1,\zeta_2, m_3,m_4;{t}_{\alpha})=& \int  ds_2 d s_{1}   \frac{d^2 s_0}{2!} \frac{\delta(s^1_0+s^2_0+\zeta_1)\delta(-s^1_0-s^2_0+s_1+\zeta_2) \sinh^2{\pi(s^1_0-s^2_0)}}{\prod_i \prod^3_{\alpha=1}\cosh{\pi(s^i_0+t_{\alpha})}} \\
&\times \frac{e^{-2\pi i m_3 s_1}}{\prod_i \cosh{\pi(s^i_0-s_1)}} \times \frac{e^{2\pi i (m_3-m_4) s_2}}{\cosh{\pi(s_1-s_2)}\cosh{\pi(s_2+t_4)}}
\end{split}
\end{equation}
Finally, integrating over $s_1$ using the delta function and shifting the remaining integration variables appropriately, we have 
\begin{equation}
\begin{split}
\widetilde{\CZ}_A(\zeta_1,\zeta_2, m_3,m_4;{t}_{\alpha})=& e^{2\pi i m_3 (\zeta_1+\zeta_2)} \int \frac{d^2 s_0}{2!} \frac{\delta(s^1_0+s^2_0)\sinh^2{\pi(s^1_0-s^2_0)}}{\prod_i \prod^3_{\alpha=1}\cosh{\pi(s^i_0+t_{\alpha}-\frac{\zeta_1}{2})}\prod_i\cosh{\pi(s^i_0 +\zeta_1+\zeta_2)}} \\
&\times \int  ds_2 \frac{e^{2\pi i (m_3-m_4)s_2}}{\cosh{\pi(s_2-t_4)}\cosh{\pi(s_2+\zeta_1+\zeta_2)}}\\
=& \widetilde{\CZ}_B(M_i,M_a;\widetilde{t_3},\widetilde{t}_4)
\end{split}
\end{equation}
The dual theory therefore splits into two parts -- an $Sp(1)$ gauge theory with 4 flavors whose partition function is given by the first line (masses labeled as $M_i$ with $i=1,2,3,4$) and a $U(1)$ gauge theory with 2 flavors whose partition function is given by the second line (masses labeled as $M_a$ with $a=5,6$). 

The mirror map for this mirror pair can then be directly read off from the above partition function.
\begin{equation}
\begin{split}
&M_a=t_a +\frac{\zeta_1}{2}\,,\quad \; a=1,2,3\\
&M_4=\frac{\zeta_1}{2}+\zeta_2\\
&\widetilde{t}_j  = m_j\,,\quad \;  \; j=3,4\\
&M_{5}=t_4 \\
&M_{6}=\zeta_1+\zeta_2\,.
\end{split}
\end{equation}
Note that the number of parameters exactly match on both sides. For the A-model, we have five  independent FI parameters -- $\{t_1,t_2,t_3,t_4\}$ with one constraint and $\{\zeta_1,\zeta_2\}$. This is matched by the 5 independent mass parameters for the B-model -- 6 mass parameters with the following constraint
\begin{equation}
M_1+M_2+M_3+M_5=3(M_6-M_4).
\end{equation}

Similarly, two mass parameters on the A-model side coincides with the two $\widetilde{t}_a$ parameters on the B-model side.

In general, in order to obtain a mirror pair in this class for $N>4$ by gauging, we start from the following linear quivers:\\
\textbf{A-model}: $(1,0)(2,1)_{1}(2,0)_2\ldots(2,0)_{N-4}(2,1)_{N-3}(1,2)_{N-2}$  and \\
\textbf{B-model}: $(2,N-1)(1,0)(1,1)$.

One needs to gauge the $U(1)$ flavor symmetries of the nodes $(2,1)_1$ and $(2,1)_{N-3}$ to obtain the appropriately framed $\hat{D}_N$ quiver. Proceeding as before, the mirror is found to consist of a $Sp(1)$ gauge theory with $N$ fundamental hypers and a decoupled $U(1)$ gauge theory with two hypers. The mirror map in this case is an obvious generalization of the $\hat{D}_4$ case.
\begin{equation}
\begin{split}
&M_a=t_a +\frac{\zeta_1}{2} \; (a=1,2,\ldots,N-1)\\
&M_N=\frac{\zeta_1}{2}+\zeta_2\\
&\widetilde{t}_j  = m_j \; (j=3,4)\\
&M_{N+1}=t_4 \\
&M_{N+2}=\zeta_1+\zeta_2\,.
\end{split}
\end{equation}

\subsubsection*{Checking Mirror Symmetry in \figref{fig:QuiverflavoredD44} by using Hilbert series}
It is instructive to check the result by computing the corresponding Hilbert series and in particular the fact that the $(b')$ quiver in \figref{fig:QuiverflavoredD44} is indeed a disjoint union of two quivers. In what follows we compute both Higgs and Coulomb branch series.
Note that in the Higgs branch Hilbert series we use
\bea
\tau = t^{1/2}~.
\eea

The Higgs branch Hilbert series of diagram (a) in \figref{fig:QuiverflavoredD44} is given by the gluing technique \cite{Benvenuti:2010pq, Hanany:2011db}:
\bea
H^{H}_{(a)} (\tau, x) &= \left(\prod_{i=1}^4 \oint_{|q_i|=1} \frac{\ud q_i}{2 \pi iq_i} \right) \left(\frac{1}{2} \prod_{i=1}^2 \oint_{|z_i|=1} \frac{\ud z_i}{2 \pi iz_i} \right) (z_1 -z_2)(z_1^{-1}-z_2^{-1})  \times \nn \\
&\qquad  \chi_{(1)-[2]}(\tau; q_1; x) \prod_{i=1}^4 \chi_{(2)-(1)_i}(\tau;q_i; \vec z)~,
\eea
where $q_1, \ldots q_4$ denote the gauge fugacities of the four $U(1)$ gauge groups, $z_1, z_2$ denote the gauge fugacities of the $U(2)$ gauge group, $(x,y)$ denotes the fugacities of the $U(2) =U(1) \times SU(2)$ flavour node, and the contributions from the hypermultiplets are
\bea
\chi_{(1)-[2]}(\tau; q_1; x,y) &= \PE \left[ \tau (q_1 x^{-1}+q_1^{-1} x) (y+y^{-1}) \right] ~, \nn \\
\chi_{(2)-(1)_i}(\tau; q_i; \vec z) &= \PE \left[ \tau (z_1+z_2) q_i^{-1} + \tau (z_1^{-1} + z_2^{-1}) q_i \right] ~,
\eea
with the plethystic exponential $\PE$ of a multivariate function $f(a_1, a_2, \ldots, a_n)$, with $f(0,0,\ldots,0)=0$, defined as
\bea
\PE [f(a_1, a_2, \ldots, a_n) ] = \exp \left( \sum_{k=1}^\infty \frac{1}{k} f(a_1^k, a_2^k, \ldots, a_n^k) \right)~.
\eea

As a result of the integrations, we find that
\bea
H^{H}_{(a)} (\tau, y) &= \frac{1-\tau^{12}}{(1-\tau^4)^2(1-\tau^6)} \sum_{m=0}^\infty [2m]_{y} \tau^{2m}
\eea
This is indeed the Hilbert series of $(\BC^2/\hat{D}_4) \times (\BC^2/\BZ_2)$ \cite{Benvenuti:2006qr}; this is in agreement with the Coulomb branch of diagram ($b'$) of \figref{fig:QuiverflavoredD44}.

The Coulomb branch Hilbert series of diagram (a) in \figref{fig:QuiverflavoredD44} is given by
\bea
H^C_{(a)}(t, \vec a, \vec b) &=  \sum_{m_1 =-\infty}^\infty \sum_{m_2 =-\infty}^{m_1} \sum_{n_1=-\infty}^\infty \cdots \sum_{n_4=-\infty}^\infty P_{U(2)} (t, m_1, m_2) P_{U(1)} (t)^4 \prod_{i=1}^2 a_i^{m_i} \prod_{j=1}^4 b_j^{n_j}\nn \\
\eea
where $m_1, m_2$ are the monopole charges associated with the $U(2)$ gauge group, and $n_1, \ldots, n_4$ are the monopole charges associated with each $U(1)$ gauge group.  Here $\Delta(\vec m, \vec n)$ is the dimension of the monopole operators:
\bea
\Delta(m_1,m_2, n_1, \ldots, n_4) = \frac{1}{2} \left( 2|n_1| + \sum_{i=1}^2 \sum_{j=1}^4 |m_i-n_j| \right) - |m_1-m_2|~,
\eea
and the functions $P_{U(2)}(t, \vec m)$ and $P_{U(1)}(t)$ are defined as
\bea
P_{U(1)}(t) &= \frac{1}{1-t}~, \nn \\
P_{U(2)}(t, \vec m) &= \begin{cases} \frac{1}{(1-t)^2}~, & \qquad m_1 \neq m_2 \\  \frac{1}{(1-t)(1-t^2)}~, &\qquad m_1 = m_2~. \end{cases}
\eea
Setting $a_i=b_j=1$ for all $i,j$, we obtain
\bea \label{CoulombSp1Sp1SO8}
H^C_{(a)}(t, \{ a_i=1 \}, \{ b_j =1 \}) &= \frac{1+19 t+83 t^2+130 t^3+83 t^4+19 t^5+t^6}{(1-t)^{12}}  \nn \\
&= \frac{(1+t)^2 \left(1+17 t+48 t^2+17 t^3+t^4\right)}{(1-t)^{12}}~.
\eea
The order of the pole at $t=1$ is 12; this is equal to the complex dimension of the Coulomb branch as expected.

Then we investigate the Higgs branch of diagram (b) in \figref{fig:QuiverflavoredD44}. The space of F-term solutions (also known as the F-flat space) of quiver in diagram (b) of \figref{fig:QuiverflavoredD44} can be decomposed into many branches.  The branch that leads to the Higgs branch after imposing the $D$-term constraints is the 18 complex dimensional branch.  The Hilbert series of this branch can be obtained using {\tt Macaulay2} \cite{M2}. The closed form is, however, too lengthy to be reported here; let us present a few terms in the series expansion:
\bea
\ff(\tau; z_1, z_2; x, \vec y) = 1+ ([1,0,0,0]_{\vec y} [1]_{z_1} + [1]_x [1]_{z_1} [1]_{z_2}) \tau+ \ldots~,
\eea
where $z_1, z_2$ are gauge fugacities for each $Sp(1)$ gauge group, $x$ is the global $SU(2)$ fugacity that transform the chiral fields in hypermultiplet in $Sp(1) \times Sp(1)$, and $y_1, \ldots, y_4$ are the fugacities of the $SO(8)$ flavour symmetry.  The corresponding unrefined Hilbert series is
\bea
\ff(\tau; z_1=z_2=1; x=1, \{y_i =1 \}) = \frac{(1+\tau)^3 (1+3 \tau)}{(1-\tau)^{18}}~.
\eea

After implementing gauge invariance, the Higgs branch Hilbert series is given by
\bea
H^H_{(b)}(\tau,x, \vec y) 
&= \oint_{|z_1|=1} \frac{\ud z_1}{2\pi i z_1} \frac{1-z_1^2}{z_1}  \oint_{|z_2|=1} \frac{\ud z_2}{2\pi i z_2} \; \frac{1-z_2^2}{z_2}\ff(\tau; z_1, z_2; x, \vec y) \nn \\
&= \sum_{m=0}^\infty [2m]_{x} \tau^{2m} \times \sum_{n=0}^\infty [0,n,0,0]_{\vec y} \tau^{2n}~.
\eea
This Hilbert series indicates that the Higgs branch of the diagram (b) of \figref{fig:QuiverflavoredD44} is indeed
\bea \label{C2Z2SU2w4flv}
\BC^2/\BZ_2 \times \CH_{\text{$SU(2)$ w/ 4 flv}}~,
\eea
where $\CH_{\text{$SU(2)$ w/ 4 flv}}$ is the Higgs branch of $SU(2)$ with $4$ flavours.  Therefore, this agrees with the Higgs branch of the $(b')$ quiver in \figref{fig:QuiverflavoredD44Lin}.

Setting $x=1$ and $y_i=1$, we obtain the unrefined Higgs branch Hilbert series
\bea \label{HiggsSp1Sp1SO8}
H^H_{(b)}(\tau,x =1, \{ y_i =1 \})  &= \frac{1-\tau^4}{(1-\tau^2)^3}  \times \frac{\left(1+\tau^2\right)\left(1+17 \tau^2+48 \tau^4+17 \tau^6+\tau^8\right)}{\left(1-\tau^2\right)^{10}}\nn \\
&= \frac{\left(1+\tau^2\right)^2 \left(1+17 \tau^2+48 \tau^4+17 \tau^6+\tau^8\right)}{\left(1-\tau^2\right)^{12}} \nn \\
&= H^C_{(a)}(\tau^2, \{ a_i=1 \}, \{ b_j =1 \})~.
\eea
Note that this is in agreement with \eref{CoulombSp1Sp1SO8}, thereby providing a very non-trivial check of the proposed mirror symmetry.

\subsubsection*{General results}
\begin{figure}[H]
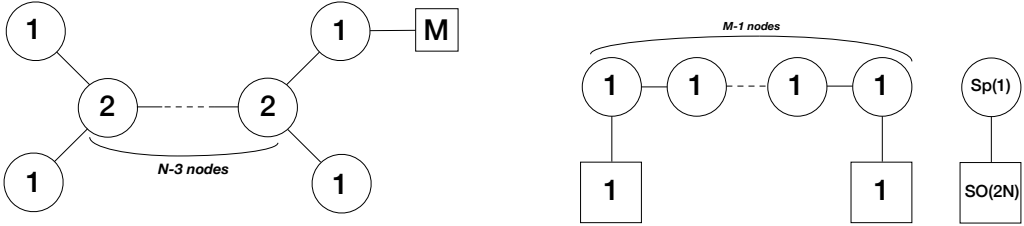

\begin{center}
\includegraphics[scale=0.3]{Dk1NfM}\qquad \qquad \includegraphics[scale=0.3]{Sp1N1111}
\caption{$\hat{D}_N$ quiver with $M$ hypermultiplets on the edge node and its mirror.
The Higgs branch of the left quiver is the moduli space of $1$ $SU(M)$ instanton on $\BC^2/\hat{D}_N$ and that of the right quiver is the moduli space of $1$ $SO(2N)$ instanton on $\BC^2/\BZ_M$.  The latter factorises into $\BC^2/\BZ_M$ times the reduced instanton moduli space of $1$ $SO(2N)$ instanton on $\BC^2$; see \cite{Dey:2013fea}.}
\label{fig:Dk1NfM}
\end{center}
\end{figure}

One can easily generalize the above computation to determine the mirror of a $\hat{D}_N$ quiver with $M>2$ fundamental hypers on one of the external nodes, as shown in \figref{fig:Dk1NfM}. 

The starting point is the mirror pair consisting of the following linear quivers:\\
\textbf{A-model}: $(1,0)(2,1)_{1}(2,0)_2\ldots(2,0)_{N-4}(2,1)_{N-3}(1,M)_{N-2}$ and \\
\textbf{B-model}: $(2,N-1)(1,0)_1\ldots(1,0)_{M-1}(1,1)_{M}$.

Again gauging the $U(1)$ flavor symmetries of the nodes $(2,1)_1$ and $(2,1)_{N-3}$ to obtain the appropriately framed $\hat{D}_N$ quiver, we find that the dual theory consists of a $Sp(1)$ gauge theory with $N$ fundamental hypers and a decoupled quiver gauge theory $(1,1)_1(1,0)_2\ldots(1,0)_{M-2}(1,1)_{M-1}$. The mirror map is an obvious generalization of the one obtained for $M=2$.

\subsubsection{Framing at two different nodes of $\hat{D}_4$ quiver}
\begin{figure}[H]
\begin{center}
\includegraphics[scale=0.6]{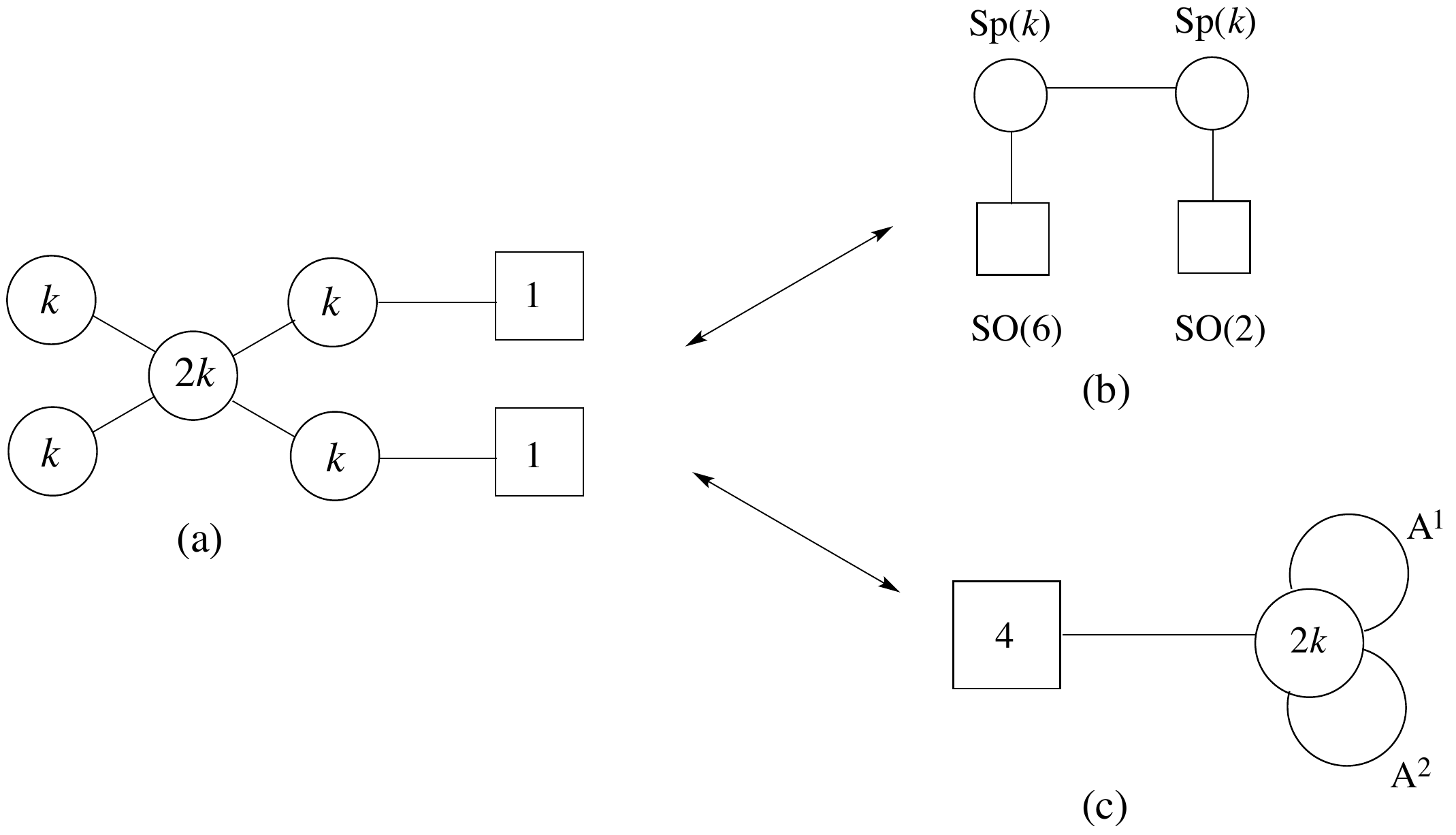}
\caption{Two possible mirrors for $\hat{D}_4$ quiver (a).}
\label{fig:QuiverflavoredD43}
\end{center}
\end{figure}

As a final example of this section let us consider a situation presented in \figref{fig:QuiverflavoredD43} for $\hat{D}_4$ quiver, when two of the boundary nodes of the tail are framed. 
Clearly, the $\hat{D}_4$ quiver shown in \figref{fig:QuiverflavoredD43} is somewhat special as it has more symmetries than a generic $\hat{D}_N$ quiver. Below we shall work in detail on two families of doubly framed $\hat{D}_N$ quivers shown in \figref{fig:QuiverflavoredDkN2duo} and \figref{fig:QuiverflavoredDkN2} which coincide for $N=4$ as we have already seen in \figref{fig:QuiverflavoredD43}

\subsubsection*{Mirror Dual with a Unitary Gauge Group}
\begin{figure}[H]
\begin{center}
\includegraphics[scale=0.35]{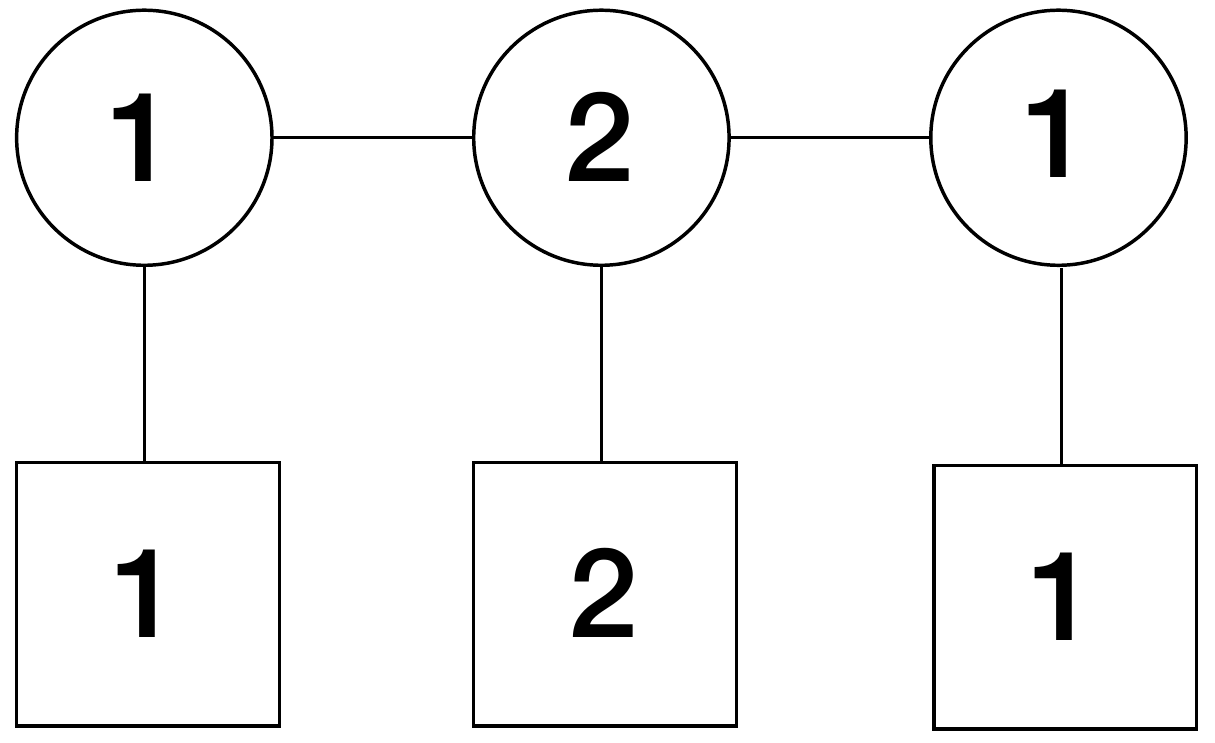}
\caption{Linear $A_3$ quiver with labels $(1,1)(2,2)(1,1)$.} 
\label{fig:quiver112211}
\end{center}
\end{figure}

Consider the mirror theory corresponding to the lower arrow first. The appropriate linear quiver in this case is $(1,1)(2,2)(1,1)$ (see \figref{fig:quiver112211}). Note that this is a self-mirror. 
Its partition functions reads
\begin{equation}
\begin{split}
\CZ_A(m_a;{t}_{a})=& \int \prod^2_{\alpha=1} d s_{\alpha}  \frac{d^2 s_0}{2!} \frac{\prod^2_{\alpha=1} e^{2\pi i s_{\alpha} \eta_{\alpha}} \prod^2_{i=1} e^{2\pi i s^i_0 \eta_0} \sinh^2{\pi(s^1_0-s^2_0)}}{\cosh{\pi(s_1+m_1)}\prod^2_{i=1}\cosh{\pi(s_1-s^i_0)}\prod^2_{a=1} \cosh{\pi(s^i_0+m_{1+a})}} \\
&\times \frac{1}{\prod^2_{i=1}\cosh{\pi(s_2-s^i_0)}\cosh{\pi(s_2+m_{4})}}\,.
\end{split}
\end{equation}
As before, we define $\eta_1=t_1-t_2$, $\eta_0=t_2-t_3$ and $\eta_2=t_3-t_4$.\\
The partition function of the mirror dual, which in this case is the same theory, simply involves the exchange of parameters $m_a \leftrightarrow t_a$. Up to some overall phase which we will ignore in this discussion, we have
\begin{equation}
\CZ_B (M_{a}; \widetilde{t}_a) =\CZ_A(t_a;m_a)\,.
\end{equation}

Now we gauge the $U(2)$ flavor symmetry of the A-model as a $U(1) \times U(1)$, which gives a $\hat{D}_4$ quiver with two hypers on a single boundary node. The partition function of this theory is 
\begin{equation}
\begin{split}
\widetilde{\CZ}_A(m_1,\zeta_2,\zeta_3, m_4;{t}_{a})=&\int dm_2 dm_3 e^{2\pi i m_2 \zeta_2} e^{2\pi i m_3 \zeta_3} \CZ_A (m_a;{t}_{a})\\
=&\int dm_2 dm_3 e^{2\pi i m_2 \zeta_2} e^{2\pi i m_3 \zeta_3} \CZ_B ({t}_{a}; m_a)\,,
\end{split}
\end{equation}
where the second equality follows from the mirror symmetry of the linear quivers. From the second equality, completing the integration over $m_2$ and $m_3$ we have
\begin{equation}
\begin{split}
\widetilde{\CZ}_A(m_1,\zeta_2,\zeta_3, m_4;{t}_{a})=& \int  ds_2 d s_{1}   \frac{d^2 s_0}{2!} \frac{\delta(s^1_0+s^2_0-s_1+\zeta_2)\delta(s^1_0+s^2_0-s_2-\zeta_3) \sinh^2{\pi(s^1_0-s^2_0)}}{\prod_i \prod^3_{\alpha=2}\cosh{\pi(s^i_0+t_{\alpha})}\cosh{\pi(s^i_0-s_1)}\cosh{\pi(s^i_0-s_2)}} \\
&\times \frac{e^{2\pi i m_1 s_1}e^{-2\pi i m_4 s_2}}{ \cosh{\pi(s_1+t_1)}\cosh{\pi(s_2+t_4)}}\,.
\end{split}
\end{equation}
Finally, integrating over $s_1$ and $s_2$ using the delta functions and shifting the remaining integration variables appropriately, we have 
\begin{equation}
\begin{split}
\widetilde{\CZ}_A(m_1,\zeta_2,\zeta_3, m_4;{t}_{a})=& \int \frac{d^2 s_0}{2!} \frac{e^{2\pi i(s^1_0+s^2_0)(m_1-m_4)}\sinh^2{\pi(s^1_0-s^2_0)}}{\prod_i \prod^3_{\alpha=2}\cosh{\pi(s^i_0+t_{\alpha})}\cosh{\pi(s^i_0 +\zeta_2)}\cosh{\pi(s^i_0-\zeta_3)}} \\
&\times \frac{1}{\cosh{\pi(s^1_0+s^2_0+\zeta_2+t_1)}\cosh{\pi(s^1_0+s^2_0-\zeta_3+t_4)}}\\
=& \widetilde{\CZ}_B(M_a;\widetilde{t_1},\widetilde{t}_4)\,.
\end{split}
\end{equation}
The dual theory is therefore  a $U(2)$ gauge theory with 4 fundamental hypers and 2 hypers in the antisymmetric representation of $U(2)$. The mirror map for this mirror pair can then be directly read off from the above partition function.
\begin{equation}
\begin{split}
&\widetilde{t}_j  = m_j \; (j=1,4)\\
&M_1=\zeta_2,\; M_2=-\zeta_3\\
&M_3=t_2, \; M_4=t_3\\
&M^{AS}_{1}=t_4-\zeta_3 \\
&M^{AS}_{2}=t_1+\zeta_2\,,\\
\end{split}
\end{equation}
which implies
\begin{equation}
M^{AS}_{1}+M^{AS}_{2}-M_1-M_2 +M_3 + M_4=0\,.
\end{equation}
Note that the number of parameters exactly match on both sides. For the A-model, we have five FI independent parameters - $\{t_1,t_2,t_3,t_4\}$ with one constraint and $\{\zeta_2,\zeta_3\}$. This is matched by the five independent mass parameters for the B-model -- six mass parameters with one constraint, namely $M^{AS}_{1}+M^{AS}_{2}-M_1-M_2 +M_3 + M_4=0$. Similarly, two mass parameters on the A-model side coincides with the two $\widetilde{t}_a$ parameters on the B-model side.\\

\begin{figure}[H]
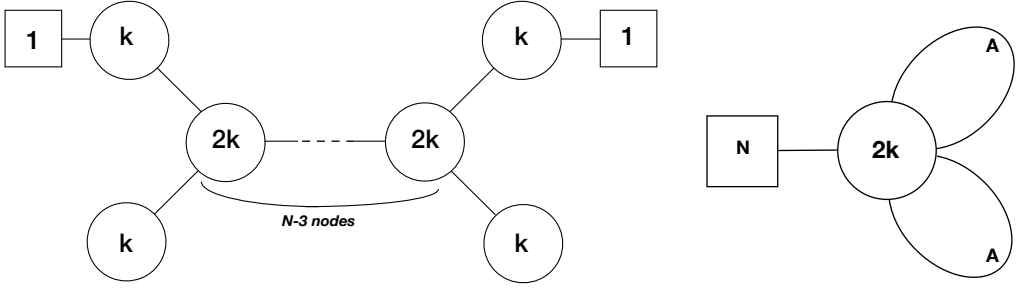

\begin{center}
\begin{tabular}{cc}
\includegraphics[scale=0.4]{DkN2} & \includegraphics[scale=0.5]{2kSO2NAA}
\end{tabular}
\end{center}
\caption{An infinite family of affine $D$-type quiver with its mirror dual}
\label{fig:QuiverflavoredDkN2}
\end{figure}

In order to obtain a generic mirror pair in this class (for rank of the quiver $N>4$) by gauging, we start from the following linear quivers:\\
\textbf{A-model}: $(1,1)(2,1)_{1}(2,0)_2\ldots(2,0)_{N-4}(2,1)_{N-3}(1,1)$ and \\
\textbf{B-model}: $(1,1)(2,N-2)(1,1)$.\\
One needs to gauge the $U(1)$ flavor symmetries of the nodes $(2,1)_1$ and $(2,1)_{N-3}$ to obtain the appropriately framed $\hat{D}_N$ quiver. Proceeding as before, the mirror is found to consist of a $U(2)$ gauge theory with $N$ fundamental hypers and two hypers in the  antisymmetric representation of $U(2)$. The mirror map in this case is an obvious generalization of the $\hat{D}_4$ case and is presented in \figref{fig:QuiverflavoredDkN2}
The corresponding mirror maps are the following 
\begin{equation}
\begin{split}
&\widetilde{t}_j  = m_j \; (j=1,4)\\
&M_1=\zeta_2,\; M_2=-\zeta_3\\
&M_i=t_{i-1}, \; (i=3,4,\ldots,N)\\
&M^{AS}_{1}=t_N-\zeta_3 \\
&M^{AS}_{2}=t_1+\zeta_2\,,
\end{split}
\end{equation}
therefore we get
\begin{equation}
M^{AS}_{1}+M^{AS}_{2}-M_1-M_2 +\sum^N_{i=3}M_i=0\,.
\end{equation}

\subsubsection*{Mirror Dual with Symplectic Gauge Groups }
\begin{figure}[H]
\begin{center}
\includegraphics[scale=0.35]{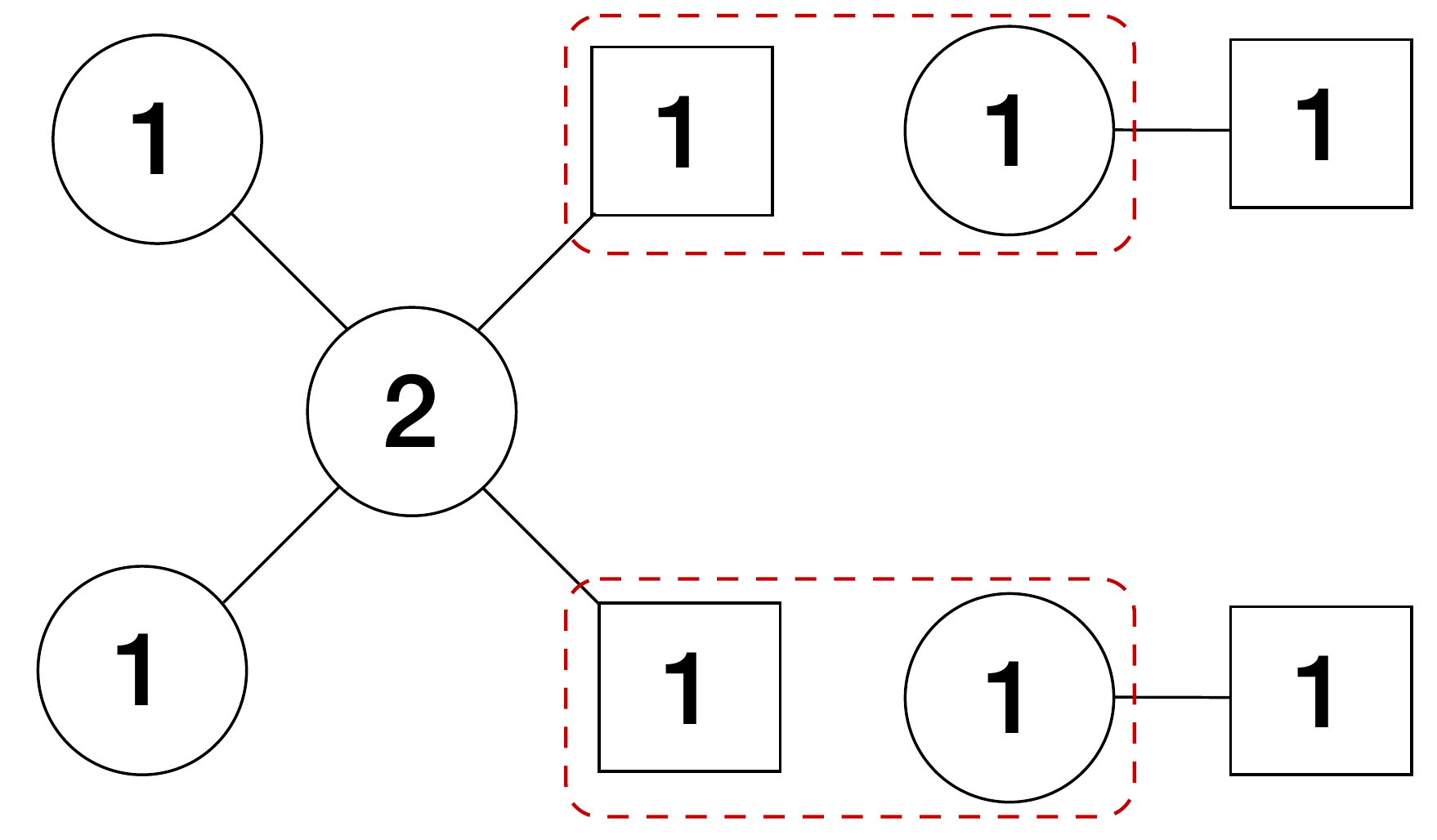}\qquad\qquad\qquad \includegraphics[scale=0.35]{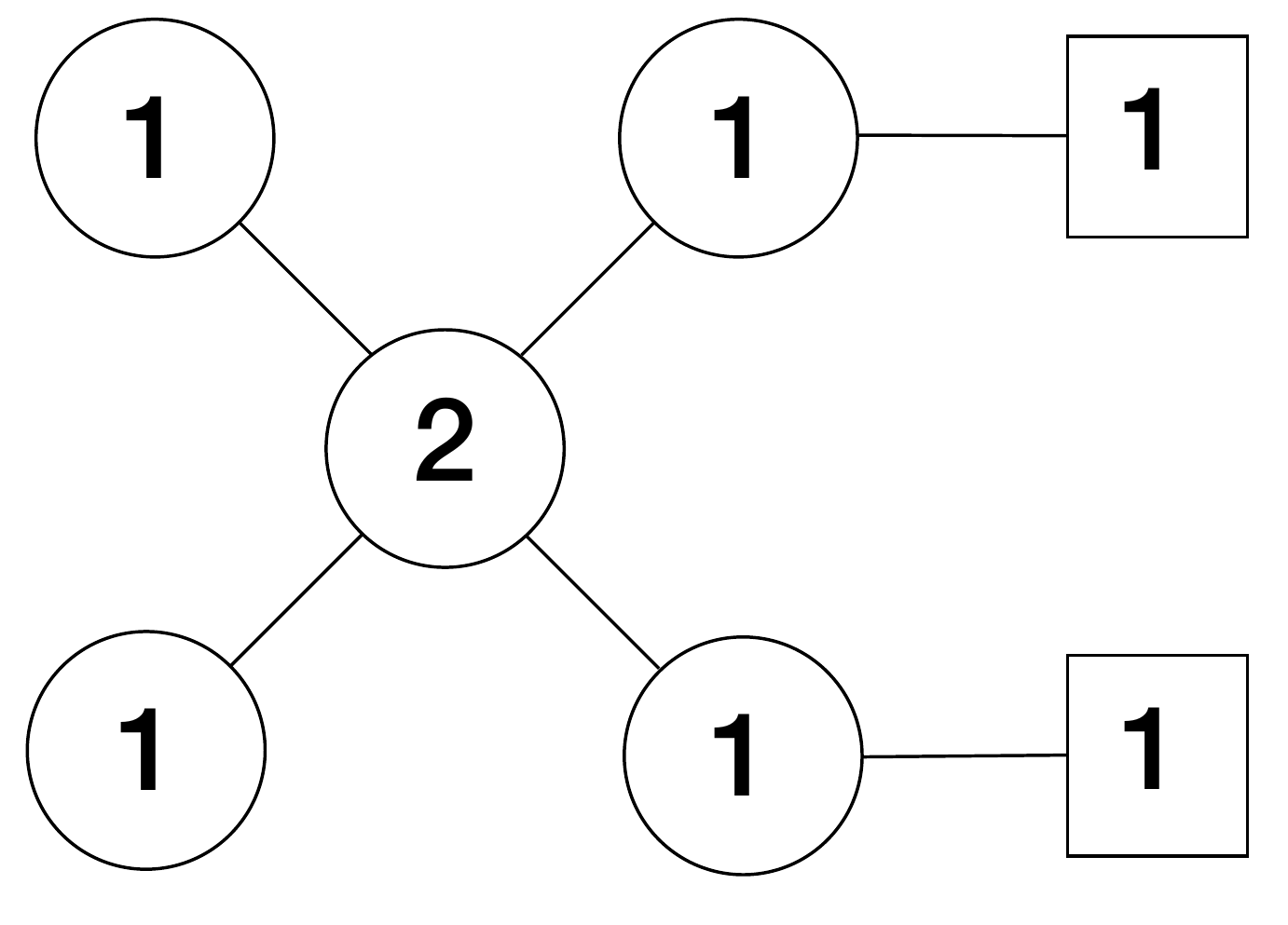}
\caption{Attaching two $(1,1)$ blocks to a $(1,0)(2,2)(1,0)$ theory (left) in order to get the $\hat{D}_4$ quiver with two framings (right).}
\label{fig:GenFraimEx}
\end{center}
\end{figure}

Consider the linear quiver $(1,0)(2,2)(1,0)$.  The partition function of this theory is given by
\begin{equation}
\begin{split}
\CZ_A(m_i;{t}_{j})=& \int \prod^2_{\alpha=1} d s_{\alpha}  \frac{d^2 s_0}{2!} \frac{\prod^2_{\alpha=1} e^{2\pi i s_{\alpha} \eta_{\alpha}} \prod^2_{i=1} e^{2\pi i s^i_0 \eta_0} \sinh^2{\pi(s^1_0-s^2_0)}}{\prod^2_{i=1}\cosh{\pi(s_1-s^i_0)}\prod^2_{a=1} \cosh{\pi(s^i_0+m_{a})}\prod^2_{i=1}\cosh{\pi(s_2-s^i_0)}}.
\end{split}
\end{equation}
As before, we define $\eta_1=t_1-t_2$, $\eta_0=t_2-t_3$ and $\eta_2=t_3-t_4$ while $m_1,m_2$ are masses of the fundamental hypers. We now attach two $(1,1)$ blocks to the globally symmetry of the quiver as shown in \figref{fig:GenFraimEx}. 
At the level of the partition function this operation can be represented in the following form
\begin{equation}
\begin{split}
\widetilde{\CZ}_A(a_1,a_2;{t}_{j}, \zeta_1,\zeta_2)= \int \prod^2_{i=1} dm_i \frac{e^{2\pi i \zeta_{i} m_{i}}}{\cosh{\pi(m_i-a_i)}} \CZ_A(m_i;{t}_{j})\,,
\end{split}
\end{equation}
where $\widetilde{\CZ}_A(a_1,a_2;{t}_{j}, \zeta_1,\zeta_2)$ is the partition function of the framed $\hat{D}_4$ quiver of interest written in terms of the partition function of the $(1,0)(2,2)(1,0)$ theory. Note that $\zeta_1,\zeta_2$ are FI parameters of the attached $U(1)$ nodes and $a_1,a_2$ are the masses of the fundamental hypers charged under those $U(1)$s.

In order to obtain the correct mirror to this theory, we will need to start with $\widetilde{\CZ}_A$ and implement S-duality at the level of the partition function in a fashion similar to \cite{Dey:2013nf,Dey:2011pt}. To see precisely how this works out, let us rewrite $\widetilde{\CZ}_A$ in the following manner,
\begin{equation}
\begin{split}
&\widetilde{\CZ}_A(a_1,a_2;{t}_{j}, \zeta_1,\zeta_2)=  \int \prod^2_{i=1} dm_i \prod^2_{\alpha=1} d s_{\alpha}  \frac{d^2 s_0}{2!} \frac{1}{\sinh{\pi(m_1-m_2)}}\prod^2_{i=1}\frac{e^{2\pi i \zeta_{i} m_{i}}}{\cosh{\pi(m_i-a_i)}} \\
&\times  \frac{\sinh{\pi(m_1-m_2)} \sinh{\pi(s^1_0-s^2_0)}}{\prod_{i,j} \cosh{\pi(s^i_0-m_j)}} \times \frac{\sinh{\pi(s^1_0-s^2_0)}\sinh{\pi(s_1-s_2)}}{\prod_{i,j} \cosh{\pi(s^i_0-s_j)}}\times \frac{\prod^2_{\alpha=1} e^{2\pi i s_{\alpha} \eta_{\alpha}} \prod^2_{i=1} e^{2\pi i s^i_0 \eta_0}}{\sinh{\pi(s_1-s_2)}}\\
&=-\int \prod^2_{i=1} dm_i dm'_i du_i dv_i dz_i dx dy \prod^2_{\alpha=1} d s_{\alpha}  \frac{d^2 s_0}{2!} \tanh{\pi x} \; e^{2\pi i x(m_1-m_2)} \prod^2_{i=1}e^{2\pi i \zeta_{i} m_{i}}\frac{e^{2\pi i (m_i-m'_i)z_i}}{\cosh{\pi(m_i-a_i)}}\\
&\times \left(\sum_{\rho}(-1)^{\rho}\prod^2_{i=1}\frac{e^{2\pi i u_i(s^i_0 - m'_{\rho(i)})}}{\cosh{\pi u_i}} \right)\times \left(\sum_{\rho'}(-1)^{\rho'}\prod^2_{i=1}\frac{e^{2\pi i v_i(s^i_0 - s_{\rho'(i)})}}{\cosh{\pi v_i}} \right)   \\
&\times  \tanh{\pi y} \; e^{2\pi i y(s_1-s_2)}  \prod^2_{\alpha=1} e^{2\pi i s_{\alpha} \eta_{\alpha}} \prod^2_{i=1} e^{2\pi i s^i_0 \eta_0}\,,
\end{split} 
\label{Amodel-FT}
\end{equation}
where $\rho, \rho'$ denote permutations over the labels $i=1,2$.

Going to the second line from the first, we have used Cauchy determinant identity and Fourier transform of hyperbolic functions to write the partition function in terms of a set of auxiliary variables $\{x,m'_i, z_i,u_i,v_i,y\}$. For example
\begin{equation}
\begin{split}
\frac{\sinh{\pi(m_1-m_2)} \sinh{\pi(s^1_0-s^2_0)}}{\prod_{i,j} \cosh{\pi(s^i_0-m_j)}}=& \sum_{\rho}(-1)^{\rho} \frac{1}{\prod^2_{i=1} \cosh{\pi (s^i_0 - m'_{\rho(i)})}}\\
=&\int \prod^2_{i=1} du_i \sum_{\rho}(-1)^{\rho}\prod^2_{i=1}\frac{e^{2\pi i u_i(s^i_0 - m'_{\rho(i)})}}{\cosh{\pi u_i}}
\end{split}
\end{equation}

In addition, we used
\begin{equation}
\begin{split}
\frac{1}{\sinh{\pi(m_1-m_2)}}=-i \int dx \; \tanh{\pi x} \; e^{2\pi i x(m_1-m_2)}
\end{split}
\end{equation}

Implementing S-duality at the level of partition function amounts to carrying out the integration over the original variables $\{s_{\alpha}, s^i_0\}$ and writing the partition function exclusively in terms of the auxiliary fields. Performing the said integrations followed by some trivial change of variables we have
\begin{equation}
\begin{split}
\widetilde{\CZ}_A(a_1,a_2;{t}_{j}, \zeta_1,\zeta_2)=&\int \prod^2_{i=1}  du_i  dz_i \; \tanh{(\pi z_1)}\tanh{(\pi u_1)} \left(\sum_{\rho}(-1)^{\rho}\prod^2_{i=1}\frac{e^{2\pi i a_i(z_i-\zeta_i +u_{\rho(i)}+ \xi_{\rho(i)})}}{\cosh{\pi(z_i-\zeta_i+u_{\rho(i)}+ \xi_{\rho(i)})}}\right) \\
&\times \left(\frac{\delta(z_1+z_2)\delta(u_1+u_2)}{\prod^2_{i=1} \cosh{\pi (u_i+\xi_i)}\cosh{\pi (u_i+\xi_i+\eta_0)}}\right)\,,
\end{split}
\end{equation}
where $\xi_1=\eta_1+\eta_0=t_1-t_3$ and $\xi_2= \eta_2+\eta_0=t_2-t_4$.

To perform the sum over permutations in $\widetilde{\CZ}_A$, we again need to use Cauchy determinant identity. However, this can only be done if the phase is independent of $\rho$, which requires that the hypermultiplet masses obey the relation
\begin{equation}\label{massconstraint}
a_1=a_2=a\,.
\end{equation}
Imposing this condition and summing over the permutations we obtain
\begin{equation}
\begin{split}
&\widetilde{\CZ}_A(a;{t}_{j}, \zeta_1,\zeta_2)\\
&=\int \frac{d^2u}{2}  \frac{d^2z}{2} \;\left( \frac{\sinh{\pi(z_1-z_2)}\sinh{\pi(z_1-z_2-\zeta_1+\zeta_2)}\sinh{\pi(u_1-u_2+\xi_1-\xi_2)}\sinh{\pi(u_1-u_2)}}{\prod_{i,j} \cosh{\pi(z_i+u_j-\zeta_i+\xi_j)}} \right)\\
&\times \left(\frac{e^{2\pi i a (\xi_{1}+\xi_{2}-\zeta_1-\zeta_2)} \delta(z_1+z_2)\delta(u_1+u_2)}{\prod^2_{i=1} \cosh{\pi (u_i+\xi_i)}\cosh{\pi (u_i+\xi_i+\eta_0)}\cosh{(\pi u_i)}\cosh{(\pi z_i)}}\right)\,.
\end{split}
\end{equation}
To interpret the numerator of the first term in parenthesis as the contribution of a $\mathcal{N}=2$ vector multiplet, we would need
\begin{equation}
\begin{split}
&\zeta_1=\zeta_2=\zeta\\
&\xi_1=\xi_2 =\xi\,,
\end{split} \label{FIconstraint}
\end{equation}
which implies
\begin{equation}
t_1+t_4=0\,,\quad t_2+t_3=0\,.
\end{equation}
The constraint equations \eqref{massconstraint} and \eqref{FIconstraint} together imply that the masses and the FI parameters obey the $\mathbb{Z}_2$ outer automorphism symmetry of the $\hat{D}_4$ quiver. Imposing these constraints, we finally have
\begin{equation}
\begin{split}
\widetilde{\CZ}_A(a;{t}_{j}, \zeta)=& \int \frac{d^2u}{2}  \frac{d^2z}{2} \;\left( \frac{\sinh^2{\pi(z_1-z_2)}\sinh^2{\pi(u_1-u_2)}}{\prod_{i,j} \cosh{\pi(z_i+u_j-\zeta+\xi)}} \right)\\
&\times \left(\frac{e^{4\pi i a (\xi-\zeta)} \delta(z_1+z_2)\delta(u_1+u_2)}{\prod^2_{i=1} \cosh{\pi (u_i+\xi)}\cosh{\pi (u_i+\xi+\eta_0)}\cosh{(\pi u_i)}\cosh{(\pi z_i)}}\right)\,.
\end{split}
\end{equation}
This is evidently the partition function of a $(Sp(1),1)(Sp(1),3)$ quiver shown in \figref{fig:QuiverflavoredD43}. The mirror map can be read off from the above formula.
\begin{equation}
\begin{split}
&M_1=t_1+t_2\\
&M_2=t_1+t_2-2t_3\\
&M_3=0\\
&M_4=0\\
&M_{\text{bif}}=t_1-t_3 +\zeta\,.
\end{split}
\end{equation}
Note that the number of non-zero mass parameters of the B model exactly match with the number of independent FI parameters of the A model.

In order to obtain a generic mirror pair in this class (for rank of the quiver $N>4$) by gauging, we start from a linear quiver $(2,2)_{1}(2,0)_2\ldots(2,2)_{N-3}$. Firstly, one needs to gauge the flavor group $U(2)_1$ as a $U(1) \times U(1)$. Then the flavor group $U(2)_{N-3}$ should be split as a $U(1) \times U(1)$ and a $(1,1)$ quiver must be attached to each $U(1)$ as we did in the $\hat{D}_4$ case. The resultant quiver can then be shown to dual to $(Sp(1),1)(Sp(1),N-1)$ using manipulations similar to the example shown above. 
\begin{figure}[H]
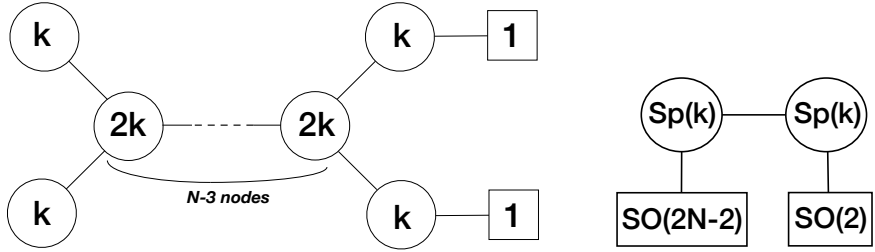

\begin{center}
\includegraphics[scale=0.35]{Dk2211}\qquad \includegraphics[scale=0.6]{Sp1Sp1SON2}
\caption{An infinite family of affine doubly framed $D$-type quivers with its mirror duals}
\label{fig:QuiverflavoredDkN2duo}
\end{center}
\end{figure}
The details of this computation and the associated mirror map can be found in \cite{Dey:2013nf}. We leave it to the enthusiastic reader as an exercise to show that, using similar manipulations and attaching $(k,1)$ blocks as we did above in the case of $k=1$, one can derive the mirror quiver to the doubly framed $\hat{D}_N$ quiver for generic $k$ (\figref{fig:QuiverflavoredDkN2duo}) directly.



\section{Flavored $E_{n}$ and $\hat{E}_{n}$ Quivers}\label{Sec:Emirrors}
In this section, we study a few examples of framed $E_{n}$ quivers (and their affine extensions) using the framework of Abelian gauging. It was known for some time already \cite{Intriligator:1996ex} (see also \cite{Benini:2010uu,Chacaltana:2010ks}) that balanced $\hat{E}_n$ quiver theories have non-Lagrangian mirror description and until recently \cite{Cremonesi:2013lqa} understanding of Higgs branches thereof was limited. The examples we are about to discuss here exclusively deal with framed $E$-type or $\hat{E}$-type quivers which have Lagrangian mirrors.

\subsection{Framed $E_6$ Theory from Linear Quiver}
\begin{figure}[!h]
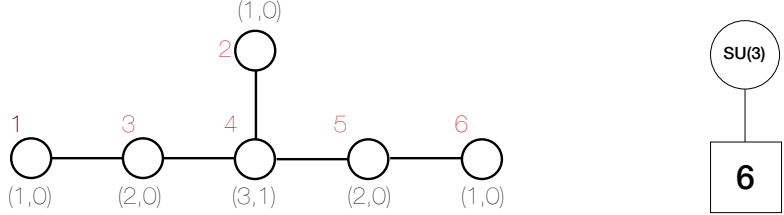

\begin{center}
\includegraphics[scale=0.9]{E6quiv} \qquad\qquad\qquad \includegraphics[scale=0.35]{SU3w6}
\caption{$E_6$ quiver with one hypermultiplet on the middle node and its mirror. Red number near the nodes of the $E_6$ quiver enumerate the nodes accruing to the Bourbaki convention.  The quiver on the left can be obtained from gluing $T_{(1,1,1)}(U(3))$, $T_{(1,1,1)}(U(3))$, $T_{(2,1)}(U(3))$ and $T_{(2,1)}(U(3))$ together via the $U(3)$ group, and the quiver on the right can be realised as as the $6d$ $(2,0)$ theory compactifying on a circle times a Riemann sphere with punctures $(1,1,1)$, $(1,1,1)$, $(2,1)$ and $(2,1)$.}
\label{fig:E6}
\end{center}
\end{figure}

Consider the mirror pair in \figref{fig:E6} \footnote{We are using Bourbaki conventions for numbering the nodes of $E$ quiver diagrams.} . This mirror pair may be obtained by the abelian gauging technique using $S^3$ partition function in a way similar to the previous sections. The starting point is again a linear quiver pair
\begin{equation}
\begin{split}
&\textbf{A-model}:\;\;(1,0)(2,0)(3,2)(2,0)(1,0) \\
&\textbf{B-model}:\;\;(3,6)\,.
\end{split}
\end{equation}
We perform the gauging trick on the global $U(2)$ symmetry of the middle node of the left quiver into $U(1)$ global and $U(1)$ gauge. From the perspective of the parameter space, this amounts to fixing one of the momenta, say $p^{(3)\,1}_\mu$, which on the mirror side results in taking out the trace part of $U(3)$.

Mirror symmetry dictates that the partition functions of the two linear quivers are related in the following manner up to an overall phase.  In \cite{Benini:2010uu} the same example was considered, however, on the A side the quiver had $SU(3)$ group in the middle instead of $U(3)$ with the overall $U(1)$ factorization. We stress again here that our computation is the correct one and only with the democratic overall $U(1)$ quotient the mirror map works correctly.
\begin{equation}
\begin{split}
&\CZ_A(m_a, t_{\alpha})= \CZ_B(M_{\alpha},\widetilde{t}_a)\\
&m_a =\widetilde{t}_a, \;\; a=1,2.\\
&t_{\alpha}=M_{\alpha},\;\; \alpha=1,2,\ldots,6
\end{split}
\end{equation}
Now consider gauging a single $U(1)$ of the $U(2)$ global symmetry in the A-model linear quiver to obtain the correct framed $E_6$ quiver. The partition function of such a theory is
\begin{equation}
\begin{split}
\widetilde{\CZ}_A(\zeta_1, m_2, t_{\alpha})=&\int dm_1 e^{2\pi i \zeta_1 m_1} {\CZ}_A(m_a, t_{\alpha}) \\
=&\int dm_1 e^{2\pi i \zeta_1 m_1} {\CZ}_B( t_{\alpha},m_a)
\end{split}
\end{equation}
where the second equality is a direct consequence of mirror symmetry. Therefore, we have
\begin{equation}
\begin{split}
\widetilde{\CZ}_A(\zeta_1, m_2, t_{\alpha})=&\int dm_1 \frac{d^3s}{3!} \frac{e^{2\pi i \zeta_1 m_1}\prod^3_{i=1} e^{2\pi i s^i (m_1-m_2)} \prod_{i<j} \sinh^2{\pi(s^i-s^j)}}{\prod^3_{i=1}\prod^6_{\alpha=1} \cosh{\pi(s^i+t_{\alpha})}}\\
=&\int \frac{d^3s}{3!}\frac{\delta(s^1+s^2+s^3) \prod_{i<j} \sinh^2{\pi(s^i-s^j)}}{\prod^3_{i=1}\prod^6_{\alpha=1} \cosh{\pi(s^i+t_{\alpha}-\frac{\zeta_1}{3})}}\\
=&\widetilde{\CZ}_B( M_{\alpha})
\end{split}
\end{equation}
The theory dual to the $E_6$ quiver with a single fundamental hyper can be read off from the partition function above -$SU(3)$ with 6 flavors. The mirror map relates the masses of the fundamental hypers of $SU(3)$ with the FI parameters of the framed $E_6$ quiver.
\begin{equation}
\begin{split}
M_{\alpha} =t_{\alpha} -\frac{\zeta_1}{3},\;\; \alpha=1,2,\ldots,6
\end{split}
\end{equation}
As expected, the 6 independent mass parameters of the B-model match with the number of independent parameters of the A-model - 6 parameters $\{t_{\alpha}\}$ with one constraint and $\zeta_1$.

Another way to realize the mirror pairs in \figref{fig:E6} is as follows. The quiver on the left can be obtained from gluing $T_{(1,1,1)}(U(3))$, $T_{(1,1,1)}(U(3))$, $T_{(2,1)}(U(3))$ and $T_{(2,1)}(U(3))$ together via the $U(3)$ group, and the mirror quiver on the right can be realised as as the $6d$ $(2,0)$ theory compactifying on a circle times a Riemann sphere with punctures $(1,1,1)$, $(1,1,1)$, $(2,1)$ and $(2,1)$ \cite{Benini:2010uu}.  Indeed, according to \cite{Argyres:2007cn} and \cite{Chacaltana:2010ks},\footnote{the diagram on page 16 of \cite{Chacaltana:2010ks}} such a mirror theory is the $SU(3)$ gauge theory with $6$ flavours.

\subsection{$\hat{E}_7$ Theory from Linear Quiver}
\begin{figure}[!h]
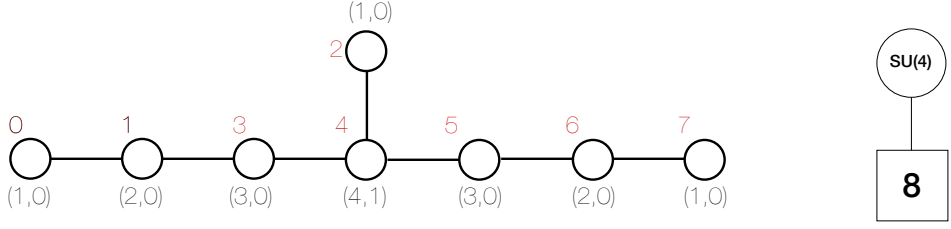

\begin{center}
\includegraphics[scale=0.9]{E7Afquiv} \qquad\qquad \includegraphics[scale=0.35]{SU4w8}
\caption{$\hat{E}_7$ quiver with one hypermultiplet on the branching node and its mirror. 
}
\label{fig:E7Aff}
\end{center}
\end{figure}

Now we consider an example of a framed $\hat{E}_7$ quiver, see \figref{fig:E7Aff}. 
In order to obtain this mirror pair via Abelian gauging, we start from the following mirror pairs.
\begin{equation}
\begin{split}
&\textbf{A-model}:\;\;(1,0)(2,0)(3,0)(4,2)(3,0)(2,0)(1,0) \\
&\textbf{B-model}:\;\;(4,8)
\end{split}
\label{eq:E7hatLin}
\end{equation}
We perform the gauging trick on the global $U(2)$ symmetry of the middle node of the left quiver into $U(1)$ global and $U(1)$ gauge. From the perspective of the parameter space, this amounts to fixing one of the momenta, which on the mirror side results in taking out the trace part of $U(4)$. 

Mirror symmetry dictates that the partition functions of the two linear quivers are related in the following manner up to an overall phase.
\begin{equation}
\begin{split}
&\CZ_A(m_a, t_{\alpha})= \CZ_B(M_{\alpha},\widetilde{t}_a)\\
&m_a =\widetilde{t}_a, \;\; a=1,2.\\
&t_{\alpha}=M_{\alpha},\;\; \alpha=1,2,\ldots,8
\end{split}
\end{equation}
Now consider gauging a single $U(1)$ of the $U(2)$ global symmetry in the A-model linear quiver to obtain the correct framed $\hat{E}_7$ quiver. The partition function of such a theory is
\begin{equation}
\begin{split}
\widetilde{\CZ}_A(\zeta_1, m_2, t_{\alpha})=&\int dm_1 e^{2\pi i \zeta_1 m_1} {\CZ}_A(m_a, t_{\alpha}) \\
=&\int dm_1 e^{2\pi i \zeta_1 m_1} {\CZ}_B( t_{\alpha},m_a)
\end{split}
\end{equation}
where the second equality is a direct consequence of mirror symmetry. Therefore, we have
\begin{equation}
\begin{split}
\widetilde{\CZ}_A(\zeta_1, m_2, t_{\alpha})=&\int dm_1 \frac{d^4s}{3!} \frac{e^{2\pi i \zeta_1 m_1}\prod^4_{i=1} e^{2\pi i s^i (m_1-m_2)} \prod_{i<j} \sinh^2{\pi(s^i-s^j)}}{\prod^4_{i=1}\prod^8_{\alpha=1} \cosh{\pi(s^i+t_{\alpha})}}\\
=&\int \frac{d^4s}{3!}\frac{\delta(s^1+s^2+s^3+s^4) \prod_{i<j} \sinh^2{\pi(s^i-s^j)}}{\prod^3_{i=1}\prod^6_{\alpha=1} \cosh{\pi(s^i+t_{\alpha}-\frac{\zeta_1}{4})}}\\
=&\widetilde{\CZ}_B( M_{\alpha})
\end{split}
\end{equation}
The theory dual to the $\hat{E}_7$ quiver with a single fundamental hyper in the middle node can now be read off from the partition function above - a $SU(4)$ with 8 flavors. The mirror map relates the masses of the fundamental hypers of $SU(4)$ with the FI parameters of the framed $\hat{E}_7$ quiver.
\begin{equation}
\begin{split}
M_{\alpha} =t_{\alpha} -\frac{\zeta_1}{4},\;\; \alpha=1,2,\ldots,8
\end{split}
\end{equation}
As expected, the 8 independent mass parameters of the B-model match with the number of independent parameters of the A-model - 8 parameters $\{t_{\alpha}\}$ with one constraint and $\zeta_1$.

Another way to realize the mirror pairs in \figref{fig:E7Aff} is as follows. The quiver on the left can be obtained from gluing $T_{(1,1,1,1)}(U(4))$, $T_{(1,1,1,1)}(U(4))$, $T_{(3,1)}(U(4))$ and $T_{(3,1)}(U(4))$ together via the $U(4)$ group, and the quiver on the right can be realized as as the $6d$ $(2,0)$ theory compactifying on a circle times a Riemann sphere with punctures $(1,1,1,1)$, $(1,1,1,1)$, $(3,1)$ and $(3,1)$ \cite{Benini:2010uu}. Indeed, according to \cite{Chacaltana:2010ks}, such a mirror theory is the $SU(4)$ gauge theory with $8$ flavors.

%

\subsection{$\hat{E}_8$ Theory from Linear Quiver}
\begin{figure}[!h]
\begin{center}
\includegraphics[scale=0.75]{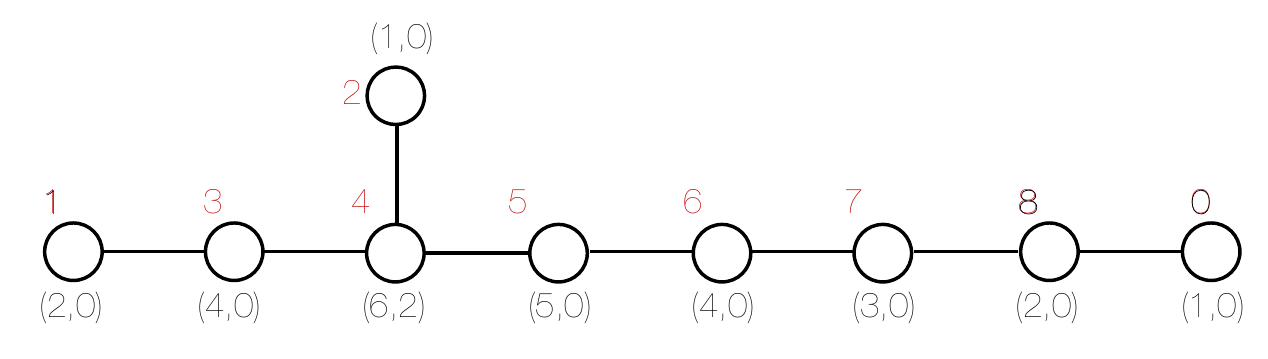} \qquad \includegraphics[scale=0.25]{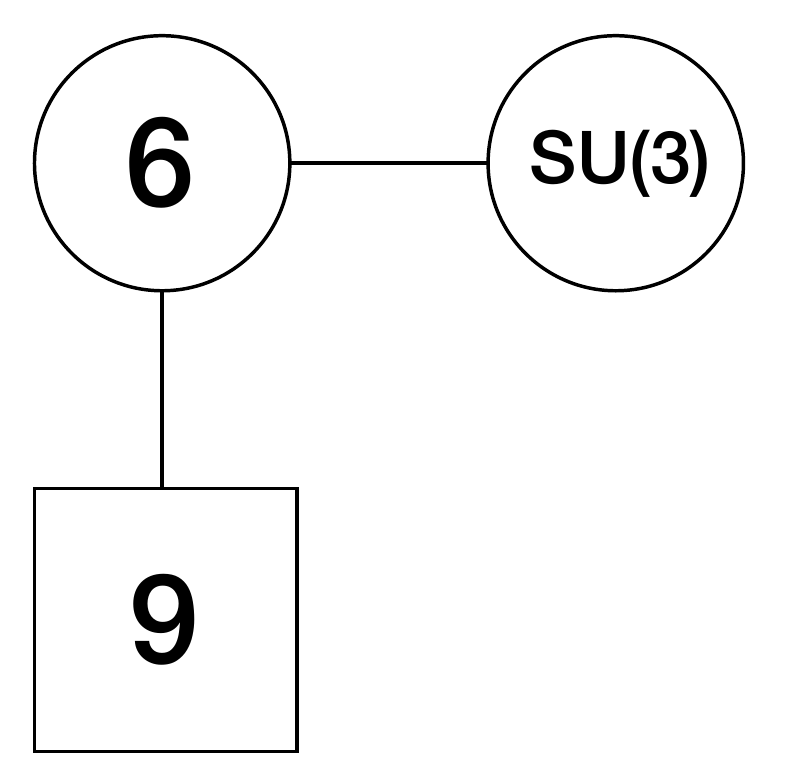}
\caption{$\hat{E}_8$ quiver with one hypermultiplet on the bifurcating node and its mirror.
}
\label{fig:E8Aff}
\end{center}
\end{figure}

\begin{figure}[!h]
\begin{center}
\includegraphics[scale=0.9]{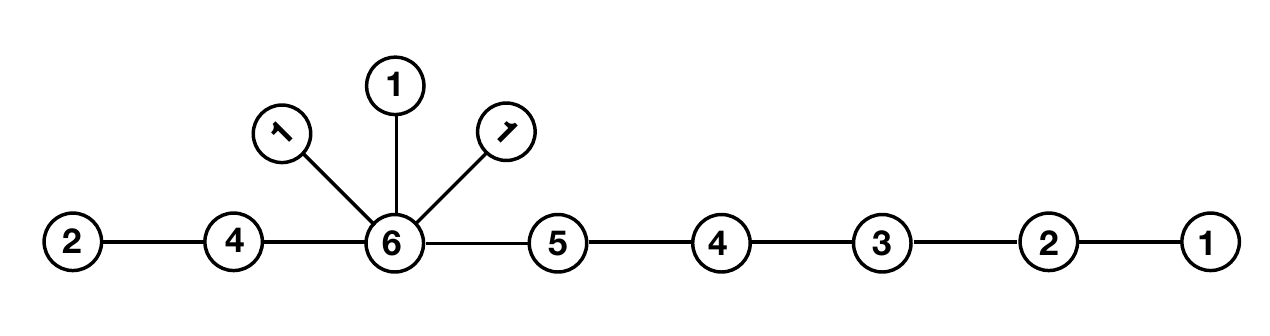}
\caption{The result of complete Abelian gauging on the $U(6)$ node. This quiver is mirror to another quiver theory: $(SU(6),9)(SU(3),0)$.}
\label{fig:E8Aff3}
\end{center}
\end{figure}

Similarly we can obtain extended $E_8$ graphs by employing Abelian gauging on the following mirror pair of linear quivers
\begin{equation}
(2,0)(4,0)(6,3)(5,0)(4,0)(3,0)(2,0)(1,0)  \qquad\qquad  (6,9)(3,0)\,.
\end{equation}
By gauging a single $U(1)$ factor on the $(6,3)$ node of the left quiver above we derive the new mirror pair, see \figref{fig:E8Aff}.
Gauging out another $U(1)$ on the bifurcating node of the left quiver in \figref{fig:E8Aff} will transform the mirror dual to $(SU(6),9)(SU(3),0)$. Finally, gauging out the remaining $U(1)$ global symmetry on the same node does not change the mirror, 
but the A-model quiver turns into the one depicted in \figref{fig:E8Aff3}. Recall that the overall $U(1)$ gauge factor decouples.

\begin{figure}[H]
\begin{center}
\includegraphics[scale=0.4]{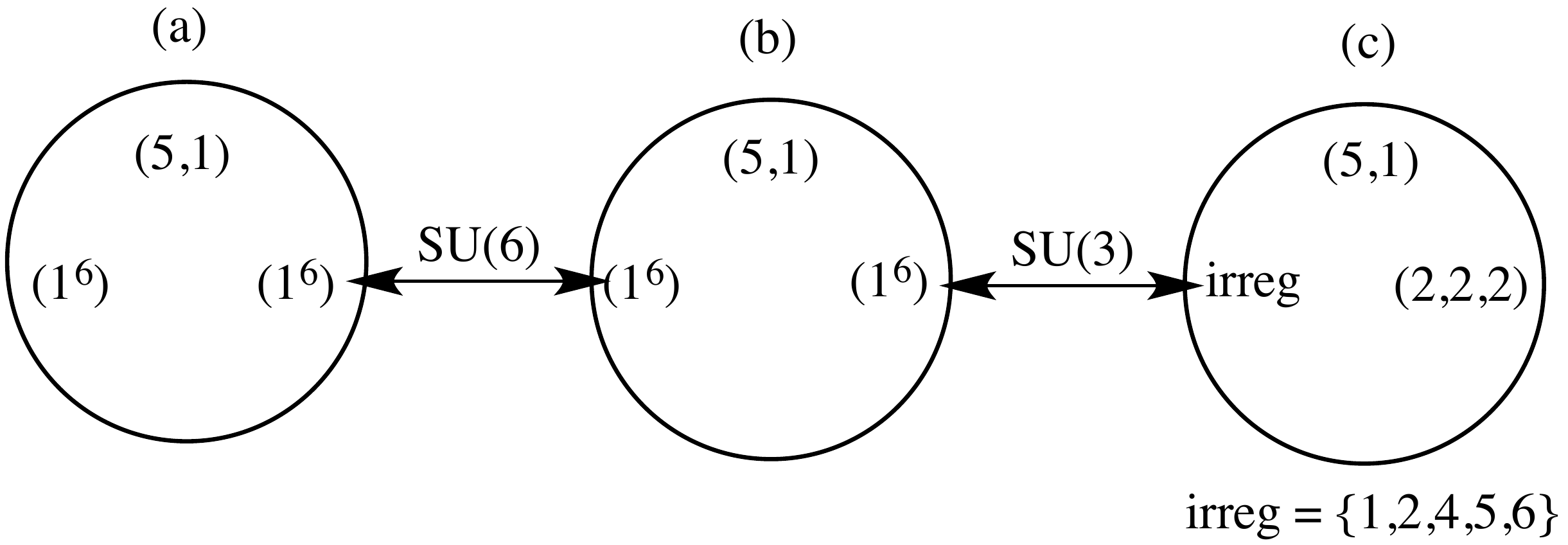}
\caption{The puncture decomposition of the mirror theory of \figref{fig:E8Aff3}.  The sequence in the round brackets corresponds to the Young diagram of the puncture, whereas the sequence in the curly bracket corresponds to the pole structure used in \cite{Chacaltana:2010ks}. The maximal puncture $(1^6)$ has the pole structure $\{1,2,3,4,5 \}$, the minimal puncture $(5,1)$ has the pole structure $\{1,1,1,1,1 \}$, and the puncture $(2,2,2)$ has the pole structure $\{ 1,2,2,3,4 \}$. The pole structure of the irregular puncture ``irreg''   is indicated in the figure.}
\label{fig:E8mir}
\end{center}
\end{figure}

It is instructive at this point to consider the six dimensional realization of the mirror theory of \figref{fig:E8Aff3}.
The quiver in \figref{fig:E8Aff3} can be constructed by gluing $T_{(2^3)}(U(6))$, 3 copies of $T_{(1^6)}(U(6))$, and  $T_{(5,1)}(U(6))$ together via the $U(6)$ group and modding out by an overall $U(1)$.  According to \cite{Benini:2010uu}, the mirror theory can be realised from the $6d$ $(2,0)$ theories compactified on a Riemann sphere with the following punctures: $(1^6)$, 3 copies of $(5,1)$ and $(2,2,2)$.  We can decompose the Riemann sphere as in \figref{fig:E8mir}.

Let us follow the prescription in \cite{Chacaltana:2010ks}.  The gauge group associated with the cylinder connecting two maximal punctures give rise to the gauge group $SU(6)$, whereas that associated with the cylinder connecting the maximal puncture and the irregular puncture has rank $2$.  There are two possibilities for the latter; it is either $SU(3)$ or $Sp(2)$.  In order to determine this, we need to compute the number of hypermultiplets associated with each fixture using Eq. (10) of \cite{Chacaltana:2010ks}: fixtures $(a)$ and $(b)$ each contains 36 hypermultiplets and fixture $(c)$ contains zero hypermultiplet.  Hence we conclude that the gauge group associated with the cylinder connecting $(b)$ and $(c)$ is $SU(3)$ with the following matter content:
\begin{table}[H]
\begin{center}
\begin{tabular}{|c|c|c|c|}
\hline
Fixture & \# hypers & $SU(6)$ & $SU(3)$ \\
\hline
$(a)$ & 6 & {\bf 6} & {\bf 1} \\
\hline
$(b)$ & 1 & {\bf 6} & {\bf 3} \\
~ & 3 &{\bf 6} & {\bf 1} \\
\hline
$(c)$ & - & - & - \\
\hline
\end{tabular}
\end{center}
\caption{Matter content of the configuration in \figref{fig:E8mir}.}
\label{default}
\end{table}%
The quiver diagram associated with this construction is therefore
\bea
\begin{tikzpicture}[font=\scriptsize]
\begin{scope}[auto,%
  every node/.style={draw, minimum size=1cm}, node distance=0.6cm];
\node[circle] (SU6) at (0, 0) {$SU(6)$};
\node[circle, right=of SU6] (SU3) {$SU(3)$};
\node[rectangle, below=of SU6] (SU3f) {$3$};
\node[rectangle, left=of SU6] (SU6f) {$6$};
\end{scope}
\draw (SU6f) -- (SU6)
(SU6) -- (SU3)
(SU6) -- (SU3f);
\end{tikzpicture}
\eea
Equivalently, this is
\bea
\begin{tikzpicture}[font=\scriptsize]
\begin{scope}[auto,%
  every node/.style={draw, minimum size=1cm}, node distance=0.6cm];
\node[rectangle] (SU9) at (0, 0) {$9$};
\node[circle, right=of SU9] (SU6b) {$SU(6)$};
\node[circle, right=of SU6b] (SU3b) {$SU(3)$};
\end{scope}
\draw (SU9) -- (SU6b)
(SU6b) -- (SU3b);
\end{tikzpicture}
\eea
as obtained using the Abelian gauging procedure.

\section{Non-Abelian Gauging: Mirrors of $Sp(N_c)$ Theories}\label{Sec:Spmirrors}
In this final section we discuss the construction of mirror duals to $Sp(N_c)$ 3d theories with $N_f$ flavors by studying parameter spaces of vacua and computing partition functions on $S^3$ for these theories.  From the discussion of \cite{Hanany:1999sj} and \cite{Feng:2000eq} we know that brane construction of mirror duals may involve O5-planes or O3-planes. In this paper, we focus on the mirror duals whose brane construction only involves O5 planes; we will refer to those as the ``O5 mirrors''. 

\subsection{Brane Construction and S-duality}
We have already studied $Sp(1)\simeq SU(2)$ theories earlier in the paper (see e.g. \secref{star-quiver}), so let us immediately proceed to more complicated examples. We will soon see that in order to understand higher rank $Sp$ theories starting from linear quivers one has to perform the \textit{non-Abelian} gauging as opposed to the Abelian gauging which we have used thus far. In this section we shall elaborate in great details on $Sp(2)$ gauge theory with six flavors. 

\begin{figure}[H]
\begin{center}
\includegraphics[scale=0.5]{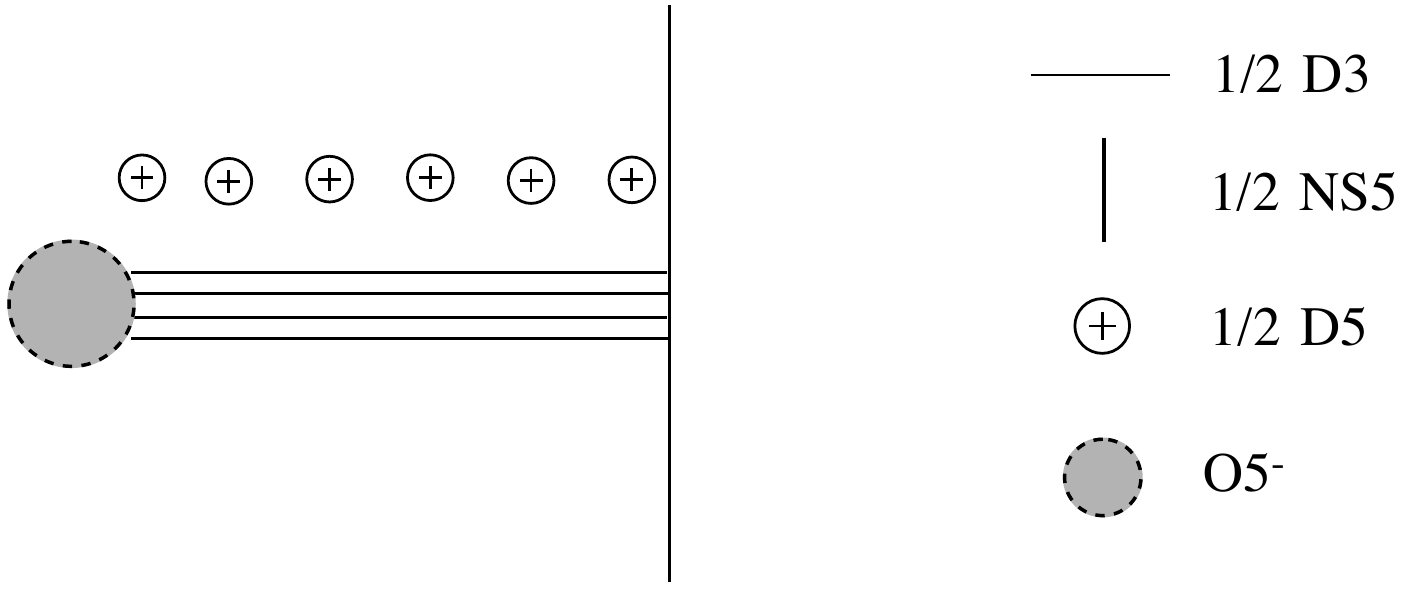}
\caption{The brane configuration of $Sp(2)$ with $6$ flavors involving an O5-plane.}
\label{fig:BraneSp26flvO5}
\end{center}
\end{figure}

The brane configuration of the $Sp(2)$ theory with $6$ flavors involving an O5-plane is presented in \figref{fig:BraneSp26flvO5} (see \cite{Hanany:1999sj} for details). 

\begin{figure}[H]
\begin{center}
\includegraphics[scale=0.5]{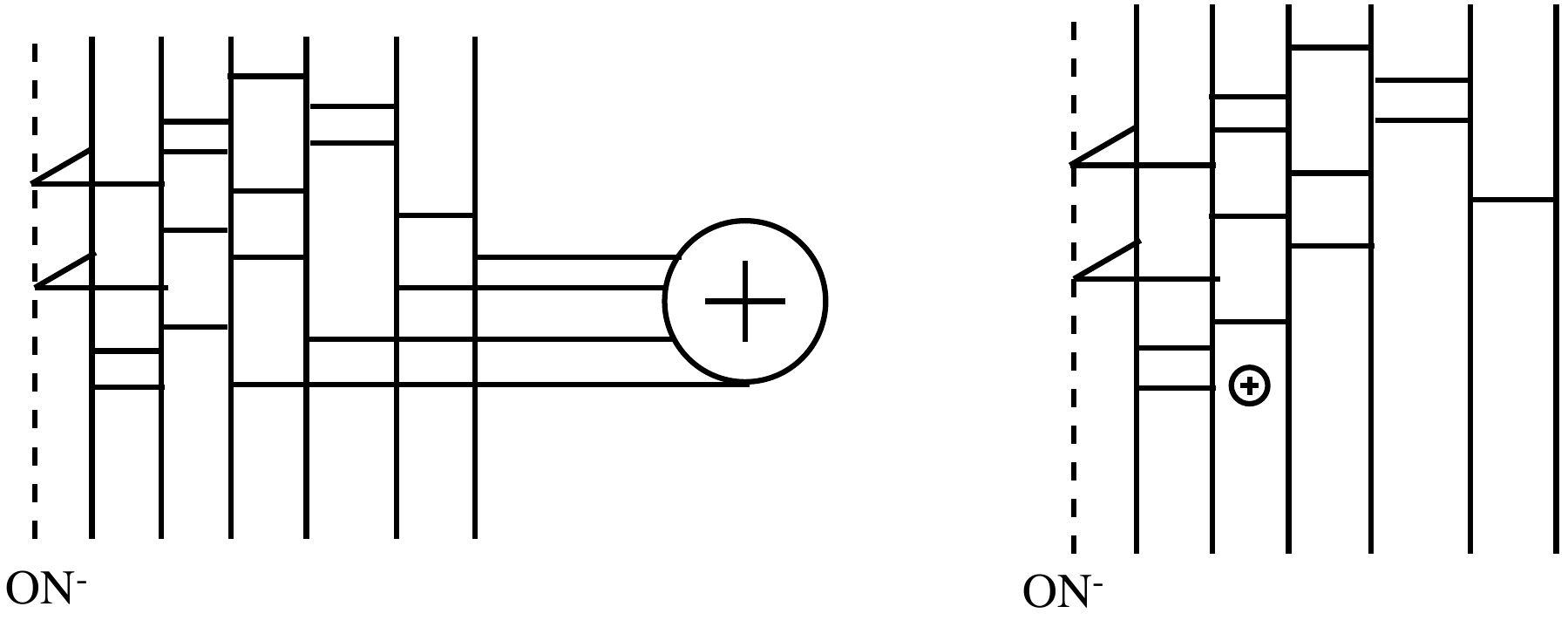}
\caption{The brane configurations of the mirror theory of $Sp(2)$ with $6$ flavors involving an O5-plane.  The notation is the same as \figref{fig:BraneSp26flvO5}, with the dashed vertical line being an $ON^-$ plane. On the left the S-duality is directly applied to \figref{fig:BraneSp26flvO5}, with the D3-branes reconnected according to the $s$-rule such that the number of D3-branes at each interval is preserved. On the right the D5-brane is moved inside so no D3 branes end on it; in this configuration the quiver data can be read off from this diagram.}
\label{fig:MirrorO5Sp26flvO5}
\end{center}
\end{figure}

Let us apply the S-duality to the brane construction from \figref{fig:BraneSp26flvO5}. Upon the S-duality the D5-branes become NS5-branes and vice versa. The $O5^-$ plane becomes the $ON^-$ plane. After all, the S-dual brane configurations is given in \figref{fig:MirrorO5Sp26flvO5}.

\begin{figure}[H]
\begin{center}
\includegraphics[scale=0.5]{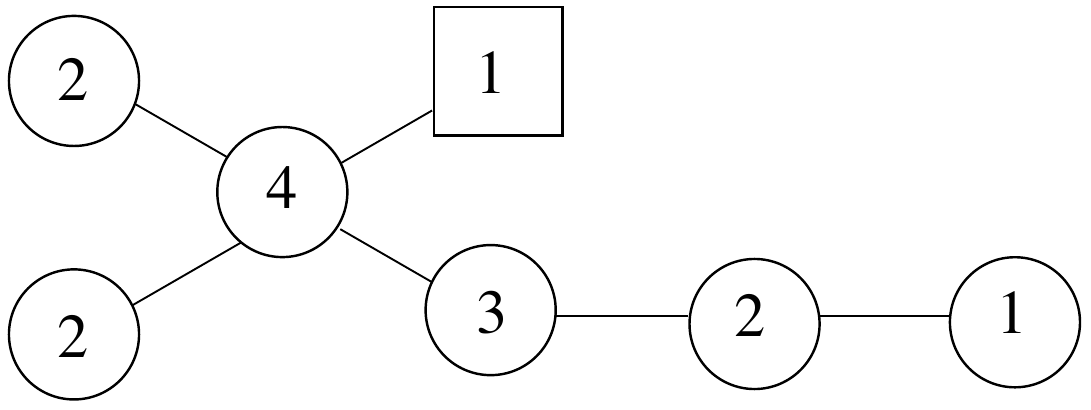}
\caption{The quiver diagram of a mirror theory of $Sp(2)$ with $6$ flavors. The Coulomb branch of this theory is $14$ quanternionic dimensional. The Higgs branch of this theory is 2-quanternionic dimensional.}
\label{fig:QuiverMirrorO5Sp26flv}
\end{center}
\end{figure}

The quiver of the mirror theory for our problem can be read off directly from the right diagram of \figref{fig:MirrorO5Sp26flvO5}; this is depicted in \figref{fig:QuiverMirrorO5Sp26flv}.

\subsection{Parameter Space Description}
\begin{figure}[H]
\begin{center}
\includegraphics[scale=0.34]{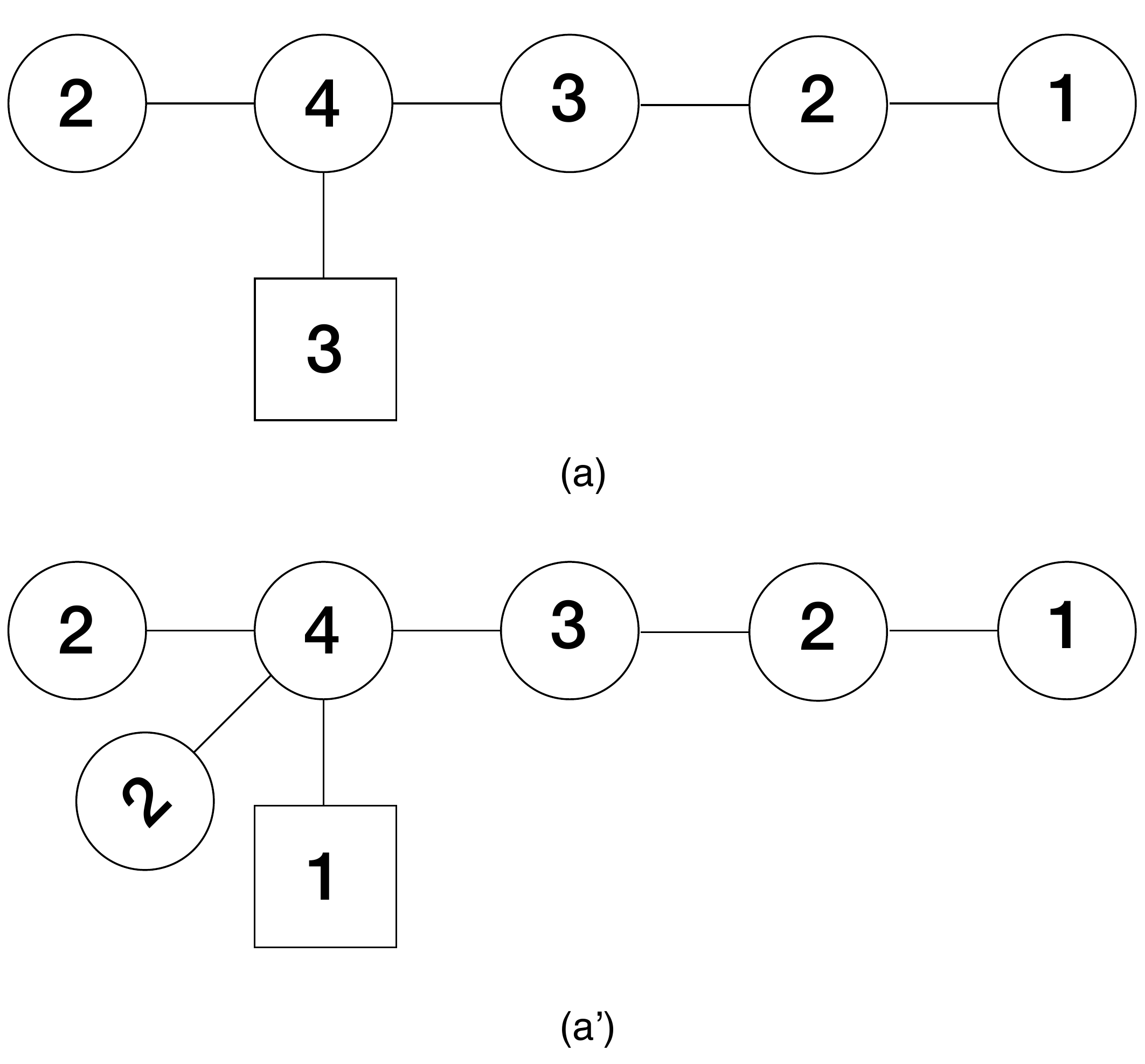}\qquad\qquad\includegraphics[scale=0.34]{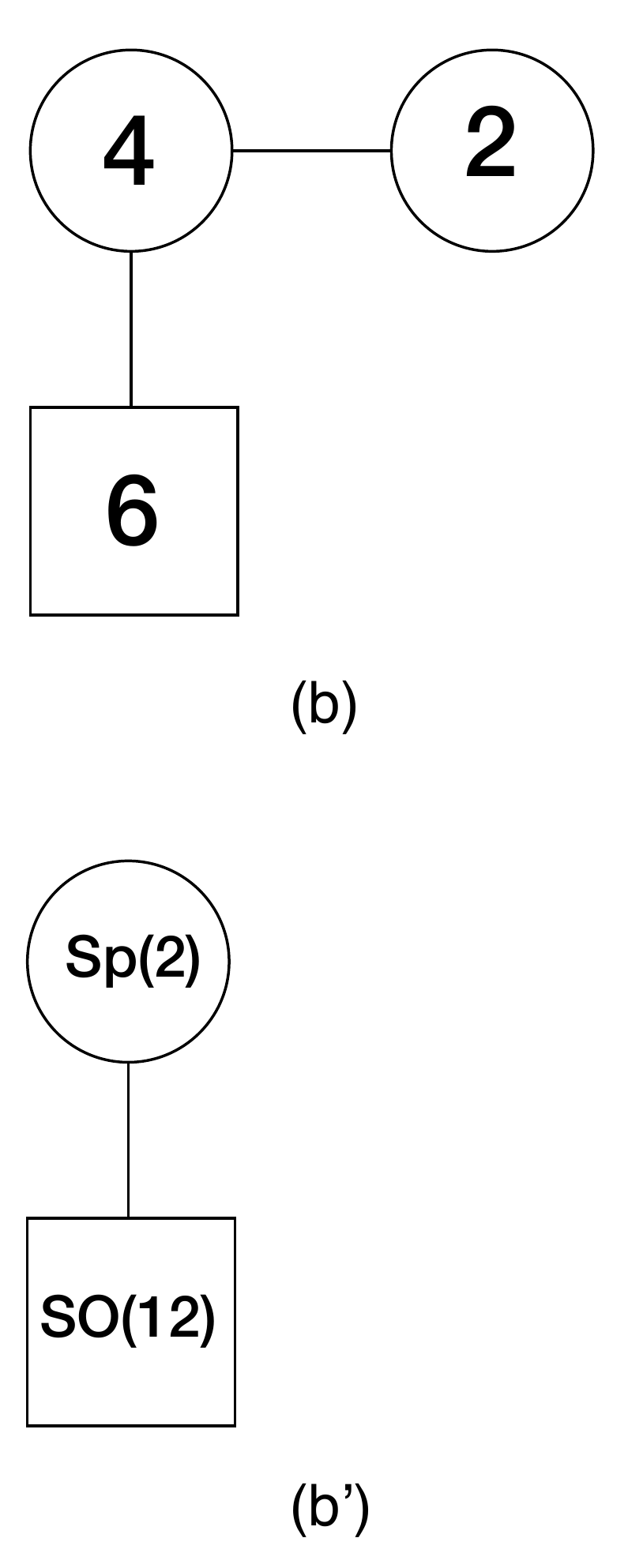}
\caption{Mirror pairs before ($a$ and $b$) and after ($a'$ and $b'$) the gauging. Dimensions of Coulomb and Higgs branches for the A-models (left) are $12$ and $6$ for the top quiver and $14$ and $2$ for the bottom quiver.}
\label{fig:Quiver42etc}
\end{center}
\end{figure}

Let us now try to derive the mirror quiver for $Sp(2)$ theory with six flavors depicted in \figref{fig:QuiverMirrorO5Sp26flv} from parameter spaces of linear quivers.
These linear quivers are the following (see top row in \figref{fig:Quiver42etc})
\begin{align}
&\textbf{A-model:}\quad (2,0)(4,3)(3,0)(2,0)(1,0)\notag\\
&\textbf{B-model:}\quad (4,6)(2,0)\,
\label{eq:ABSp}
\end{align}
We then gauge the $U(2)$ subgroup of the $U(3)$ global symmetry on the second node of the A-quiver. This procedure leaves behind $U(1)$ flavor symmetry. Now we need to understand the consequences of gauging on the mirror side.
We see that the dimension of the Higgs branch of the bottom-left quiver in \figref{fig:Quiver42etc} has decreased by $2^2=4$, therefore we expect the same to happen for the Coulomb branch of the mirror quiver. Also adding a $U(2)$ gauge node on the A-side increases its Coulomb branch dimension by two, therefore the Higgs branch of the mirror has to be fourteen-dimensional. Clearly the $Sp(2)$ theory with six flavors or with $SO(12)$ global symmetry (bottom-right of \figref{fig:Quiver42etc}) is a good candidate since the dimensions of branches match perfectly. However, matching of the dimensions alone is simply not enough to claim victory and a robust derivation of our result is due. 

Note that in \figref{fig:Quiver42etc} quiver ($a'$) can be constructed by gluing $T_{(2,2)}(U(4))$, $T_{(2,2)}(U(4))$, $T_{(3,1)}(U(4))$ and $T_{(1,1,1,1)}(U(4))$ via the $U(4)$ group and modding out by the overall $U(1)$; its mirror \cite{Benini:2010uu}, quiver ($b'$), can be realized as the $6d$ $(2,0)$ theory compactified on $S^1$ times a Riemann surface with punctures $(2,2)$, $(2,2)$, $(3,1)$ and $(1,1,1,1)$.  Note also that this particular theory belongs to the classification of \cite{Chacaltana:2010ks}\footnote{See top diagram on page 22 of of \cite{Chacaltana:2010ks}}. However, \cite{Chacaltana:2010ks} classifies theories only up to rank four, whereas here we are interested in constructing mirrors for any $N_c$ and $N_f$.

Let us begin with the Bethe equations for the original linear quivers on top of \figref{fig:Quiver42etc}. Vacua equations of the $U(4)$ node of the A-model quiver in \eqref{eq:ABSp} read
\begin{align}
\label{eq:AmodelU4eq}
\frac{\tau_3}{\tau_2} \prod_{a=1}^3\frac{\eta\sigma^{(2)}_i-\mu^{(2)}_a}{\eta\mu^{(2)}_a-\sigma^{(2)}_i}\cdot\prod_{a=1}^2\frac{\eta\sigma^{(2)}_i-\sigma^{(1)}_a}{\eta\sigma^{(1)}_a-\sigma^{(2)}_i} \cdot \prod_{j\neq i}^4\frac{\eta\sigma^{(2)}_i-\eta^{-1}\sigma^{(2)}_j}{\eta\sigma^{(2)}_j-\eta^{-1}\sigma^{(2)}_i}\cdot\prod_{a=1}^3\frac{\eta\sigma^{(2)}_i-\sigma^{(3)}_a}{\eta\sigma^{(3)}_a-\sigma^{(2)}_i}&=1\,,
\end{align}
together with the corresponding momenta
\begin{equation}
p_\mu^{(2)\,a}=\tau_1\tau_2\prod_{j=1}^4\frac{\eta\mu_a-\sigma^{(2)}_j}{\eta\sigma^{(2)}_j-\mu_a}\,,\quad a=1,2,3\,.
\label{eq:PmuSp2}
\end{equation}
After the gauging an extra node with $U(2)$ gauge group is added to the quiver, let's call it $0$th node. Vacua equations for this node read
\begin{equation}
\prod_{j=1}^4\frac{\eta\sigma^{(0)}_a-\sigma^{(2)}_j}{\eta\sigma^{(2)}_j-\sigma^{(0)}_a}\cdot \prod_{b\neq a}\frac{\eta\sigma^{(0)}_a-\eta^{-1}\sigma^{(0)}_b}{\eta\sigma^{(0)}_b-\eta^{-1}\sigma^{(0)}_a}=1\,,\quad a=1,2\,.
\end{equation}
If we multiply the above two equations for $a=1$ and $a=2$ we immediately arrive to the following constraint $p_\sigma^{(0)\,1} p_\sigma^{(0)\,2}=1$ which can be also written as 
\begin{equation}
p_\mu^{(2)\,1} p_\mu^{(2)\,2}=1\,,
\label{eq:pmuconstrsp2}
\end{equation}
if we relabel $\sigma^{(0)}$s with $\mu$s.

Meanwhile, on the mirror side we get 
\begin{align}
\label{eq:BmodelU4eq}
\frac{\tau^\vee_2}{\tau^\vee_1} \prod_{a=1}^6\frac{\eta^{-1}\sigma^{\vee\,(1)}_i-\mu^{\vee\,(1)}_a}{\eta^{-1}\mu^{\vee\,(1)}_a-\sigma^{\vee\,(1)}_i}\cdot \prod_{j\neq i}^4\frac{\eta\sigma^{\vee\,(1)}_i-\eta^{-1}\sigma^{\vee\,(1)}_j}{\eta\sigma^{\vee\,(1)}_j-\eta^{-1}\sigma^{\vee\,(1)}_i}\cdot\prod_{a=1}^2\frac{\eta^{-1}\sigma^{\vee\,(1)}_i-\sigma^{\vee\,(2)}_a}{\eta^{-1}\sigma^{\vee\,(2)}_a-\sigma^{\vee\,(1)}_i} &=1\,,\notag\\
\frac{\tau^\vee_3}{\tau^\vee_2} \prod_{a=1}^4\frac{\eta^{-1}\sigma^{\vee\,(2)}_i-\sigma^{\vee\,(1)}_a}{\eta^{-1}\sigma^{\vee\,(1)}_a-\sigma^{\vee\,(2)}_i} \cdot \prod_{j\neq i}^2\frac{\eta^{-1}\sigma^{\vee\,(2)}_i-\eta\sigma^{\vee\,(2)}_j}{\eta^{-1}\sigma^{\vee\,(2)}_j-\eta\sigma^{\vee\,(2)}_i}&=1\,.
\end{align}
The mirror analogue of \eqref{eq:pmuconstrsp2} is 
\begin{equation}
p_\tau^{\vee\,1} p_\tau^{\vee\, 2}=1\,,
\label{eq:ptauBconstr}
\end{equation}
or
\begin{equation}
\tau^\vee_1(\tau^\vee_2)^2\frac{1}{\sigma^{\vee\,(2)}_1 \sigma^{\vee\,(2)}_2}=1\,.
\end{equation}
The latter condition, up to a constant, (which we shall fix soon) provides an embedding of $Sp(1)\subset U(2)$ for the second node of the B-side quiver. 

Note that there is an ambiguity in the choice of \eqref{eq:ptauBconstr} which is due to the breaking of the $U(3)$ flavor symmetry on the A side, in other words, one needs to chose which masses (or FI terms on the mirror side) to pick. For a different choice of masses, say $\mu_2$ and $\mu_3$ \eqref{eq:ptauBconstr} would imply
\begin{equation}
\tau^\vee_1\tau^\vee_2\frac{1}{\sigma^{\vee\,(1)}_1 \sigma^{\vee\,(1)}_2 \sigma^{\vee\,(1)}_3 \sigma^{\vee\,(1)}_4}=1\,
\end{equation}
instead. In order to provide the remaining constraint to ensure the projection of $U(4)$ onto $Sp(2)$ we need to solve Bethe equations and express the solution in terms of momenta \eqref{eq:PmuSp2}. Thus we put 
\begin{equation}
\sigma^{\vee\,(2)}_1=\eta \sigma^{\vee\,(1)}_3=\frac{\eta}{\sigma^{\vee\,(1)}_2}\,\quad \sigma^{\vee\,(2)}_2=\eta \sigma^{\vee\,(1)}_4=\frac{\eta}{\sigma^{\vee\,(1)}_1}\,,
\end{equation}
and observe that the first equation of \eqref{eq:BmodelU4eq} telescopes down to the Bethe equation for $Sp(2)$ theory with six flavors.

In general, if one splits a $U(2N+1)$ flavor symmetry on the A side into two $U(N)$ gauge groups and $U(1)$ global symmetry, one imposes $N$ constraints in total on momenta $p_\mu^a$. Those constraints, translated into the mirror side provide a canonical embedding of $Sp(N)$ gauge group into $U(2N)$ group which appeared in the original mirror construction. 

\subsection{Partition Function Description}
Let us now derive the mirror of $Sp(2)$ with $6$ flavors using the technique of non-Abelian gauging using, as before, the $S^3$ partition function as a tool. We shall see that working with partition functions will turn out to be a very powerful tool and can be used in gauging of arbitrary quiver theories.

Consider again the pair of mirror quivers \eqref{eq:ABSp}. Their partition functions read as follows
\begin{equation}
\begin{split}
\CZ_A\left(m_a, t_{\alpha}\right)&=\int  \frac{d^{2}s_{1}}{2!} \frac{d^{4}s_{2}}{4!} \frac{\prod_{i<j}\sinh^2{\pi(s^i_1-s^j_1)}\prod_{p<l}\sinh^2{\pi(s^p_2-s^l_2)}}{\prod_{i,p}\cosh{\pi(s^i_1-{s}^p_{2})}\prod_{p}\prod^3_{a=1}\cosh{\pi(s^p_2-m_a)}}\\
&\times \prod^2_{i=1} e^{2\pi i s^i_1(t_1-t_2)} \prod^4_{p=1} e^{2\pi i s^p_2(t_2-t_3)} \CZ_{T(U(4))}\left(s^p_2; t_3,t_4,t_5,t_6\right)\,,
\end{split}
\end{equation}
 \begin{equation}
\begin{split}
\CZ_B\left(M_{\alpha}, \widetilde{t}_{a}\right)&=\int  \frac{d^{2}s_{1}}{2!} \frac{d^{4}s_{2}}{4!} \frac{\prod_{i<j}\sinh^2{\pi(s^i_1-s^j_1)}\prod_{p<l}\sinh^2{\pi(s^p_2-s^l_2)}}{\prod_{i,p}\cosh{\pi(s^i_1-{s}^p_{2})}\prod_{p}\prod^6_{\alpha=1}\cosh{\pi(s^p_2-M_{\alpha})}}\\
&\times \prod^2_{i=1} e^{2\pi i s^i_1(\widetilde{t}_1-\widetilde{t}_2)} \prod^4_{p=1} e^{2\pi i s^p_2(\widetilde{t}_2-\widetilde{t}_3)}\,.
\end{split}
\end{equation}
The mirror symmetry implies that $\CZ_A(m_a;{t}_{\alpha})=\CZ_B (M_{\alpha}; \widetilde{t}_a)$ up to some overall phase provided the parameters are related as follows:
\begin{equation}
\begin{split}
&M_{\alpha} = {t}_{\alpha}\\
&\widetilde{t}_a  = m_a\,.
\end{split}
\end{equation}
Now, we gauge a $U(2)$ subgroup of the $U(3)$ flavor symmetry of the A-model, which gives the mirror theory of $Sp(2)$ with $6$ flavors. The partition function of this theory is 
\begin{equation}
\begin{split}
\widetilde{\CZ}_A(\zeta, m_3;{t}_{\alpha})=&\int dm_1 dm_2 e^{2\pi i (m_1+m_2) \zeta}  \sinh^2{\pi(m_1-m_2)} \CZ_A (m_a;{t}_{\alpha})\\
=&\int dm_1 dm_2  e^{2\pi i (m_1+m_2) \zeta}  \sinh^2{\pi(m_1-m_2)} \CZ_B (t_{\alpha}; m_a)\,,
\end{split}
\end{equation}
where the second equality follows from the mirror symmetry of the linear quivers. From the second equality, completing the integration over $m_1$ and $m_2$ we have
\begin{equation}
\begin{split}
&\widetilde{\CZ}_A(\zeta, m_3;{t}_{\alpha})= \int  \frac{d^{2}s_{1}}{2!} \frac{d^{4}s_{2}}{4!} \frac{\prod^4_{p=1} e^{-2\pi m_3 s^p_2}\prod_{i<j}\sinh^2{\pi(s^i_1-s^j_1)}\prod_{p<l}\sinh^2{\pi(s^p_2-s^l_2)}}{\prod_{i,p}\cosh{\pi(s^i_1-{s}^p_{2})}\prod_{p}\prod^6_{\alpha=1}\cosh{\pi(s^p_2-t_{\alpha})}}\\
&\times \Big(-2\delta(\zeta+\sum_i s^i_1)\delta(\zeta-\sum_i s^i_1+\sum_p s^p_2)+\delta(\zeta+\sum_i s^i_1+i)\delta(\zeta-\sum_i s^i_1+\sum_p s^p_2-i)\\
&+\delta(\zeta+\sum_i s^i_1-i)\delta(\zeta-\sum_i s^i_1+\sum_p s^p_2+i)\Big) \equiv T_1+T_2+T_3\,.
\end{split}
\end{equation}
It is useful to divide up the partition function into three parts $T_1,T_2,T_3$ as follows
\begin{equation}
\begin{split}
&T_1=-2 \int  \frac{d^{2}s_{1}}{2!} \frac{d^{4}s_{2}}{4!} f(s^i_1,s^p_2) g(s^p_2)\delta(\zeta+\sum_i s^i_1)\delta(\zeta-\sum_i s^i_1+\sum_p s^p_2)\,,\\
&T_2=\int  \frac{d^{2}s_{1}}{2!} \frac{d^{4}s_{2}}{4!} f(s^i_1,s^p_2) g(s^p_2)\delta(\zeta+\sum_i s^i_1+i)\delta(\zeta-\sum_i s^i_1+\sum_p s^p_2-i)\,,\\
&T_3=\int  \frac{d^{2}s_{1}}{2!} \frac{d^{4}s_{2}}{4!} f(s^i_1,s^p_2) g(s^p_2)\delta(\zeta+\sum_i s^i_1-i)\delta(\zeta-\sum_i s^i_1+\sum_p s^p_2+i)\,,\\
&f(s^i_1,s^p_2)= \frac{\prod_{i<j}\sinh^2{\pi(s^i_1-s^j_1)}}{\prod_{i,p}\cosh{\pi(s^i_1-{s}^p_{2})}}\,,\\
&g(s^p_2)=\frac{\prod^4_{p=1} e^{-2\pi m_3 s^p_2} \prod_{p<l}\sinh^2{\pi(s^p_2-s^l_2)}}{\prod_{p}\prod^6_{\alpha=1}\cosh{\pi(s^p_2-t_{\alpha})}}\,.
\end{split}
\end{equation}

Let us consider the term $T_2$ first. Note that the integrand of $T_2$ (like $T_1$ and $T_3$) has poles at $s^1_1=s^p_2 \pm (2k_p-1)\frac{i}{2}, \;s^2_1=s^p_2 \pm (2k_p-1)\frac{i}{2}$ with $k_p \in \mathbb{Z}_{+}$  - the residues of only one half of these poles contribute to the integral depending on whether one closes the contour in the upper half plane or the lower half plane.

In order to rewrite the partition function $\widetilde{\CZ}_A(\zeta, m_3;{t}_{\alpha})$ in a form where the dual gauge theory can be read off, one needs to remove the imaginary contributions in the delta functions which may be done, for example, by shifting the integration variable $s^1_1 \to s^1_1 -i $. But this amounts to shifting the contour of the $s^1_1$ integration and therefore the integral after and before the shift will differ by residues of poles which are included (or excluded) by this change of contour. Keeping this in mind, the matrix integral in $T_2$ may be written as
\begin{equation}
\begin{split}
T_2 =-\frac{T_1}{2} + \mathcal{C}_1\int ds^1_2 \frac{d^{4}s_{2}}{4!}\frac{\cosh{\pi(s^2_1-s^1_2)}\delta(2\zeta+2s^2_1+2s^1_2+i)\delta(2\sum_{p\neq 1} s^p_2+2\zeta-2s^2_1-i)}{\prod_{p \neq 1}\sinh{\pi(s^1_2-s^p_2)}\cosh{\pi(s^2_1-s^p_2)}}g(s^p_2)\,,
\end{split}
\end{equation}
where $\mathcal{C}_1$ is a combinatorial and/or phase factor which can be ignored for our discussion. The above expression can be further groomed by shifting the integration variable $s^2_1 \to s^2_1 -i/2 $. Taking into account the residues of the poles that are affected by this change of contour, we get
\begin{equation}
\begin{split}
T_2 =& -\frac{T_1}{2} - \mathcal{C}_1\int ds^1_2 \frac{d^{4}s_{2}}{4!}\frac{\sinh{\pi(s^2_1-s^1_2)}\delta(2\zeta+2s^2_1+2s^1_2)\delta(2\sum_{p\neq 1} s^p_2+2\zeta-2s^2_1)}{\prod_{p \neq 1}\sinh{\pi(s^1_2-s^p_2)}\sinh{\pi(s^2_1-s^p_2)}}g(s^p_2)\\
&+ \frac{\mathcal{C}_2}{2}\int \frac{d^{4}s_{2}}{2^2 2!}\frac{\delta(s^1_2+s^2_2+\zeta)\delta(s^3_2+s^4_2+\zeta)}{\prod_{i=1,2}\sinh{\pi(s^i_2-s^{i+2}_2)}}g(s^p_2)\,,
\end{split}
\end{equation}
where $\mathcal{C}_2$ is a phase factor. Manipulating with $T_3$ in exactly the same way we obtain 
\begin{equation}
\begin{split}
T_3 =& -\frac{T_1}{2} + \mathcal{C}_1 \int ds^1_2 \frac{d^{4}s_{2}}{4!}\frac{\sinh{\pi(s^2_1-s^1_2)}\delta(2\zeta+2s^2_1+2s^1_2)\delta(2\sum_{p\neq 1} s^p_2+2\zeta-2s^2_1)}{\prod_{p \neq 1}\sinh{\pi(s^1_2-s^p_2)}\sinh{\pi(s^2_1-s^p_2)}}g(s^p_2)\\
&+ \frac{\mathcal{C}_2}{2}\int \frac{d^{4}s_{2}}{2^2 2!}\frac{\delta(s^1_2+s^2_2+\zeta)\delta(s^3_2+s^4_2+\zeta)}{\prod_{i=1,2}\sinh{\pi(s^i_2-s^{i+2}_2)}}g(s^p_2)\,.
\end{split}
\end{equation}
Summing up the contributions of $T_1,T_2$ and $T_3$ one obtains the following formula for $\widetilde{\CZ}_A(\zeta, m_3;{t}_{\alpha})$ (up to some phase factor)
\begin{equation}
\begin{split}
\widetilde{\CZ}_A(\zeta, m_3;{t}_{\alpha})=&\int \frac{d^{4}s_{2}}{2^2 2!}\frac{\delta(s^1_2+s^2_2+\zeta)\delta(s^3_2+s^4_2+\zeta)}{\prod_{i=1,2}\sinh{\pi(s^i_2-s^{i+2}_2)}}g(s^p_2)\\
=& \int \frac{d^{4}s_{2}}{2^2 2!}\frac{\delta(s^1_2+s^2_2+\zeta)\delta(s^3_2+s^4_2+\zeta)}{\prod_{i=1,2}\sinh{\pi(s^i_2-s^{i+2}_2)}}\frac{\prod^4_{p=1} e^{-2\pi m_3 s^p_2} \prod_{p<l}\sinh^2{\pi(s^p_2-s^l_2)}}{\prod_{p}\prod^6_{\alpha=1}\cosh{\pi(s^p_2-t_{\alpha})}}\\
=& \int \frac{d^{4}s_{2}}{2^2 2!}\frac{\sinh^2{\pi(s^1_2-s^3_2)}\sinh^2{\pi(s^1_2+s^3_2)} \sinh^2{\pi(2s^1_2)}\sinh^2{\pi(2s^3_2)}}{\prod_{p=1,3}\prod^6_{\alpha=1}\cosh{\pi(s^p_2-t_{\alpha}-\zeta/2)}\cosh{\pi(s^p_2+t_{\alpha}+\zeta/2)}}\,.
\end{split}
\end{equation}
This is precisely the partition function of a $Sp(2)$ gauge theory with $6$ fundamental hypers. The corresponding mirror map is given by
\begin{equation}
M_{\alpha}=t_{\alpha}+\zeta/2\,.
\end{equation}

\begin{figure}[H]
\begin{center}
\includegraphics[scale=0.8]{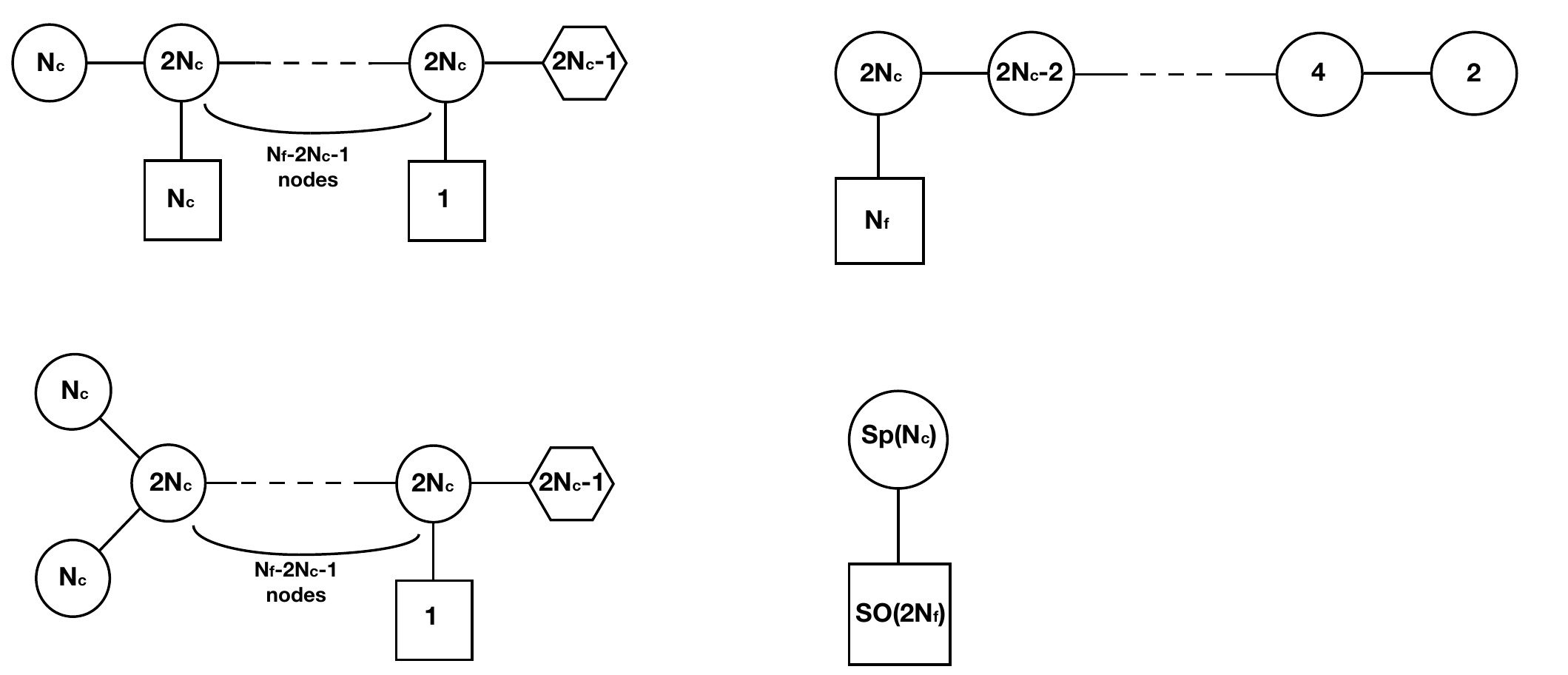}
\caption{Gauging $U(N_c)$ global symmetry of the top-left linear quiver produces framed D-shaped quiver on the bottom left. As a result on the mirror side the tail of top-right quiver collapses onto single $Sp(N_c)$ node (bottom-right). Hexagons on the right ends of the two left quivers denote $T[U(2N_c-1)]$ tails: $(2N_c-1)-(2N_c-2)-\dots-(2)-(1)$.}
\label{fig:QuiverSpGen}
\end{center}
\end{figure}

Our analysis can be easily extended to the generic case in \figref{fig:QuiverSpGen}. We refrain from repeating the partition function analysis for this generic case, however we present the result and support it by dimension counting. We start with the two linear quivers on top of the figure and gauge global $U(N_c)$ symmetry on the second node of the left quiver.
This procedure changes the dimensions of the Coulomb and Higgs branches of the left quiver by $+N_c$ and $-N_c^2$ respectively. 
On the mirror side we therefore should see the annihilation of $N_c^2$ parameters on its Coulomb branch. They disappear as a result of the collapse of the ``double tail'' $(2N_c-2)-(2N_c-4)-\dots-(4)-(2)$ of the top-right quiver in \figref{fig:QuiverSpGen}. Let us count how many Cartan generators this tail has. Indeed, the counting works properly since 
\begin{equation}
2+4+\dots+2N_c-2 = N_c^2-N_c\,,
\label{eq:DimensionsMatch}
\end{equation}
and the remaining $N_c$ Cartan elements are extracted from the projection of $U(2N_c)$ to $Sp(N_c)$.

\subsection{A Remark on non-Abelian Gauging}
At this point, let us quickly clarify an important issue regarding the program of non-Abelian gauging which we have demonstrated in the previous subsection for a special case. Suppose we start with some linear quiver which includes framed nodes. We take one of those framed nodes, assume it has labels $(N_i,M_i)$ and $M_i>1$, so the global symmetry is genuinely non-Abelian. The node may be connected to other nodes via bifundamental hypermultiplets, but their existence is irrelevant for the argument we are about to make. 

Now we gauge a non-Abelian subgroup of $U(M_i)$, which may also be the $U(M_i)$ itself. As we discussed in the end of the last computation around formula \eqref{eq:DimensionsMatch}, this procedure decreases the dimension of the Higgs branch by the dimension of that subgroup we have just gauged. On the mirror side the Coulomb branch will suffer the same loss in dimension. However, there is a potential obstacle for this to happen -- the Coulomb branch of the mirror quiver may be too small to sustain this deformation! 

Framed $E_n$ quivers, which we have discussed in \secref{Sec:Emirrors} provide us with perfect illustrations of this fact. Consider, e.g. the mirror pair which we used in the construction of  $\hat{E}_7$ quiver \eqref{eq:E7hatLin}. Only this time, instead of gauging the $U(1)$ subgroup of the middle node of the A-model $A_7$ quiver, we shall try to gauge the whole $U(2)$, which has dimension four, in order to get a fully balanced $\hat{E}_7$ quiver. However, the Coulomb branch of the mirror $U(4)$ theory is only four-dimensional! Therefore
we conclude that the mirror of the $\hat{E}_7$ cannot be presented as a Lagrangian quiver theory of any kind, which confirms the fact we know from compactifications of six dimensional $(2,0)$ theory, viz. all extended balanced $E_n$ quivers have \textit{non-Lagrangian} mirrors. 

\section*{Acknowledgements}
We are very much grateful to the Simons Center for Geometry and Physics at Stony Brook University and especially to Cumrun Vafa and Martin Ro$\check{\text{c}}$ek for organizing the 2013 Summer Workshop ``Defects'', where this project started. We are also thankful to Chulalongkorn University in Bangkok and the organizers of ``3rd Bangkok Workshop on High Energy Theory" for kind hospitality during the completion of the manuscript. AD would like to thank Jacques Distler for discussion and useful comments. AD would also like to thank the Theory Group at Imperial College London, where part of the work was done, for kind hospitality. PK would also like to thank the Theoretical Physics group at Imperial College in London for kind hospitality during his visit. In addition PK thanks Davide Gaiotto for many fruitful discussions. We would like to thank Stefano Cremonesi and Alberto Zaffaroni for a close collaboration and sharing many ideas.
AH and PK would like to thank the organizers of the workshop ``Quiver Varieties'' at the Simons Center for creating a productive atmosphere during the completion of this work. Some results of this paper were presented at that workshop. PK thanks W. Fine Institute for Theoretical Physics at University of Minnesota and Kavli Institute for Theoretical Physics at University of California Santa Barbara, where part of his work was done, for kind hospitality. NM thanks Raffaele Savelli and Jasmina Selmic for their very kind hospitality during the completion of this paper. The research of AD is supported by the National Science Foundation under Grant Numbers PHY-1316033 and PHY-0969020. The research of PK was supported by the Perimeter Institute for Theoretical Physics. Research at Perimeter Institute is supported by the Government of Canada through Industry Canada and by the Province of Ontario through the Ministry of Economic Development and Innovation. This research was supported in part by the National Science Foundation under Grant No. NSF PHY11-25915.

\appendix
\section{Mirror of $Sp(N_c)$ with $N_f$ Flavors: Checking the Duality}\label{Sec:SpGen}
\begin{figure}[H]
\begin{center}
\includegraphics[scale=0.55]{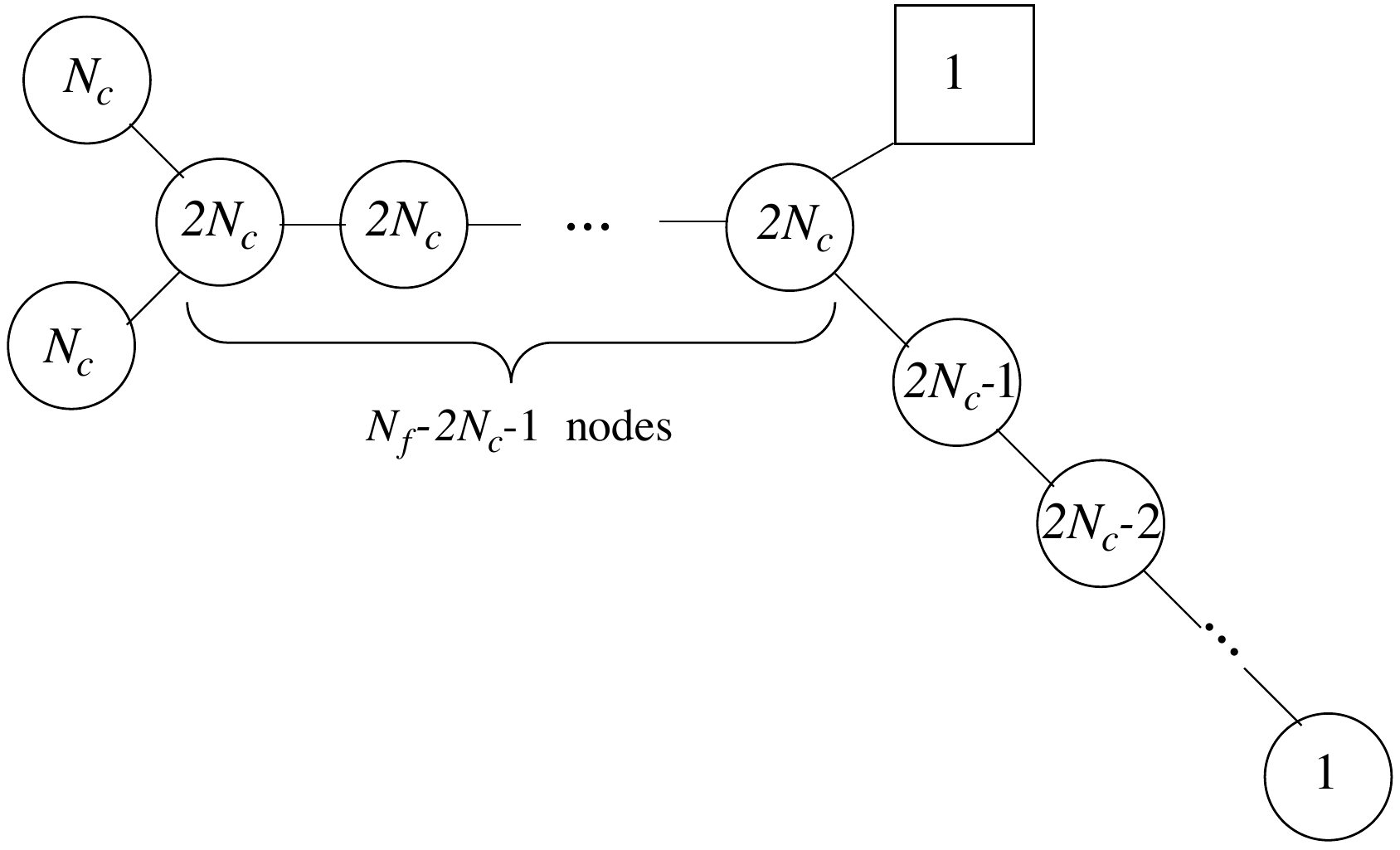}
\caption{The quiver for the mirror theory of $Sp(N_c)$ with $N_f$ flavors involving O5-planes.}
\label{fig:QuiverMirrorO5SpNcNfflv}
\end{center}
\end{figure}

In this section, we present an explicit check for the ``O5 mirror" of an $Sp(N_c)$ gauge theory with $N_f$ flavors for arbitrary $N_c$ and $N_f$.
The quiver diagram for the $O5$ mirror to the $Sp(N_c)$ theory with $N_f$ flavors is depicted in \figref{fig:QuiverMirrorO5SpNcNfflv}.
The Coulomb branch of \figref{fig:QuiverMirrorO5SpNcNfflv} is $2N_c N_f - \frac{1}{2}(2N_c)(2N_c+1)$ quanternionic dimensional, agreeing with the Higgs branch of $Sp(N_c)$ with $N_f$ flavors using sphere partition function.  The Higgs branch is $N_c$ quanternionic dimensional, agreeing with the Coulomb branch of $Sp(N_c)$ theory with $N_f$ flavors. 

The $S^3$ partition functions of the dual theories can be explicitly written ($k=N_c$ in the following formulae) as functions of the FI parameters and the fundamental masses. For the A-model, one has
\begin{equation}
\begin{split}
\CZ_A&=\int \frac{d^{N_c}s}{(2^{N_c} N_c!)} \frac{\prod_{i<j}\sinh^2{\pi(s^i-s^j)}\sinh^2{\pi(s^i+s^j)}\prod_i \sinh^2{\pi(2s^i)}}{\prod^{N_c}_{i=1}\prod^{N_f}_{a=1}\cosh{\pi(s^i+m_a)}\prod^3_{a=1}\cosh{\pi(s^i-m_a)}}\\
\end{split}
\end{equation}

\begin{equation}
\begin{split}
\CZ_B&=\int  \prod^{2}_{\alpha=1}\frac{d^{N_c}s_{\alpha}}{N_c!} \prod^{L}_{\beta=1}\frac{d^{2N_c}\widetilde{s}_{\beta}}{2N_c!} \prod^{2N_c-1}_{\gamma=1} \frac{d^{k_{\gamma}} u_{\gamma}}{k_{\gamma} !} \prod^{N_c}_{i=1,\alpha}e^{2\pi i \eta_{\alpha}s^i_{\alpha}} \prod^{2N_c}_{p=1,\beta}e^{2\pi i \widetilde{\eta}_{\beta}\widetilde{s}^p_{\beta}} \prod^{2N_c-1}_{\gamma=1} \prod^{k_{\gamma}}_{p=1} e^{2\pi i {\zeta}_{\gamma} u^p_{\gamma}}\\
&\times\frac{\prod_{i<j}\sinh^2{\pi(s^i_1-s^j_1)}\sinh^2{\pi(s^i_2-s^j_2)}}{\prod_{i,p}\cosh{\pi(s^i_1-\widetilde{s}^p_{1}+m_1)}\cosh{\pi(s^i_2-\widetilde{s}^p_{1}+m_2)} } \frac{\prod^{L}_{\beta=1}\prod_{p<l}\sinh^2{\pi(\widetilde{s}^p_{\beta}-\widetilde{s}^l_{\beta})}}{\prod^{L-1}_{\beta=1}\prod_{p,l} \cosh{\pi(\widetilde{s}^p_{\beta}-\widetilde{s}^l_{\beta+1}+\widetilde{M}_{\beta})} }\\
&\times \frac{1}{\prod^{2N_c}_{p=1}\cosh{\pi(\widetilde{s}_{L}^p+m^f)}} \times \frac{1}{\prod^{2N_c}_{p} \prod^{2N_c-1}_{l}\cosh{\pi(\widetilde{s}_{L}^p - u^l_{2N_c-1}+M_{2N_c-1})}}\\
&\times  \frac{\prod^{2N_c-1}_{\gamma=1}\prod^{k_{\gamma}}_{p<l}\sinh^2{\pi({u}^p_{\gamma}-{u}^l_{\gamma})}}{\prod^{2N_c-2}_{\gamma=1}\prod^{k_{\gamma}}_{p,l} \cosh{\pi({u}^p_{\gamma}-{u}^l_{\gamma+1}+{M}_{\gamma})} }
\end{split}
\end{equation}
In the above equation, the integer $L=N_f -2 N_c -1>0$ and the set $\{k_\gamma\}=\{k_1,k_2,..., k_{2N_c-1}\}=\{1,2,......,2N_c-2,2N_c-1\}$. For convenience, we set $\eta_1=\eta_2 $ (which we will label as $\widetilde{\eta}_0$) -- note that it is not necessary to assume this as the $N_c=2, N_f=6$ example above demonstrates. 

We also label $\widetilde{s}^p_0=(s^i_1,s^i_2)$. Note that all the masses in the partition function $Z_B$ can be eliminated by simply shifting the integration variables by constants. Therefore, we can ignore all masses in $\CZ_B$ from here on. One obtains
\begin{equation}
\begin{split}
\CZ_B=&\frac{i^{N_c}}{N_c!N_c!}\int  d^{N_c} \tau \prod^{L+1}_{\beta=0}\frac{d^{2N_c}\widetilde{s}_{\beta}}{2N_c!} \prod^{L}_{\beta=0}d^{2N_c}\widetilde{\tau}_{\beta}\\
&\times\prod^{L}_{\beta=0}\prod^{2N_c}_{p=1}e^{2\pi i \widetilde{\eta}_{\beta}\widetilde{s}^p_{\beta}} \left(\sum_{\rho} (-1)^{\rho} \prod^{N_c}_{i=1} \tanh{\pi \tau^i}e^{2\pi i \tau^i (\widetilde{s}^i_0-\widetilde{s}^{k+\rho(i)}_0)}\right)\\
&\times \prod^{L-1}_{\beta=0} \left(\sum_{\widetilde{\rho}_{\beta}}(-1)^{\widetilde{\rho}_{\beta}} \prod^{2N_c}_{p=1}\frac{e^{2\pi i \widetilde{\tau}^p_{\beta}(\widetilde{s}^p_{\beta}-\widetilde{s}^{\widetilde{\rho}_{\beta}(p)}_{\beta+1})}}{\cosh{\pi \widetilde{\tau}^p_{\beta}}}\right)  \left(\sum_{\widetilde{\rho}_{L}}(-1)^{\widetilde{\rho}_{L}} \prod^{2N_c}_{p=1}\frac{e^{2\pi i \widetilde{\tau}^p_{L}(\widetilde{s}^p_{L}-\widetilde{s}^{\widetilde{\rho}_{L}(p)}_{L+1})}}{\cosh{\pi \widetilde{s}^p_{L}}}\right)\\
&\times \prod^{2N_c}_{p<l} \sinh{\pi(\widetilde{s}^p_{L+1}-\widetilde{s}^l_{L+1})} Z_{T(U(2N_c))} \left(\widetilde{s}^p_{L+1}, \zeta_{\gamma}\right)
\end{split}
\end{equation}
where we have decomposed the partition function of the B-quiver into two parts --the truncated $D$-quiver (+ 1 fundamental hyper) and $T_{U(2N_c)}$ tail with the $U(2N_c)$ flavor symmetry gauged. The latter contribution $Z_{T(U(2N_c))} \left(\widetilde{s}^p_{L+1}, \zeta_{\gamma}\right)$ may be explicitly obtained in terms of $\{\widetilde{s}^p_{L+1}, \zeta_{\gamma}\}$. However, it is easier to write the answer in terms $\{e_{\gamma}\}$ defined in the usual way as $\zeta_{\gamma}=e_{\gamma}-e_{\gamma+1}$ \cite{Benvenuti:2011ga} 
\begin{equation}
\begin{split}
\CZ_{T(U(2N_c))} \left(\widetilde{\sigma}^p_{L+1}, e_{\gamma}\right)=&\int \prod^{2N_c-1}_{\gamma=1} \frac{d^{k_{\gamma}} u_{\gamma}}{k_{\gamma} !} \prod^{2N_c-1}_{\gamma=1} \prod^{k_{\gamma}}_{p=1} e^{2\pi i {\zeta}_{\gamma} u^p_{\gamma}}\frac{\prod^{2N_c-1}_{\gamma=1}\prod^{k_{\gamma}}_{p<l}\sinh^2{\pi({u}^p_{\gamma}-{u}^l_{\gamma})}}{\prod^{2N_c-2}_{\gamma=1}\prod^{k_{\gamma}}_{p,l} \cosh{\pi({u}^p_{\gamma}-{u}^l_{\gamma+1})} }\\
&\times  \frac{1}{\prod^{2N_c}_{p} \prod^{2N_c-1}_{l}\cosh{\pi(\widetilde{s}_{L+1}^p - u^l_{2N_c-1})}}\\
=& \sum_{\widetilde{\rho}_{L+1}} (-1)^{\widetilde{\rho}_{L+1}}\frac{i^{-N_c(2N_c-1)}e^{2\pi i \widetilde{s}^{\widetilde{\rho}_{L+1}(p)}_{L+1}\left(e_p-e_{2N_c}\right)}}{\prod^{2N_c}_{p<l} \sinh{\pi(\widetilde{s}^{p}_{L+1}-\widetilde{s}^{l}_{L+1})} \prod^{2N_c}_{p<l}\sinh{\pi(e_p-e_l)}  }\,.
\end{split}
\end{equation}
Putting together the two results we have
\begin{equation}
\begin{split}
\CZ_B=&\frac{1}{N_c!N_c!}\int  d^{N_c} \tau \prod^{L+1}_{\beta=0}\frac{d^{2N_c}\widetilde{s}_{\beta}}{2N_c!} \prod^{L}_{\beta=0}d^{2N_c}\widetilde{\tau}_{\beta}\\
&\times\prod^{L}_{\beta=0}\prod^{2N_c}_{p=1}e^{2\pi i \widetilde{\eta}_{\beta}\widetilde{s}^p_{\beta}} \left(\sum_{\rho} (-1)^{\rho} \prod^{N_c}_{i=1} \tanh{\pi \tau^i}e^{2\pi i \tau^i (\widetilde{s}^i_0-\widetilde{s}^{k+\rho(i)}_0)}\right)\\
&\times \prod^{L-1}_{\beta=0} \left(\sum_{\widetilde{\rho}_{\beta}}(-1)^{\widetilde{\rho}_{\beta}} \prod^{2N_c}_{p=1}\frac{e^{2\pi i \widetilde{\tau}^p_{\beta}(\widetilde{s}^p_{\beta}-\widetilde{s}^{\widetilde{\rho}_{\beta}(p)}_{\beta+1})}}{\cosh{\pi \widetilde{\tau}^p_{\beta}}}\right)  \left(\sum_{\widetilde{\rho}_{L}}(-1)^{\widetilde{\rho}_{L}} \prod^{2N_c}_{p=1}\frac{e^{2\pi i \widetilde{\tau}^p_{L}(\widetilde{s}^p_{L}-\widetilde{s}^{\widetilde{\rho}_{L}(p)}_{L+1})}}{\cosh{\pi \widetilde{s}^p_{L}}}\right)\\
&\times \left(\sum_{\widetilde{\rho}_{L+1}} (-1)^{\widetilde{\rho}_{L+1}}\frac{e^{2\pi i \widetilde{s}^{\widetilde{\rho}_{L+1}(p)}_{L+1}\left(e_p-e_{2N_c}\right)}}{\prod^{2N_c}_{p<l}\sinh{\pi(e_p-e_l)}}\right)
\end{split}
\end{equation}
Integrating the variables $\widetilde{s}_{\beta}$ and imposing the resulting delta functions, we obtain the following form for $Z_B$ after applying Cauchy's determinant identity
\begin{equation}
\begin{split}
\CZ_B=& \int \frac{d^{2N_c}\widetilde{\tau}_0}{(2^{N_c} N_c!)} \frac{\prod^{N_c}_{i=1} \sinh{\pi 2\widetilde{\tau}^i_0}\prod^{2N_c}_{p<l} \sinh{\pi(\widetilde{\tau}^p_0-\widetilde{\tau}^l_0)} }{\prod^{2N_c}_{p=1} \cosh{\pi \widetilde{\tau}^p_0}\cosh{\pi (\widetilde{\tau}^p_0-\widetilde{\eta}_0)} \ldots \cosh{\pi (\widetilde{\tau}^p_0-\widetilde{\eta}_0-\widetilde{\eta}_1-\ldots -\widetilde{\eta}_{L-1})}}\\
&\times \frac{1}{\prod^{2N_c}_{p,l} \cosh{\pi(\widetilde{\tau}^p_0 -(e_l-e_{2N_c} +\widetilde{\eta}_0+\widetilde{\eta}_1+\ldots+\widetilde{\eta}_{L}))}} \prod^{N_c}_{i=1} \delta(\widetilde{\tau}^i_0+\widetilde{\tau}^{k+i}_0)\\
=\CZ_A\,.
\end{split}
\end{equation}
For uniformity of notation, we define $\widetilde{\eta}_i =t_i -t_{i+1}$ where $i=0,1,..,L$.  Therefore the $N_f=L+1+2N_c$ masses of the A-model can be written in terms of the FI parameters of the A-model as follows,
\begin{equation}
\begin{split}
\begin{gathered}m_{a} = t_{a}-t_0 \; (a=0,1,2,\ldots,L),\\
m_{L+l}=e_l -e_{2N_c} +t_0 -t_{L}\; (l=1,2,\ldots,2N_c)\,.
\end{gathered}
\end{split}
\end{equation}
Note that the mirror map closely resembles that of linear quivers, i.e. up to an additive constant we have $m_a \leftrightarrow t_a$, $m_{L+l} \leftrightarrow e_l$.

\bibliography{cpn1}
\bibliographystyle{JHEP}

\end{document}